\documentclass[journal,twocolumn]{IEEEtran}
\usepackage{graphicx,comment}

\usepackage{array,graphicx,epsfig,amssymb,amsfonts,multirow}
\usepackage{amsmath,amsthm}
\usepackage{epsfig,eufrak,eucal,hhline}
\usepackage{colortbl,color,nonfloat,twoopt,stmaryrd}
\usepackage[usenames,dvipsnames,svgnames,table]{xcolor}
\usepackage{graphics}
\usepackage[geometry]{ifsym}
\usepackage{mathrsfs}
\usepackage{nohyperref,url}
\usepackage{cite}%\usepackage[style=numeric-comp]{biblatex}
\usepackage{epstopdf}
\usepackage{subfig,float}
\floatstyle{plaintop}
\restylefloat{table}
\usepackage{caption}
\usepackage{enumitem}
\usepackage[normalem]{ulem}

\usepackage{tikz}
\usetikzlibrary{arrows,shapes,trees,patterns,snakes}
\usetikzlibrary{decorations.markings}
\usetikzlibrary{calc}
\tikzstyle{vecArrow} = [thick, decoration={markings,mark=at position
   1 with {\arrow[semithick]{open triangle 60}}},
   double distance=1.4pt, shorten >= 5.5pt,
   preaction = {decorate},
   postaction = {draw,line width=1.4pt,white,shorten >= 4.5pt}]
\tikzstyle{innerWhite} = [semithick, white,line width=1.4pt, shorten >= 4.5pt]
\tikzstyle{vecArrowDotted} = [thick, dotted, decoration={markings,mark=at position
   1 with {\arrow[semithick]{open triangle 60}}},
   double distance=1.4pt, shorten >= 5.5pt,
   preaction = {decorate},
   postaction = {draw,line width=1.4pt, white,shorten >= 4.5pt}]
\tikzstyle{innerWhite} = [semithick, white,line width=1.4pt, shorten >= 4.5pt]
\makeatletter
\tikzset{
    rect/.style n args={4}{
        draw=none,
        rectangle,
        append after command={
            \pgfextra{%
                \pgfkeysgetvalue{/pgf/outer xsep}{\oxsep}
                \pgfkeysgetvalue{/pgf/outer ysep}{\oysep}
                \def\arg@one{#1}
                \def\arg@two{#2}
                \def\arg@three{#3}
                \def\arg@four{#4}
                \begin{pgfinterruptpath}
                    \ifx\\#1\\\else
                        \draw[draw,#1] ([xshift=-\oxsep,yshift=+\pgflinewidth]\tikzlastnode.south east) edge ([xshift=-\oxsep,yshift=0\ifx\arg@two\@empty-\pgflinewidth\fi]\tikzlastnode.north east);
                    \fi\ifx\\#2\\\else
                        \draw[draw,#2] ([xshift=-\pgflinewidth,yshift=-\oysep]\tikzlastnode.north east) edge ([xshift=0\ifx\arg@three\@empty+\pgflinewidth\fi,yshift=-\oysep]\tikzlastnode.north west);
                    \fi\ifx\\#3\\\else
                        \draw[draw,#3] ([xshift=\oxsep,yshift=0-\pgflinewidth]\tikzlastnode.north west) edge ([xshift=\oxsep,yshift=0\ifx\arg@four\@empty+\pgflinewidth\fi]\tikzlastnode.south west);
                    \fi\ifx\\#4\\\else
                        \draw[draw,#4] ([xshift=0+\pgflinewidth,yshift=\oysep]\tikzlastnode.south west) edge ([xshift=0\ifx\arg@one\@empty-\pgflinewidth\fi,yshift=\oysep]\tikzlastnode.south east);
                    \fi
                \end{pgfinterruptpath}
            }
        }
    },
    rect'/.style n args={4}{
        rectangle,
        append after command={
            \pgfextra{%
                \pgfkeysgetvalue{/pgf/outer xsep}{\oxsep}
                \pgfkeysgetvalue{/pgf/outer ysep}{\oysep}
                \begin{pgfinterruptpath}
                    \ifx\\#1\\\else
                        \draw[draw,#1] ([xshift=-\oxsep,yshift=0]\tikzlastnode.south east) edge ([xshift=-\oxsep,yshift=0]\tikzlastnode.north east);
                    \fi\ifx\\#2\\\else
                        \draw[draw,#2] ([xshift=-\pgflinewidth,yshift=-\oysep]\tikzlastnode.north east) edge ([xshift=0+\pgflinewidth,yshift=-\oysep]\tikzlastnode.north west);
                    \fi\ifx\\#3\\\else
                        \draw[draw,#3] ([xshift=\oxsep,yshift=0-\pgflinewidth]\tikzlastnode.north west) edge ([xshift=\oxsep,yshift=0+\pgflinewidth]\tikzlastnode.south west);
                    \fi\ifx\\#4\\\else
                        \draw[draw,#4] ([xshift=0+\pgflinewidth,yshift=\oysep]\tikzlastnode.south west) edge ([xshift=0-\pgflinewidth,yshift=\oysep]\tikzlastnode.south east);
                    \fi
                \end{pgfinterruptpath}
            }
        }
    },
    dontshortenme/.style={
        shorten >=0pt,
        shorten <=0pt
    },
    rect''/.style n args={4}{
        draw=none,
        rectangle,
        append after command={
            \pgfextra{%
                \pgfkeysgetvalue{/pgf/outer xsep}{\oxsep}
                \pgfkeysgetvalue{/pgf/outer ysep}{\oysep}
                \def\my@path{\path[shorten >=\pgflinewidth,shorten <=\pgflinewidth] ([xshift=-\oxsep]\tikzlastnode.south east) edge}
                \def\arg@{#1}
                \ifx\arg@\@empty
                    \def\arg@{draw=none}
                \fi
                \eappto\my@path{[\arg@] }
                \appto\my@path{ ([xshift=-\oxsep]\tikzlastnode.north east)
                                          ([yshift=-\oysep]\tikzlastnode.north east) edge }
                \def\arg@{#2}
                \ifx\arg@\@empty
                    \def\arg@{draw=none}
                \fi
                \eappto\my@path{[\arg@] }
                \appto\my@path{ ([yshift=-\oysep]\tikzlastnode.north west)
                                          ([xshift=\oxsep] \tikzlastnode.north west) edge }
                \def\arg@{#3}
                \ifx\arg@\@empty
                    \def\arg@{draw=none}
                \fi
                \eappto\my@path{[\arg@] }
                \appto\my@path{ ([xshift=\oxsep]\tikzlastnode.south west)
                                          ([yshift=\oysep] \tikzlastnode.south west) edge }
                \def\arg@{#4}
                \ifx\arg@\@empty
                    \def\arg@{draw=none}
                \fi
                \eappto\my@path{[\arg@] }
                \appto\my@path{ ([yshift=\oysep]\tikzlastnode.south east);}
                \begin{pgfinterruptpath}
                    \my@path
                \end{pgfinterruptpath}
            }
        }
    }
}
\makeatother

\newcommand{\appsection}[1]{\let\oldthesection\thesection\renewcommand{\thesection}{Appendix
\oldthesection}\section{#1}\let\thesection\oldthesection}

\theoremstyle{remark}

\newtheorem{prop}{Proposition}
\newtheorem{lem}{Lemma}

\newtheorem{mydef}{Definition}

\newcommand{\Ref}[1]{(\ref{#1})}

\newcommand{\PropRef}[1]{Proposition~\ref{#1}}
\newcommand{\PropsRef}[2]{Propositions~\ref{#1} and~\ref{#2}}

\newcommand{\LemRef}[1]{Lemma~\ref{#1}}
\newcommand{\LemsRef}[2]{Lemmas~\ref{#1} and~\ref{#2}}

\newcommand{\DefRef}[1]{Definition~\ref{#1}}

\newcommand{\SecRef}[1]{Section~\ref{#1}}

\newcommand{\AppRef}[1]{Appendix~\ref{#1}}

\def \TReq{\triangleq}

\def \nN{\nonumber}

\makeatletter
\renewcommand*\env@matrix[1][c]{\hskip -\arraycolsep
  \let\@ifnextchar\new@ifnextchar
  \array{*\c@MaxMatrixCols #1}}
\makeatother

\newcommandtwoopt{\MONO}[2][][]{m_{\!#1}(#2)}                            % Monomial function in G
\newcommandtwoopt{\MONOL}[2][][]{\mathring{m}_{#1}(#2)}                  % Monomial function in L(G)
\newcommandtwoopt{\EDGE}[2][][]{e_{#1}^{#2}}                             % Edge
\newcommandtwoopt{\NNNN}[2][][]{N_{#2#1}}                                %%%%% Nji
\newcommandtwoopt{\NNNNh}[2][][]{\hat{N}_{#2#1}}                         %%%%% \hat{N}ji
\newcommandtwoopt{\NNNNt}[2][][]{\tilde{N}_{#2#1}}                       %%%%% \tilde{N}ji
\newcommandtwoopt{\mincut}[2][][]{{\small\mathsf{mincut}}(#1;#2)}       % Minimum Edge Cut
\newcommandtwoopt{\onecut}[2][][]{{\small\mathsf{1cut}}(#1;#2)}         % 1-edge Cut
\newcommandtwoopt{\EC}[2][][]{{\small\mathsf{EC}}(#1;#2)}               % Edge Cut
\newcommandtwoopt{\VC}[2][][]{{\small\mathsf{VC}}(#1;#2)}               % Vertex Cut
\newcommandtwoopt{\GCD}[2][][]{{\small\mathsf{GCD}}(\,#1,#2)}           % Greatest Common Divisor
\newcommandtwoopt{\FF}[2][][]{\mathbb{F}_{#1}^{#2}}                       % Finite Field

\def \ET#1{\mathbb{E}\left[#1\right]}

\newcommandtwoopt{\V}[2][][]{\textrm{V}_{#1}^{#2}}

\def\msf#1{\mathtt{#1}}

\def\b{{\mathbf{b}}}
\def\NETori{{\small\mathsf{FTs}}}

\def\pnumb#1{x_\b^{(#1)}}

\def\DIGIT#1{\overline{\b}_{#1}}

\def\RK#1{\mathsf{rank}(#1)}

\def \p#1#2{p_{#1\rightarrow#2}}

\def \prOa#1#2#3{p_{#1\rightarrow\overline{#2}#3}}
\def \prOb#1#2#3{p_{#1\rightarrow#2\overline{#3}}}

\def \PROB{\mathsf{Prob}}

\def \SP#1{S_{#1}}
\def \SPt#1{S_{#1}(t)}
\def \SPn#1{S_{#1}(n)}
\def \SPtT#1#2{S_{#1}(#2)}
\def \SPtPREV#1{S_{#1}(t-1)}

\def \Mtot{\Omega}

\def \SPAN{\mathsf{span}}
\def \ELEM#1{\bold{e}_{#1}}

\def \boldWB#1{\bold{W}_{\!#1}}

\def \boldWtotV{\bold{W}}

\def \boldZ{\bold{Z}}
\def \boldZt{\boldZ(t)}

\def \Z#1#2{Z_{#1\rightarrow#2}}
\def \Zt#1#2{\Z{#1}{#2}(t)}

\def \Y#1#2{Y_{#1\rightarrow#2}}
\def \Yt#1#2{\Y{#1}{#2}(t)}

\def \X#1{X_{#1}}
\def \Xt#1{\X{#1}(t)}

\def \Ct{\bold{c}_t}
\def \CtT#1{\bold{c}_{#1}}

\def \SCH{\sigma}
\def \SCHt{\SCH(t)}

\def \Rsum{R_{\Sigma}}

\def \CtT#1{\bold{c}_{#1}}
\def \Ct{\CtT{t}}

\def \tableSTARTreduce{\vspace*{-\baselineskip}\vspace*{-\baselineskip}}
\def \tableENDreduce{\vspace*{-\baselineskip}\vspace*{-\baselineskip}\vspace*{+5pt}}

\def \RL#1{\mathsf{RL}_{\{#1\}}}

\def \boldWBtot{\bold{W}}
\def \boldWB#1{\bold{W}_{\!#1}}
\def \boldZn#1{\bold{Z}_{#1}}
\def \boldZnt#1{\boldZn#1(t)}
\def \nSRP{\{s,r\}}
\def \MB#1{\Omega_{#1}}
\def \RB#1{R_{#1}}
\def \SRPTYPEb#1{{\small\mathsf{TYPE}^{(#1)}_\b}}

\def \NEToriR{r{\small\mathsf{FTs}}}
\def \prOc#1#2#3#4{p_{#1\rightarrow#2\overline{#3}#4}}
\def \prOd#1#2#3#4{p_{#1\rightarrow\overline{#2}#3\overline{#4}}}
\def \pSRPsim#1{p_{#1}}
\def \pSRPsimT#1#2{\pSRPsim{#1}(#2)}

\def \snumUC#1{s_{{\small\mathsf{UC}}}^{#1}}

\def \snumPM#1{s_{{\small\mathsf{PM1}}}^{#1}}
\def \snumAM#1{s_{{\small\mathsf{PM2}}}^{#1}}

\def \snumRC#1{s_{{\small\mathsf{RC}}}^{#1}}
\def \snumCX#1{s_{{\small\mathsf{CX}};#1}}
\def \snumDX#1#2{s_{{\small\mathsf{DX}}}^{#1#2}}
\def \snumSX#1#2{s_{{\small\mathsf{SX};#2}}^{#1}}

\def \rnumUC#1{r_{{\small\mathsf{UC}}}^{#1}}
\def \rnumRC{r_{{\small\mathsf{RC}}}}
\def \rnumOX#1{r_{{\small\mathsf{XT}}}^{#1}}
\def \rnumCX#1{r_{{\small\mathsf{CX}}}^{#1}}
\def \rnumDX#1#2{r_{{\small\mathsf{DT}}}^{#1#2}}

\def \wnumUC#1#2{w_{{\small\mathsf{UC}}}^{(#1):#2}}
\def \wnumRC#1{w_{{\small\mathsf{RC}}}^{(#1)}}
\def \wnumOX#1#2{w_{{\small\mathsf{XT}}}^{(#1)#2}}
\def \wnumCX#1#2{w_{{\small\mathsf{CX}}}^{(#1)#2}}
\def \wnumDX#1#2#3{w_{{\small\mathsf{DT}}}^{(#1):#2#3}}

\def \SRPschemeFIVE{Intra-Flow Network Coding only}
\def \SRPschemeFOUR{Always Relaying with NC}
\def \SRPschemeTHREE{Always Relaying with routing}
\def \SRPschemeTWO{\cite{GeorgiadisTassiulas:NetCod09} without Relaying}
\def \SRPschemeONE{Routing without Relaying}

\def \SRPQe#1{Q^{#1}_{\phi}}
\def \SRPQr#1{Q^{#1}_{\!\{\!r\!\}}}
\def \SRPQb#1#2{Q^{m|#2}_{\!\{\!d_{#2\!}\}|\{\!r\!\}}}
\def \SRPQd#1{Q^{#1}_{\footnotesize\mathsf{dec}}}

\def \SRPQm{Q_{\small\mathsf{mix}}}
\def \SRPQx#1#2{Q^{[#1]}_{\!\{\!rd_{#2}\!\}}}
\def \SRPQsx#1#2#3{Q^{#1#3}_{\!\{\!d_{#2}\!\}}}
\def \SRPQsxSP#1#2{Q^{(#1)|#1}_{\!\{\!d_{#2}\!\}|\{\!r\!\}}}
\def \SRPQstar{Q^{m_{\mathsf{CX}}}_{\!\{\!r\!\}}}

\def \SRPA#1{A_{#1}}
\def \SRPAt#1{\SRPA{#1}(t)}
\def \SRPAtT#1#2{\SRPA{#1}(#2)}
\def \SRPAn#1{\SRPA{#1}(n)}

\usepackage{tikz}%use PDFtexify(Ctrl+Shift+P) for PDF conversion

\def \Rsum{(\RB{1}\!+\!\RB{2})}

\begin{document}
\IEEEoverridecommandlockouts
\title{Linear Network Coding Capacity Region of The Smart Repeater with Broadcast Erasure Channels}

\author{\IEEEauthorblockN{Jaemin Han and Chih-Chun Wang}
%Center of Wireless Systems and Applications (CWSA)\\
\\School of Electrical and Computer Engineering, Purdue University, USA\\
\IEEEauthorblockA{\{han83,chihw\}@purdue.edu}
}

\maketitle

% No citation in Abstract but names such as Jaemin et al.
% Usually, there is no sentence starting with We or You.

\begin{abstract} This work considers the smart repeater network where a single source $s$ wants to send two independent packet streams to destinations $\{d_1,d_2\}$ with the help of relay $r$. The transmission from $s$ or $r$ is modeled by packet erasure channels: For each time slot, a packet transmitted by $s$ may be received, with some probabilities, by a random subset of $\{d_1,d_2,r\}$; and those transmitted by $r$ will be received by a random subset of $\{d_1,d_2\}$. Interference is avoided by allowing at most one of $\{s,r\}$ to transmit in each time slot. One example of this model is any cellular network that supports two cell-edge users when a relay in the middle uses the same downlink resources for throughput/safety enhancement.

In this setting, we study the capacity region of $(\RB{1},\RB{2})$ when allowing linear network coding (LNC). The proposed LNC inner bound introduces more advanced packing-mixing operations other than the previously well-known butterfly-style XOR operation on overheard packets of two co-existing flows. A new LNC outer bound is derived by exploring the inherent algebraic structure of the LNC problem. Numerical results show that, with more than 85\% of the experiments, the relative sum-rate gap between the proposed outer and inner bounds is smaller than 0.08\% under the strong-relaying setting and 0.04\% under arbitrary distributions, thus effectively bracketing the LNC capacity of the smart repeater problem.
\end{abstract}


%------------------------------------------------------------------------------------------%
%------------------------------------------------------------------------------------------%
%-------------------------- Section I: Introduction ---------------------------------------%
%------------------------------------------------------------------------------------------%
%------------------------------------------------------------------------------------------%
\section{Introduction}\label{sec:SRP:Intro}

Increasing throughput/connectivity within scarce resources has been the main motivation for modern wireless communications. Among the various proposed techniques, the concept of {\em relaying} has attracted much attention as a cost-effective enabler to extend the network coverage and capacity. In recent 5G discussions, relaying became one of the core parts for the future cellular architecture including techniques of small cell managements and device-to-device communications between users \cite{IMCOMM:5G:14}.

In network information theory, many intelligent and cooperative relaying strategies have been devised such as decode-and-forward/compress-and-forward for relay networks \cite{CoverGamal:IT79,KramerGastparKupta:IT05}, network coding for noiseless networks \cite{AhlswedeCaiLiYeung:IT00,LiYeungCai:IT03}, and general noisy network coding for discrete memoryless networks \cite{LimKimElGamalChung:IT11}. Among them, network coding has emerged as a promising technique for a practical wireless networking solution, which models the underlying wireless channels by a simple but non-trivial random packet erasure network. That is, each node is associated with its own broadcast packet erasure channel (PEC). Namely, each node can choose a symbol $X\in\FF[q]$ from some finite field $\FF[q]$, transmits $X$, and a random subset of receivers will receive the packet. In this setting, \cite{DanaEffros:IT06} proved that the {\em linear network coding} (LNC), operating only by ``linear" packet-mixings, suffices to achieve the single-multicast capacity. Moreover, recent wireless testbeds have also demonstrated substantial LNC throughput gain for multiple-unicasts over the traditional store-and-forward 802.11 routing protocols \cite{KattiKatabiMedard:SIGCOMM06,KoutsonikolasWangHu:TON11}.

\begin{figure}
        \begin{tikzpicture}
        %%------ 1st figure
        \node[draw,circle,inner sep=0pt,minimum size=12pt] (s) at (0,0) {$s$};
        \draw[densely dotted] ([shift={(-60:0.3)}]s) arc (-60:60:0.3);
        \draw[densely dotted] ([shift={(-60:0.4)}]s) arc (-60:60:0.4);
        \node[draw,circle,inner sep=0pt,minimum size=12pt] (r) at ([shift={(1,0,0)}]s) {$r$};
        \draw[densely dotted] ([shift={(-60:0.3)}]r) arc (-60:60:0.3);
        \draw[densely dotted] ([shift={(-60:0.4)}]r) arc (-60:60:0.4);
        \node[draw,circle,inner sep=0pt,minimum size=12pt] (d1) at ([shift={(15:2.25)}]s) {$d_1$};
        \node[draw,circle,inner sep=0pt,minimum size=12pt] (d2) at ([shift={(-15:2.25)}]s){$d_2$};
        \node[below=0.8cm, align=flush left, text width=5cm] at ([shift={(1.6,0)}]s) {{\small (a) $4$-node $2$-hop relay network}};
        %%------ 2nd figure
        \node[draw,circle,inner sep=0pt,minimum size=12pt] (s) at (3.3,0) {$s$};
        \node[draw,rectangle,inner sep=1pt, minimum size=12pt] (sPEC) at ([shift={(1.0,0)}]s) {{\footnotesize\textsf{s-PEC}}};
        \node[draw,circle,inner sep=0pt,minimum size=12pt] (r) at ([shift={(2,0,0)}]s) {$r$};
        \node[draw,rectangle,inner sep=1pt, minimum size=12pt] (rPEC) at ([shift={(1.0,0)}]r) {{\footnotesize\textsf{r-PEC}}};
        \node[draw,circle,inner sep=0pt,minimum size=12pt] (d1) at ([shift={(15:2.25)}]r) {$d_1$};
        \node[draw,circle,inner sep=0pt,minimum size=12pt] (d2) at ([shift={(-15:2.25)}]r){$d_2$};
        \draw[ultra thick,->] (s) -- (sPEC);
        \draw[->] (sPEC) .. controls([shift={(-2.1,0)}]d1) .. (d1);
        \draw[->] (sPEC) .. controls([shift={(-2.1,0)}]d2) .. (d2);
        \draw[->] (sPEC) -- (r);
        \draw[ultra thick,->] (r) -- (rPEC);
        \draw[->] (rPEC) .. controls([shift={(0.7,0)}]rPEC) .. (d1);
        \draw[->] (rPEC) .. controls([shift={(0.7,0)}]rPEC) .. (d2);
        \node[below=0.8cm, align=flush left, text width=5cm] at ([shift={(2.65,0)}]s) {{\small (b) The PEC network model}};
        \end{tikzpicture}
        \\
        \begin{tikzpicture}
        %%------ 3rd figure
        \node[draw,circle,inner sep=0pt,minimum size=12pt] (n1) at (0,0) {$s$};
        \node[draw,rectangle,inner sep=1pt, minimum size=5pt] (PEC1) at ([shift={(0.75,0)}]n1) {};
        \node[draw,circle,inner sep=0pt,minimum size=12pt] (n2) at ([shift={(25:1.5)}]n1) {$d_1$};
        \node[draw,circle,inner sep=0pt,minimum size=12pt] (n3) at ([shift={(-25:1.5)}]n1) {$d_2$};
        \draw[ultra thick,->] (n1) -- (PEC1);
        \draw[->] (PEC1) .. controls([shift={(0.9,0)}]n1) .. (n2);
        \draw[->] (PEC1) .. controls([shift={(0.9,0)}]n1) .. (n3);
        \node[anchor=center] at ([shift={(0.2,0.4)}]n1) {{\footnotesize $[X\!\!+\!\!Y]$}};
        \node[anchor=west] at ([shift={(0.2,0)}]n2) {{\footnotesize $Y$}};
        \node[anchor=west] at ([shift={(0.2,0)}]n3) {{\footnotesize $X$}};
        \node[below=1.2cm, anchor=west] at ([shift={(-1.2,0)}]n1) {{\small (c) A $1$-to-$2$ PEC without relay}};
        %%------ 4th figure
        \node[draw,circle,inner sep=0pt,minimum size=12pt] (s1) at (4,0.5) {$s_1$};
        \node[draw,rectangle,inner sep=1pt, minimum size=5pt] (s1PEC) at ([shift={(0.75,0)}]s1) {};
        \node[draw,circle,inner sep=0pt,minimum size=12pt] (s2) at (4,-0.5) {$s_2$};
        \node[draw,rectangle,inner sep=1pt, minimum size=5pt] (s2PEC) at ([shift={(0.75,0)}]s2) {};
        \node[draw,circle,inner sep=0pt,minimum size=12pt] (r) at ([shift={(1.6,-0.5)}]s1) {$r$};
        \node[draw,rectangle,inner sep=1pt, minimum size=5pt] (rPEC) at ([shift={(0.75,0)}]r) {};
        \node[draw,circle,inner sep=0pt,minimum size=12pt] (d1) at ([shift={(25:1.5)}]r) {$d_1$};
        \node[draw,circle,inner sep=0pt,minimum size=12pt] (d2) at ([shift={(-25:1.5)}]r){$d_2$};
        \draw[ultra thick,->] (s1) -- (s1PEC);
        \draw[->] (s1PEC) .. controls([shift={(-2.1,0.2)}]d1) and ([shift={(-0.5,0.2)}]d1) .. (d1);
        \draw[->] (s1PEC) .. controls([shift={(-2.0,0)}]d2) .. (d2);
        \draw[->] (s1PEC) -- (r);
        \draw[ultra thick,->] (s2) -- (s2PEC);
        \draw[->] (s2PEC) .. controls([shift={(-2.0,0)}]d1) .. (d1);
        \draw[->] (s2PEC) .. controls([shift={(-2.1,-0.2)}]d2) and ([shift={(-0.5,-0.2)}]d2) .. (d2);
        \draw[->] (s2PEC) -- (r);
        \draw[ultra thick,->] (r) -- (rPEC);
        \draw[->] (rPEC) .. controls([shift={(0.2,0)}]rPEC) .. (d1);
        \draw[->] (rPEC) .. controls([shift={(0.2,0)}]rPEC) .. (d2);
        \node[anchor=west] at ([shift={(-0.7,0)}]s1) {{\footnotesize $X$}};
        \node[anchor=west] at ([shift={(-0.7,0)}]s2) {{\footnotesize $Y$}};
        \node[anchor=center] at ([shift={(0.2,0.4)}]r) {{\footnotesize $[X\!\!+\!\!Y]$}};
        \node[anchor=west] at ([shift={(0.2,0)}]d1) {{\footnotesize $Y$}};
        \node[anchor=west] at ([shift={(0.2,0)}]d2) {{\footnotesize $X$}};
        \node[below=1.2cm, anchor=west] at ([shift={(-0.6,-0.5)}]s1) {{\small (d) A 2-flow wireless butterfly}};
        \end{tikzpicture}
    \caption{\footnotesize The $2$-flow Smart Repeater Network and its subset scenarios}
    \label{fig:SRP:model}
\end{figure}
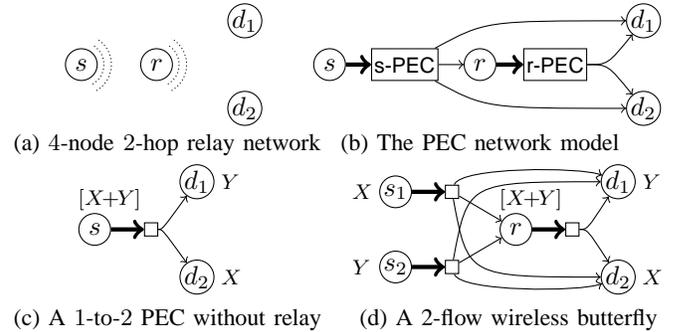

Motivated by these results, we are interested in finding an optimal or near-optimal LNC strategy for wireless relaying networks. To simplify the analysis, we consider a $4$-node $2$-hop network with one source $s$, two destinations $\{d_1,d_2\}$, and a common relay $r$ inter-connected by two broadcast PECs. See Fig.~\ref{fig:SRP:model}(a-b) for details. We assume time-sharing between $s$ and $r$ so that interference is fully avoided, and assume the causal packet ACKnowledgment feedback \cite{KattiKatabiMedard:SIGCOMM06,KoutsonikolasWangHu:TON11,GeorgiadisTassiulas:NetCod09,KuoWang:ISIT11,Wang:IT12a,Wang:IT12b,WangLove:ISIT12,Wang:ITW12,GatzianasGeorgiadisTassiulas:IT13,Wang:ISIT13,KuoWang:IT13,WangHan:IT14,KuoWang:INFOCOM14,PapadopoulosGeorgiadis:Allerton14,HanWang:ISIT15}.
In this way, we can concentrate on how the relay $r$ and source $s$ can jointly exploit the broadcast channel diversity within the network.

When relay $r$ is not present, Fig.~\ref{fig:SRP:model}(b) collapses to Fig.~\ref{fig:SRP:model}(c), the $2$-receiver broadcast PEC. It was shown in \cite{GeorgiadisTassiulas:NetCod09} that a simple LNC scheme is capacity-achieving. The idea is to exploit the wireless diversity created by random packet erasures, i.e., overhearing packets of other flows. Whenever a packet $X$ intended for $d_1$ is received only by $d_2$ and a packet $Y$ intended for $d_2$ is received only by $d_1$, $s$ can transmit their linear mixture $[X+Y]$ to benefit both receivers simultaneously. This simple but elegant ``butterfly-style" LNC operation achieves the Shannon capacity of Fig.~\ref{fig:SRP:model}(c) \cite{GeorgiadisTassiulas:NetCod09}. Another related scenario is a $2$-flow wireless butterfly network in Fig.~\ref{fig:SRP:model}(d) that contains two separate sources $s_1$ and $s_2$ instead of a single source $s$ as in our setting. In this butterfly scenario, two separate sources are not coordinating with each other and thus each source can only mix packets of their own flow. \cite{KuoWang:IT13} showed that the same butterfly-style LNC is no longer optimal but very close to optimal. In contrast, in our setting of Fig.~\ref{fig:SRP:model}(b), the two flows are originating from the same source $s$. Therefore, $s$ can perform ``inter-flow NC'' to further improve the performance. As we will see, relay $r$ should not just ``forward'' the packets it has received and need to actively perform coding in order to approach the capacity. This is why we call such a scenario the smart repeater problem.

{\bf Contributions:} This work investigates the LNC capacity region $(\RB{1},\RB{2})$ of the smart repeater network. The outer bound is proposed by leveraging upon the algebraic structure of the underlying LNC problem. For the achievability scheme, we show that the classic butterfly-style is far from optimality and propose new LNC operations that lead to close-to-optimal performance. By numerical simulations, we demonstrate that the proposed outer/inner bounds are very close, thus effectively bracketing the LNC capacity of the smart repeater problem.

\section{Problem Definition And Useful Notations}\label{sec:SRP:ProblemDef-UsefulNot}

\subsection{Problem Formation for The Smart Repeater Network}\label{sec:SRP:ProblemDef}

The $2$-flow wireless smart repeater network with broadcast PECs, see Fig.~\ref{fig:SRP:model}(b), can be modeled as follows. Consider two traffic rates $(\RB{1},\RB{2})$ and assume slotted transmissions. Within a total budget of $n$ time slots, source $s$ would like to send $n \RB{k}$ packets, denoted by a row vector $\boldWB{k}$, to destination $d_k$ for all $k\!\in\!\{1,2\}$ with the help of relay $r$. Each packet is chosen uniformly randomly from a finite field $\FF[q]$ with size $q\!>\!0$. To that end, we denote $\boldWtotV\triangleq(\boldWB{1}, \boldWB{2})$ as an $n\Rsum$-dimensional row vector of all the packets, and define the linear space $\Mtot\triangleq(\FF[q])^{n\Rsum}$ as the {\em overall message/coding space}.

To represent the reception status, for any time slot $t\in\{1,\cdots,n\}$, we define two {\em channel reception status vectors}:
\begin{align*}
\boldZnt{s} & = (\Zt{s}{d_1},\Zt{s}{d_2},\Zt{s}{r})\in\{1,\ast\}^3, \\
\boldZnt{r} & = (\Zt{r}{d_1},\Zt{r}{d_2})\in\{1,\ast\}^2,
\end{align*}
where ``$1$" and ``$\ast$" represent successful reception and erasure, respectively. For example, $\Zt{s}{d_1} = 1$ and $\ast$ represents whether $d_1$ can receive the transmission from source $s$ or not at time slot $t$. We then use $\boldZt \TReq (\boldZnt{s}, \boldZnt{r})$ to describe the $5$-dimensional channel reception status vector of the entire network. We also assume that $\boldZt$ is memoryless and stationary, i.e., $\boldZt$ is independently and identically distributed over the time axis $t$.

We assume that either source $s$ or relay $r$ can transmit at each time slot, and express the {\em scheduling decision} by $\SCHt\!\in\!\nSRP$. For example, if $\SCHt=s$, then source $s$ transmits a packet $\Xt{s}\in\FF[q]$; and only when $\Zt{s}{h}=1$, node $h$ (one of $\{d_1,d_2,r\}$) will receive $\Yt{s}{h}=\Xt{s}$. In all other cases, node $h$ receives an erasure $\Yt{s}{h}=\ast$. The reception $\Yt{r}{h}$ of relay $r$'s transmission is defined similarly.

Assuming that the $5$-bit $\boldZt$ vector is broadcast to both $s$ and $r$ after each packet transmission through a separate control channel, a {\em linear network code} contains $n$ scheduling functions
\begin{equation}\label{eq:SRP:sch}
\forall\,t\in\{1,\cdots,n\},\;\;\SCHt = f_{\sigma,t}([\boldZ]_{1}^{t-1}),
\end{equation}
where we use brackets $[\,\cdot\,]_{1}^{\tau}$ to denote the collection from time $1$ to $\tau$. Namely, at every time $t$, scheduling is decided based on the network-wide channel state information (CSI) up to time $(t-1)$. If source $s$ is scheduled, then it can send a linear combination of any packets. That is,
\begin{equation}\label{eq:SRP:lin_enc_src}
\text{If $\SCHt=s$, then}~\Xt{s} = \Ct\boldWBtot^{\top} \text{ for some } \Ct\in\Mtot,
\end{equation}
where $\Ct$ is a row coding vector in $\Mtot$. The choice of $\Ct$ depends on the past CSI vectors $[\boldZ]_{1}^{t-1}$, and we assume that $\Ct$ is known causally to the entire network.\footnote{Coding vector $\Ct$ can either be appended in the header or be computed by the network-wide causal CSI feedback $[\boldZ]_1^{t\!-\!1}$.} Therefore, decoding can be performed by simple Gaussian elimination.

We now define two important linear space concepts: The {\em individual message subspace} and the {\em knowledge subspace.} To that end, we first define $\ELEM{l}$ as an $n\Rsum$-dimensional elementary row vector with its $l$-th coordinate being one and all the other coordinates being zero. Recall that the $n\Rsum$ coordinates of a vector in $\Mtot$ can be divided into $2$ consecutive ``intervals'', each of them corresponds to the information packets $\boldWB{k}$ for each flow from source to destination $d_k$. We then define the {\em individual message subspace} $\MB{k}$:
\begin{equation}\label{def:SRP:MP}
\MB{k}\triangleq\SPAN\{\ELEM{l} : l\in\text{``interval'' associated to $\boldWB{k}$} \},
\end{equation}
That is, $\MB{k}$ is a linear subspace corresponding to any linear combination of $\boldWB{k}$ packets. By \eqref{def:SRP:MP}, each $\MB{k}$ is a linear subspace of the overall message space $\Mtot$ and $\RK{\MB{k}}=n\RB{k}$.

We define the knowledge space for $\{d_1,d_2,r\}$. The knowledge space $\SPt{h}$ in the end of time $t$ is defined by
\begin{align}
& \SPt{h}\triangleq \SPAN\{\bold{c}_{\tau}\!: \forall\tau\!\leq\!t \text{ s.t. node } h \text{ receives the linear} \nonumber \\
& \qquad\;\;\; \text{combination } (\bold{c}_{\tau}\!\cdot\!\boldWtotV^{\top}) \text{ successfully in time }\tau \} \label{def:SRP:SP}
\end{align}
where $h\!\in\!\{d_1,d_2,r\}$. For example, $\SPt{r}$ is the linear space spanned by the packets successfully delivered from source to relay up to time $t$. $\SPt{d_1}$ is the linear space spanned by the packets received at destination $d_1$ up to time $t$, either transmitted by source or by relay.

For shorthand, we use $\SPt{1}$ and $\SPt{2}$ instead of $\SPt{d_1}$ and $\SPt{d_2}$, respectively. Then, by the above definitions, we quickly have that destination $d_k$ can decode the desired packets $\boldWB{k}$ as long as $\SPn{k}\supseteq \MB{k}$. That is, when the knowledge space in the end of time $n$ contains the desired message space.

%%-- Back up for the original writings
%We define the knowledge space for $\{d_1,d_2,r\}$. To that end, we first define the {\em reception subspace} in the end of time $t$ by \vspace{-13pt}
%\begin{align}
%& \!\!\!\RPt{h}\triangleq \SPAN\{\bold{c}_{\tau}\!: \forall\tau\!\leq\!t \text{ s.t. } \SCHtT{\tau}\!\neq\!h,\;Z_{\SCHtT{\tau}\rightarrow h}(\tau)\!= \!1, \nonumber \\
%& \qquad\qquad\quad\;\;\; \text{ and } Y_{\SCHtT{\tau}\rightarrow h}(\tau)\!=\!X_{\SCHtT{\tau}}(\tau)\!=\!\bold{c}_{\tau}\boldWtotV^{\top} \}. \label{def:SRP:RP}
%\end{align}
%That is, $\RPt{h}$, $h\in\{d_1,d_2,r\}$ is the linear subspace spanned by the coding vectors $\bold{c}_\tau$ corresponding to the packets that are sent by node $\SCHtT{\tau}\neq h$ and have successfully arrived at node $h$ by the end of time $t$. For example, $\RPt{r}$ is the linear space spanned by the packets successfully delivered from source to relay up to time $t$. $\RPt{d_1}$ is the linear space spanned by the packets received at destination $d_1$ up to time $t$, either from source or from relay. The {\em knowledge space}\footnote{The knowledge space $\SPt{h}$ is a superordinate concept that contains not only the reception subspace $\RPt{h}$ but also the messages originated from node $h$, if any. In our problem of interest, the messages are originated only from source and thus its meaning is identical to the reception subspace.} $\SPt{h}$ for $h\in\{d_1,d_2,r\}$ can be simply defined as $\SPt{h} \triangleq \RPt{h}$.

With the above linear space concepts, we now can describe the packet transmission from relay. Recall that, unlike the source where the packets are originated, relay can only send a linear mixture of {\em the packets that it has known.} Therefore, the encoder description from relay can be expressed by
\begin{equation}\label{eq:SRP:lin_enc_rly}
\text{If } \SCHt\!=\!r,\,\text{then } \Xt{r}\!=\Ct\boldWBtot^{\top}\,\text{for some }\Ct\!\in\SPtPREV{r}.
\end{equation}

For comparison, in \Ref{eq:SRP:lin_enc_src}, the source $s$ chooses $\Ct$ from $\Mtot$. We can now define the LNC capacity region.
\begin{mydef}\label{def:SRP:LNCcapacity} Fix the distribution of $\boldZt$ and finite field $\FF[q]$. A rate vector $(\RB{1},\RB{2})$ is achievable by LNC if for any $\epsilon>0$ there exists a joint scheduling and LNC scheme with sufficiently large $n$ such that $\PROB( \SPn{k} \supseteq \MB{k} ) > 1 - \epsilon$ for all $k\in\{1,2\}$. The LNC capacity region is the closure of all LNC-achievable $(\RB{1},\RB{2})$.
\end{mydef}

\subsection{A Useful Notation}\label{sec:SRP:UsefulNot}

In our network model, there are two broadcast PECs associated with $s$ and $r$. For shorthand, we call those PECs the $s$-PEC and the $r$-PEC, respectively. The distribution of the network-wide channel status vector $\boldZt = (\boldZnt{s}, \boldZnt{r})$ can be described by the probabilities $\prOb{s}{T}{\{d_1,d_2,r\}\backslash T}$ for all $T\subseteq\{d_1,d_2,r\}$, and $\prOb{r}{U}{\{d_1,d_2\}\backslash U}$ for all $U\subseteq\{d_1,d_2\}$. In total, there are $8 + 4 = 12$ channel parameters.\footnote{By allowing some coordinates of $\boldZt$ to be correlated (i.e., spatially correlated as it is between coordinates, not over the time axis), our setting can also model the scenario in which $d_1$ and $d_2$ are situated in the same physical node and thus have perfectly correlated channel success events.}

For notational simplicity, we also define the following two probability functions $\pSRPsimT{s}{\cdot}$ and $\pSRPsimT{r}{\cdot}$, one for each PEC. The input argument of $\pSRPsim{s}$ is a collection of the elements in $\{d_1, d_2, r, \overline{d_1}, \overline{d_2}, \overline{r}\}$. The function $\pSRPsimT{s}{\cdot}$ outputs the probability that the reception event is compatible to the specified collection of $\{d_1, d_2, r, \overline{d_1}, \overline{d_2}, \overline{r}\}$. For example,
\begin{align}
\pSRPsimT{s}{d_2\overline{r}} = \prOd{s}{d_1}{d_2}{r} + \prOb{s}{d_1 d_2}{r} \label{eq:SRP:CHex}
\end{align}
is the probability that the input of the source-PEC is successfully received by $d_2$ but not by $r$. Herein, $d_1$ is a don’t-care receiver and $\pSRPsimT{s}{d_2\overline{r}}$ thus sums two joint probabilities together ($d_1$ receives it or not) as described in \Ref{eq:SRP:CHex}. Another example is $\pSRPsimT{r}{d_2} = \p{r}{d_1 d_2} + \prOa{r}{d_1}{d_2}$, which is the marginal success probability that a packet sent by $r$ is heard by $d_2$. To slightly abuse the notation, we further allow $\pSRPsimT{s}{\cdot}$ and $\pSRPsimT{r}{\cdot}$ to take multiple input arguments separated by commas. With this new notation, they can represent the probability that the reception event is compatible to at least one of the input arguments. For example,
\begin{align*}
\pSRPsimT{s}{d_1\overline{d_2},r} & = \prOb{s}{d_1}{d_2 r} + \prOc{s}{d_1}{d_2}{r} + \p{s}{d_1 d_2 r} \\
& \quad + \prOa{s}{d_1}{d_2 r} + \prOa{s}{d_1 d_2}{r}.
\end{align*}
That is, $\pSRPsimT{s}{d_1\overline{d_2},r}$ represents the probability that $(\Z{s}{d_1},$ $\Z{s}{d_2},\Z{s}{r})$ equals one of the following $5$ vectors $(1, \ast, \ast)$, $(1, \ast, 1)$, $(1, 1, 1)$, $(\ast, 1, 1)$, and $(\ast, \ast, 1)$. Note that these $5$ vectors are compatible to either $d_1\overline{d_2}$ or $r$ or both. Another example of this $\pSRPsimT{s}{\cdot}$ notation is $\pSRPsimT{s}{d_1,d_2,r}$, which represents the probability that a packet sent by $s$ is received by at least one of the three nodes $d_1$, $d_2$, and $r$.

The indicator function and taking expectation is denoted by $1_{\{\cdot\}}$ and $\ET{\cdot}$, respectively.

\section{LNC Capacity Outer Bound}\label{sec:SRP:CAPouter}

Since the coding vector $\Ct$ has $n\Rsum$ number of coordinates, there are exponentially many ways of jointly designing the scheduling $\SCHt$ and the coding vector $\Ct$ choices over time when sufficiently large $n$ and $\FF[q]$ are used. Therefore, we will first simplify the aforementioned design choices by comparing $\Ct$ to the knowledge spaces $\SPtPREV{h}$, $h\in\{d_1,d_2,r\}$. Such a simplification allows us to derive \PropRef{prop:SRP:LP-outer}, which uses a linear programming (LP) solver to exhaustively search over the entire coding and scheduling choices and thus computes an LNC capacity outer bound. %An LNC capacity inner bound will later be derived in \SecRef{sec:SRP:CAPinner} by proposing an elegant LNC solution and analyze its performance.
%Finally, we prove that the inner and outer bounds match.

To that end, we use $\SP{k}$ as shorthand for $\SPtPREV{k}$, the knowledge space of destination $d_k$ in the end of time $t\!-\!1$.
We first define the following $7$ linear subspaces of $\Mtot$.
\begin{align}
& \SRPAt{1} \TReq \SP{1}, \;\;\qquad\qquad\quad\;\;\, \SRPAt{2} \TReq \SP{2}, \label{def:SRP:Aset1} \\
& \SRPAt{3} \TReq \SP{1}\oplus\MB{1}, \;\quad\quad\quad\;\; \SRPAt{4} \TReq \SP{2}\oplus\MB{2}, \label{def:SRP:Aset2} \\
& \SRPAt{5} \TReq \SP{1}\oplus\SP{2}, \label{def:SRP:Aset3} \\
& \SRPAt{6} \TReq \SP{1}\oplus\SP{2}\oplus\MB{1}, \;\quad\; \SRPAt{7} \TReq \SP{1}\oplus\SP{2}\oplus\MB{2}, \label{def:SRP:Aset4}
\end{align}
where $A\oplus B \TReq\SPAN\{\bold{v} : \bold{v}\in A\cup B\}$ is the {\em sum space} of any $A, B\subseteq\Mtot$. In addition, we also define the following eight additional subspaces involving $\SPtPREV{r}$:

\noindent\begin{align}
\SRPAt{i+7} & \TReq \SRPAt{i} \oplus \SP{r} \quad \text{ for all } i=1,\cdots,7, \label{def:SRP:AsetR1} \\
\SRPAt{15} & \TReq \SP{r}, \label{def:SRP:AsetR2}
\end{align}
where $\SP{r}$ is a shorthand notation for $\SPtPREV{r}$, the knowledge space of relay $r$ in the end of time $t\!-\!1$.

In total, there are $7+8 = 15$ linear subspaces of $\Mtot$. We then partition the overall message space $\Mtot$ into $2^{15}$ disjoint subsets by the {\em Venn diagram} generated by these $15$ subspaces. That is, at any time $t$, we can place any coding vector $\Ct$ in exactly one of the $2^{15}$ disjoint subsets by testing whether it belongs to which $A$-subspaces. In the following discussion, we often drop the input argument ``$(t)$" when the time instant of interest is clear in the context.
%%-- Back up
%That is, for any given coding vector $\Ct$, we can place it in exactly one of the $2^{15}$ disjoint subsets by testing whether it belongs to which $A$-subspaces. This is always true regardless of the time index $t$, i.e., any coding vector $\Ct$ transmitted by either source or relay always lies in one of the $2^{15}$ disjoint subsets while the regions of disjoint subsets may change over the course of time.

We now use $15$ bits to represent each disjoint subset in $\Mtot$. For any $15$-bit string $\b=b_1 b_2 \cdots b_{15}$, we define ``the coding type-$\b$" by
\begin{align}
\SRPTYPEb{s} \TReq \bigg(\bigcap_{l:b_l=1} \SRPA{l}\bigg)\,\backslash\,\bigg(\bigcup_{l:b_l=0} \SRPA{l}\bigg). \label{def:SRP:TYPE}
\end{align}

\noindent where the regions of these $2^{15}$ disjoint coding types may vary at every time instant as the $15$ $A$-subspaces defined in \Ref{def:SRP:Aset1} to \Ref{def:SRP:AsetR2} will evolve over the course of time. The superscript ``(s)'' indicates the source, meaning that $s$ can send $\Ct$ in any coding type since source $s$ knows all $\boldWB{1}$ and $\boldWB{2}$ packets to begin with. Note that not all $2^{15}$ disjoint subsets are feasible. For example, any $\SRPTYPEb{s}$ with $b_7=1$ but $b_{14}=0$ is always empty because any coding vector that lies in $\SRPA{7}=\SP{1}\oplus\SP{2}\oplus\MB{2}$ cannot lie outside the larger $\SRPA{14}=\SP{1}\oplus\SP{2}\oplus\SP{r}\oplus\MB{2}$, see \Ref{def:SRP:Aset4} and \Ref{def:SRP:AsetR1}, respectively. We say those always empty subsets {\em infeasible coding types} and the rest is called {\em feasible coding types} ($\NETori$). By exhaustive computer search, we can prove that out of $2^{15}\!\!=\!32768$ subsets, only $\!154$ of them are feasible. Namely, the entire coding space $\Omega$ can be viewed as a union of $154$ disjoint coding types. Source $s$ can choose a coding vector $\Ct$ from one of these $154$ types. See \Ref{eq:SRP:lin_enc_src}.

For coding vectors that relay $r$ can choose, we can further reduce the number of possible placements of $\Ct$ in the following way. By \Ref{eq:SRP:lin_enc_rly}, we know that when $\SCHt=r$, the $\Ct$ sent by relay must belong to its knowledge space $\SPtPREV{r}$. Hence, such $\Ct$ must always lie in $\SPtPREV{r}$, which is $\SRPAt{15}$, see \Ref{def:SRP:AsetR2}. As a result, any coding vector $\Ct$ sent by relay $r$ must lie in those $154$ subsets $\NETori$ that satisfy:
\begin{align}
\SRPTYPEb{r} \TReq \{ \SRPTYPEb{s} : \b\in\NETori \text{ such that } b_{15}=1 \}. \label{def:SRP:TYPEr}
\end{align}

\noindent Again by computer search, there are $18$ such coding types out of 154 subsets $\NETori$. We call those $18$ subsets as {\em relay's feasible coding types} ($\NEToriR$). Obviously, $\NEToriR \subseteq \NETori$. See \AppRef{app:SRP:NETori} for the enumeration of those $\NETori$ and $\NEToriR$.

We can then derive the following upper bound.

\begin{prop}\label{prop:SRP:LP-outer} A rate vector $(\RB{1},\RB{2})$ is in the LNC capacity region only if there exists $154$ non-negative variables $\pnumb{s}$ for all $\b\in\NETori$, $18$ non-negative variables $\pnumb{r}$ for all $\b\in\NEToriR$, and $14$ non-negative $y$-variables, $y_1$ to $y_{14}$, such that jointly they satisfy the following three groups of linear conditions:

\noindent$\bullet$ Group~1, termed the {\em time-sharing condition}, has $1$ inequality:
\begin{equation}\label{prop:SRP:LP-outer:TS}
\bigg(\sum_{\forall \b\in\NETori} \pnumb{s} \bigg) + \bigg(\sum_{\forall \b\in\NEToriR} \pnumb{r} \bigg) \leq 1.
\end{equation}

\noindent$\bullet$ Group~2, termed the {\em rank-conversion conditions}, has $14$ equalities:
\begin{align}
& y_1 =
\left( \hspace{-1.65cm} \sum_{\qquad\qquad\quad\forall\b\in\NETori \textrm{ s.t. }b_1=0 } \hspace{-1.65cm} \pnumb{s} \!\cdot \pSRPsimT{s}{d_1} \right) +
\left( \hspace{-1.75cm} \sum_{\qquad\qquad\quad\;\,\forall\b\in\NEToriR \textrm{ s.t. }b_1=0} \hspace{-1.75cm} \pnumb{r} \!\cdot \pSRPsimT{r}{d_1} \right) \!, \label{prop:SRP:LP-outer:A1} \\
& y_2 =
\left( \hspace{-1.65cm} \sum_{\qquad\qquad\quad\forall\b\in\NETori \textrm{ s.t. }b_2=0 } \hspace{-1.65cm} \pnumb{s} \!\cdot \pSRPsimT{s}{d_2} \right) +
\left( \hspace{-1.75cm} \sum_{\qquad\qquad\quad\;\,\forall\b\in\NEToriR \textrm{ s.t. }b_2=0} \hspace{-1.75cm} \pnumb{r} \!\cdot \pSRPsimT{r}{d_2} \right) \!, \label{prop:SRP:LP-outer:A2} \\
& y_3 =
\left( \hspace{-1.65cm} \sum_{\qquad\qquad\quad\forall\b\in\NETori \textrm{ s.t. }b_3=0 } \hspace{-1.65cm} \pnumb{s} \!\cdot \pSRPsimT{s}{d_1} \right) +
\left( \hspace{-1.75cm} \sum_{\qquad\qquad\quad\;\,\forall\b\in\NEToriR \textrm{ s.t. }b_3=0} \hspace{-1.75cm} \pnumb{r} \!\cdot \pSRPsimT{r}{d_1} \right) + \RB{1}, \label{prop:SRP:LP-outer:A3} \\
& y_4 =
\left( \hspace{-1.65cm} \sum_{\qquad\qquad\quad\forall\b\in\NETori \textrm{ s.t. }b_4=0 } \hspace{-1.65cm} \pnumb{s} \!\cdot \pSRPsimT{s}{d_2} \right) +
\left( \hspace{-1.75cm} \sum_{\qquad\qquad\quad\;\,\forall\b\in\NEToriR \textrm{ s.t. }b_4=0} \hspace{-1.75cm} \pnumb{r} \!\cdot \pSRPsimT{r}{d_2} \right) + \RB{2}, \label{prop:SRP:LP-outer:A4} \\
& y_5 =
\left( \hspace{-1.65cm} \sum_{\qquad\qquad\quad\forall\b\in\NETori \textrm{ s.t. }b_5=0 } \hspace{-1.65cm} \pnumb{s} \!\cdot \pSRPsimT{s}{d_1,d_2} \!\right) +
\left( \hspace{-1.75cm} \sum_{\qquad\qquad\quad\;\,\forall\b\in\NEToriR \textrm{ s.t. }b_5=0} \hspace{-1.75cm} \pnumb{r} \!\cdot \pSRPsimT{r}{d_1,d_2} \!\right) \!, \label{prop:SRP:LP-outer:A5} \\
& y_6 =
\left( \hspace{-1.65cm} \sum_{\qquad\qquad\quad\forall\b\in\NETori \textrm{ s.t. }b_6=0 } \hspace{-1.65cm} \pnumb{s} \!\cdot \pSRPsimT{s}{d_1,d_2} \!\right) +
\left( \hspace{-1.75cm} \sum_{\qquad\qquad\quad\;\,\forall\b\in\NEToriR \textrm{ s.t. }b_6=0} \hspace{-1.75cm} \pnumb{r} \!\cdot \pSRPsimT{r}{d_1,d_2} \!\right) \!+ \RB{1}, \label{prop:SRP:LP-outer:A6} \\
& y_7 =
\left( \hspace{-1.65cm} \sum_{\qquad\qquad\quad\forall\b\in\NETori \textrm{ s.t. }b_7=0 } \hspace{-1.65cm} \pnumb{s} \!\cdot \pSRPsimT{s}{d_1,d_2} \!\right) +
\left( \hspace{-1.75cm} \sum_{\qquad\qquad\quad\;\,\forall\b\in\NEToriR \textrm{ s.t. }b_7=0} \hspace{-1.75cm} \pnumb{r} \!\cdot \pSRPsimT{r}{d_1,d_2} \!\right) \!+ \RB{2}, \label{prop:SRP:LP-outer:A7} \\
& y_8 =
\left( \hspace{-1.65cm} \sum_{\qquad\qquad\quad\forall\b\in\NETori \textrm{ s.t. }b_8=0 } \hspace{-1.65cm} \pnumb{s} \!\cdot \pSRPsimT{s}{d_1,r} \right) \!, %\label{prop:SRP:LP-outer:A8}
\quad y_9 =
\left( \hspace{-1.65cm} \sum_{\qquad\qquad\quad\forall\b\in\NETori \textrm{ s.t. }b_9=0 } \hspace{-1.65cm} \pnumb{s} \!\cdot \pSRPsimT{s}{d_2,r} \right) \!, \label{prop:SRP:LP-outer:A9} \\
& y_{10} =
\left( \hspace{-1.65cm} \sum_{\qquad\qquad\quad\forall\b\in\NETori \textrm{ s.t. }b_{10}=0 } \hspace{-1.65cm} \pnumb{s} \!\cdot \pSRPsimT{s}{d_1,r} \right) + \RB{1}, \label{prop:SRP:LP-outer:A10} \\
& y_{11} =
\left( \hspace{-1.65cm} \sum_{\qquad\qquad\quad\forall\b\in\NETori \textrm{ s.t. }b_{11}=0 } \hspace{-1.65cm} \pnumb{s} \!\cdot \pSRPsimT{s}{d_2,r} \right) + \RB{2}, \label{prop:SRP:LP-outer:A11} \\
& y_{12} =
\left( \hspace{-1.65cm} \sum_{\qquad\qquad\quad\forall\b\in\NETori \textrm{ s.t. }b_{12}=0 } \hspace{-1.65cm} \pnumb{s} \!\cdot \pSRPsimT{s}{d_1,d_2,r} \right) \!, \label{prop:SRP:LP-outer:A12} \\
& y_{13} =
\left( \hspace{-1.65cm} \sum_{\qquad\qquad\quad\forall\b\in\NETori \textrm{ s.t. }b_{13}=0 } \hspace{-1.65cm} \pnumb{s} \!\cdot \pSRPsimT{s}{d_1,d_2,r} \right) + \RB{1}, \label{prop:SRP:LP-outer:A13} \\
& y_{14} =
\left( \hspace{-1.65cm} \sum_{\qquad\qquad\quad\forall\b\in\NETori \textrm{ s.t. }b_{14}=0 } \hspace{-1.65cm} \pnumb{s} \!\cdot \pSRPsimT{s}{d_1,d_2,r} \right) + \RB{2}, \label{prop:SRP:LP-outer:A14}
\end{align}

\noindent$\bullet$ Group~3, termed the {\em decodability conditions}, has $5$ equalities:
\begin{align}
& y_1=y_3, \quad y_2=y_4, \quad y_8=y_{11}, \quad y_{9}=y_{11}, \label{prop:SRP:LP-outer:dec1} \\
& y_5=y_6=y_{7}=y_{12}=y_{13}=y_{14}=\Rsum. \label{prop:SRP:LP-outer:dec2}
\end{align}
\end{prop}

%%-- Back up for the short ISIT'16 version
%The very high-level intuition is as follows. Since we are partitioning $\Mtot$ (the entire coding space) and $\SP{r}$ (the knowledge space of $r$), any LNC scheme one can think of can be classified as sending different coding types at each time instant. When treating the given LNC scheme as a black box, we use $\pnumb{s}$ (resp.\ $\pnumb{r}$) to denote the fraction of time that $s$ (resp.\ $r$) spends on sending coding type-$\b$. Therefore, jointly they must satisfy Group 1 condition. Group 2 conditions quantify the impact of sending $\Ct$ of different coding types to $\RK{\SRPAt{l}}$, $l\!=\!1$ to $14$, until the end of time $n$, which is possible due to definition \Ref{def:SRP:TYPE}. E.g., $y_2$ and $y_4$ represents the normalized $\RK{\SRPAn{2}}$ and $\RK{\SRPAn{4}}$, respectively. Finally, since the given scheme is ``decodable'', we must have $\RK{\SRPAn{2}}=\RK{\SPn{1}}=\mathsf{rank}( \SPn{1}\oplus\MB{1})=\RK{\SRPAn{4}}$, implying $y_2=y_4$ in \Ref{prop:SRP:LP-outer:dec1}.

The intuition is as follows. Since we are partitioning $\Mtot$ (the entire coding space) and $\SP{r}$ (the knowledge space of $r$) into $154$ feasible coding types $\NETori$ and $18$ subsets $\NEToriR$, any LNC scheme can be classified as either $s$ or $r$ sending a coding vector $\Ct$ in certain coding type at each time instant. More specifically, consider an achievable rate vector $(\RB{1},\RB{2})$ and the associated LNC scheme. In the beginning of any time $t$, we can always compute the knowledge spaces $\SPtPREV{1}$, $\SPtPREV{2}$, and $\SPtPREV{r}$ by \Ref{def:SRP:SP} and use them to compute the $A$-subspaces in \Ref{def:SRP:Aset1}--\Ref{def:SRP:AsetR2}. Then suppose that for some specific time $\tau$, the given scheme chooses the source $s$ to transmit a coding vector $\CtT{\tau}$. By the previous discussions, we can classify which coding type $\SRPTYPEb{s}$ this $\CtT{\tau}$ belongs to, by comparing it to those computed $15$ $A$-subspaces. After running the given scheme from time $1$ to $n$, we can thus compute the variable $\pnumb{s} \TReq$ $\frac{1}{n}\ET{\sum_{t=1}^{n} 1_{\{\Ct\in\SRPTYPEb{s}\}}}$ for each $\b\in\NETori$ as the {\em frequency} of scheduling source $s$ with the chosen coding vectors being in $\SRPTYPEb{s}$. Similarly for the relay $r$, we can compute the variable $\pnumb{r} \TReq$ $\frac{1}{n}\ET{\sum_{t=1}^{n} 1_{\{\Ct\in\SRPTYPEb{r}\}}}$ for each $\b\in\NEToriR$ as the {\em frequency} of scheduling relay $r$ with the chosen coding vectors being in $\SRPTYPEb{r}$. Obviously, the computed variables $\{\pnumb{s}, \pnumb{r}\}$ satisfy the time-sharing inequality \Ref{prop:SRP:LP-outer:TS}.

We then compute the $y$-variables by
\begin{align}
& y_{l}  \TReq \frac{1}{n}\ET{\mathsf{rank}\big(\SRPAn{l}\big)}, \;\forall\,l\in\{1,2,\cdots,14\},\label{eq:SRP:yi}
\end{align}
as the normalized expected ranks of $A$-subspaces in the end of time $n$. We now claim that these variables satisfy \Ref{prop:SRP:LP-outer:A1} to \Ref{prop:SRP:LP-outer:dec2}. This claim implies that for any LNC-achievable $(\RB{1},\RB{2})$, there exists $\pnumb{s}$, $\pnumb{r}$, and $y$-variables satisfying \PropRef{prop:SRP:LP-outer}, thus constituting an outer bound on the LNC capacity.

To prove that \Ref{prop:SRP:LP-outer:A1} to \Ref{prop:SRP:LP-outer:A14} are true,\footnote{For rigorous proofs, we need to invoke the law of large numbers and take care of the $\epsilon$-error probability. For ease of discussion, the corresponding technical details are omitted when discussing the intuition of \PropRef{prop:SRP:LP-outer}.} consider an $A$-subspace, say
$\SRPAt{3}=\SPtPREV{1}\oplus\MB{1}$ as defined in \Ref{def:SRP:Aset2} and \Ref{def:SRP:SP}. In the beginning of time $1$, destination $d_1$ has not received any packet yet, i.e., $\SPtT{1}{0}=\{\bold{0}\}$. Thus the rank of $\SRPAtT{3}{1}$ is $\RK{\MB{1}}=n\RB{1}$.

The fact that $\SPtPREV{1}$ contributes to $\SRPAt{3}$ implies that $\RK{\SRPAt{3}}$ will increase by one whenever the destination $d_1$ receives a packet $\Ct\boldWtotV^{\top}$ satisfying $\Ct\not\in\SRPAt{3}$. Specifically, whenever source $s$ sends a $\Ct$ in $\SRPTYPEb{s}$ with $b_3\!=\!0$, such $\Ct$ is not in $\SRPAt{3}$, and whenever $d_1$ receives it, $\RK{\SRPAt{3}}$ increases by $1$. Moreover, whenever relay $r$ sends a $\Ct$ in $\SRPTYPEb{r}$ with $b_3\!=\!0$ and $d_1$ receives it, $\RK{\SRPAt{3}}$ also increases by $1$. Therefore, in the end of time $n$, we have
\begin{equation}\label{eq:SRP:evproc}\begin{split}
\RK{\SRPAn{3}} = & \sum_{t=1}^{n} 1_{\left\{\substack{\textrm{source } s \textrm{ sends }\Ct\in\SRPTYPEb{s} \;\!\textrm{with } b_3=0, \\ \textrm{and destination } d_1 \textrm{ receives it}}\right\}} \\
+ & \sum_{t=1}^{n} 1_{\left\{\substack{\textrm{relay } r \textrm{ sends }\Ct\in\SRPTYPEb{r}  \;\!\textrm{with } b_3=0,\\ \textrm{and destination } d_1 \textrm{ receives it}}\right\}} \\
+ & \;\RK{\SRPAtT{3}{0}}.
\end{split}\end{equation}

\noindent  Taking the normalized expectation of \Ref{eq:SRP:evproc}, we have proven \Ref{prop:SRP:LP-outer:A3}. By similar {\em rank-conversion} arguments, \Ref{prop:SRP:LP-outer:A1} to \Ref{prop:SRP:LP-outer:A14} can be shown to be true.

In the end of time $n$, since the given scheme is ``decodable'' (i.e.,  both $d_1$ and $d_2$ can decode the desired packets $\boldWB{1}$ and $\boldWB{2}$, respectively), we must have $\SPn{1}\supseteq\MB{1}$ and $\SPn{2}\supseteq\MB{2}$, or equivalently $\SPn{k}=\SPn{k}\oplus\MB{k}$ for all $k\in\{1,2\}$. This implies that the ranks of $\SRPAn{1}$ and $\SRPAn{3}$, and the ranks of $\SRPAn{2}$ and $\SRPAn{4}$ are equal, respectively. Together with \Ref{eq:SRP:yi}, we thus have the first two equalities in \Ref{prop:SRP:LP-outer:dec1}. Similarly, one can prove that the remaining equalities in \Ref{prop:SRP:LP-outer:dec1} and \Ref{prop:SRP:LP-outer:dec2} are satisfied as well. The claim is thus proven.

\section{LNC Capacity Inner Bound}\label{sec:SRP:CAPinner}

\subsection{LNC Inner Bound of the Strong-Relaying Scenario}\label{sec:SRP:NonDegenerate}

In the smart repeater problem, $s$ can always take over relay's operations, and thus $r$ becomes useless when the $r$-PEC is weaker than the $s$-PEC. To fully fetch the coding and diversity benefits using relay, we first focus on the following assumption.

\begin{mydef}\label{def:SRP:strong-relaying} The smart repeater network with $\{d_1,d_2\}$ is {\em strong-relaying} if $\pSRPsimT{r}{T\overline{\{d_1,d_2\}\backslash T}}\!>\!\pSRPsimT{s}{T\overline{\{d_1,d_2\}\backslash T}}$ for all $T\subseteq\{d_1,d_2\}\backslash\emptyset$. That is, the given $r$-PEC is stronger than the given $s$-PEC for all non-empty subsets of $\{d_1,d_2\}$.
\end{mydef}

We describe our capacity-approaching achievability scheme based on the strong-relaying scenario. The general inner bound that works in arbitrary $s$-PEC and $r$-PEC distributions, and introduces more advanced LNC operations will be described in \PropRef{prop:SRP:sim-innerV2}.

\begin{prop}\label{prop:SRP:sim-inner} A rate vector $(\RB{1},\RB{2})$ is LNC-achievable if there exist $2$ non-negative variables $t_s$ and $t_r$, $(6 \times 2 + 8)$ non-negative $s$-variables:
\begin{align*}
& \big\{\snumUC{k},\; \snumPM{k},\; \snumAM{k},\; \snumRC{k},\; \snumDX{k}{},\; \snumDX{(\!k\!)}{} \;:\; \text{for all}\;k\in\{1,2\} \big\}, \\
& \big\{\snumCX{l} \,(l\!=\!1,\cdots\!,8) \big\},
\end{align*}

\noindent and $(3 \times 2 + 3)$ non-negative $r$-variables:
\begin{align*}
\big\{ \rnumUC{k},\; \rnumDX{(\!k\!)}{},\; \rnumDX{[k]}{} \;:\; \text{for all}\;k\in\{1,2\} \big\}, \;\; \big\{ \rnumRC,\; \rnumOX{},\; \rnumCX{} \big\},
\end{align*}

\noindent such that jointly they satisfy the following five groups of linear conditions:
\end{prop}

\noindent$\bullet$ Group~1, termed the {\em time-sharing conditions}, has $3$ inequalities:
\begin{align}
1 & > t_s + t_r, \label{prop:SRP:sim-inner:TS1} \\
t_s & \geq \!\!\! \sum_{k\in\{1,2\}} \!\!\!\! \left( \snumUC{k} \!+\! \snumPM{k} \!+\! \snumAM{k} \!+\!  \snumRC{k} \!+\! \snumDX{k}{} \!+\! \snumDX{(\!k\!)}{} \right) + \sum_{l=1}^{8} \snumCX{l}, \label{prop:SRP:sim-inner:TS2} \\
t_r & \geq \!\!\! \sum_{k\in\{1,2\}} \!\!\!\! \left( \rnumUC{k} + \rnumDX{(\!k\!)}{} + \rnumDX{[k]}{}\right) + \rnumRC + \rnumOX{} + \rnumCX{}. \label{prop:SRP:sim-inner:TS3}
\end{align}

\noindent$\bullet$ Group~2, termed the {\em packets-originating condition}, has $2$ inequalities: Consider any $i,j\in\{1,2\}$
satisfying $i\neq j$. For each $(i,j)$ pair (out of the two choices $(1,2)$ and $(2,1)$),
%For $k\in\{1,2\}$,
\begin{align}
\RB{i} \geq \left(\snumUC{i} + \snumPM{i}\right)\cdot\pSRPsimT{s}{d_i,d_j,r}, \tag{E}\label{prop:SRP:sim-inner:E}
\end{align}
%%-- Back up (longer version without symmetry)
%\begin{align}
%\RB{1} \geq \left( \snumUC{1} + \snumPM{1} \right) \cdot\pSRPsimT{s}{d_1,d_2,r}, \tag{E1}\label{prop:SRP:sim-inner:E1} \\
%\RB{2} \geq \left( \snumUC{2} + \snumPM{2} \right) \cdot\pSRPsimT{s}{d_1,d_2,r}. \tag{E2}\label{prop:SRP:sim-inner:E2}
%\end{align}

\noindent$\bullet$ Group~3, termed the {\em packets-mixing condition}, has $4$ inequalities: For each $(i,j)$ pair,
\begin{align}
\begin{split}
\left(\snumUC{i}+\snumPM{i}\right)\cdot\prOa{s}{d_i d_j}{r} & \geq (\snumPM{j} + \snumAM{i}) \cdot \pSRPsimT{s}{d_i,d_j} \\
& \quad + \rnumUC{i} \cdot \pSRPsimT{r}{d_i,d_j},
\end{split} \tag{A}\label{prop:SRP:sim-inner:A} \\
\snumPM{i}\cdot\prOd{s}{d_i}{d_j}{r} & \geq \snumRC{i} \cdot \pSRPsimT{s}{d_i,d_j,r}, \tag{B}\label{prop:SRP:sim-inner:B}
\end{align}
%%-- Back up (longer version without symmetry)
%\begin{align}
%%%
%& \left(\snumUC{1}+\snumPM{1}\right)\cdot\prOa{s}{d_1 d_2}{r} \geq \left\{ \begin{aligned} & (\snumPM{2} + \snumAM{1}) \cdot \pSRPsimT{s}{d_1,d_2} \\ & + \rnumUC{1} \cdot \pSRPsimT{r}{d_1,d_2} \end{aligned} \right\}, \tag{A1}\label{prop:SRP:sim-inner:A1} \\
%%%
%& \left(\snumUC{2}+\snumPM{2}\right)\cdot\prOa{s}{d_1 d_2}{r} \geq \left\{ \begin{aligned} & (\snumPM{1} + \snumAM{2}) \cdot \pSRPsimT{s}{d_1,d_2} \\ & + \rnumUC{2} \cdot \pSRPsimT{r}{d_1,d_2} \end{aligned} \right\}, \tag{A2}\label{prop:SRP:sim-inner:A2} \\
%%%
%& \qquad\quad\;\;\; \snumPM{1}\cdot\prOd{s}{d_1}{d_2}{r} \geq \snumRC{1} \cdot \pSRPsimT{s}{d_1,d_2,r}, \tag{B1}\label{prop:SRP:sim-inner:B1} \\
%%%
%& \qquad\quad\;\;\; \snumPM{2}\cdot\prOb{s}{d_1}{d_2 r} \geq \snumRC{2} \cdot \pSRPsimT{s}{d_1,d_2,r}, \tag{B2}\label{prop:SRP:sim-inner:B2}
%\end{align}
and the following one inequality:
\begin{align}
& \!\!\!\!\! \snumPM{1}\!\cdot\!\pSRPsimT{s}{d_1,d_2 r} + \snumPM{2}\!\cdot\!\pSRPsimT{s}{d_2,d_1 r} + \snumAM{1}\!\cdot\!\pSRPsimT{s}{\overline{d_1}{d_2}} \,+ \nN \\
& \!\!\!\!\! \snumAM{2}\!\cdot\!\pSRPsimT{s}{d_1\overline{d_2}} \!+\! \left( \snumRC{1} \!+\! \snumRC{2} \right) \!\cdot\! \prOa{s}{d_1d_2}{r} \geq \rnumRC\!\cdot\!\pSRPsimT{r}{d_1,d_2}. \tag{M}\label{prop:SRP:sim-inner:M}
\end{align}
%%-- Back up (longer version without symmetry)
%\begin{align}
%\left\{ \begin{aligned} & \snumPM{1}\cdot\pSRPsimT{s}{d_1,d_2 r} + \snumPM{2}\cdot\pSRPsimT{s}{d_2,d_1 r} \\ & + \snumAM{1}\cdot\pSRPsimT{s}{\overline{d_1}{d_2}} + \snumAM{2}\cdot\pSRPsimT{s}{d_1\overline{d_2}} \\ & + \left( \snumRC{1} + \snumRC{2} \right) \cdot \prOa{s}{d_1d_2}{r} \end{aligned}\right\} \geq \rnumRC\cdot\pSRPsimT{r}{d_1,d_2}.
%\tag{M}\label{prop:SRP:sim-inner:M}
%\end{align}

\noindent$\bullet$ Group~4, termed the {\em classic XOR condition by source only}, has $4$ inequalities: For each $(i,j)$ pair,
\begin{align}
& \left( \snumUC{i} + \snumRC{i} \right) \prOd{s}{d_i}{d_j}{r} \geq \left(\snumAM{j} + \snumDX{i}{}\right)\cdot\pSRPsimT{s}{d_i,r}\,+ \nN \\
& \qquad \left( \snumCX{1} + \snumCX{1+i} \right)\cdot\,\pSRPsimT{s}{d_i,r} + \snumCX{4+i}\cdot\pSRPsimT{s}{d_i,r}, \tag{S}\label{prop:SRP:sim-inner:S} \\
& \snumRC{j}\cdot\prOd{s}{d_i}{d_j}{r} \geq \snumDX{(\!i\!)}{}\cdot\pSRPsimT{s}{d_i,r} + \rnumDX{(\!i\!)}{}\cdot\pSRPsimT{r}{d_i,d_j}\,+ \nN \\
& \qquad \left( \snumCX{1+j} + \snumCX{4} \right)\cdot \, \pSRPsimT{s}{d_i,r} + \snumCX{6+i}\cdot\pSRPsimT{s}{d_i,r}. \tag{T}\label{prop:SRP:sim-inner:T}
\end{align}
%%-- Back up (longer version without symmetry)
%\begin{align}
%%%
%& \left( \snumUC{1} + \snumRC{1} \right) \prOd{s}{d_1}{d_2}{r} \geq \left\{ \begin{aligned} & \left(\snumAM{2} + \snumDX{1}{}\right)\cdot\pSRPsimT{s}{d_1,r} \\ & + \left( \snumCX{1} + \snumCX{2} \right)\cdot\pSRPsimT{s}{d_1,r} \\ & + \snumCX{5}\cdot\pSRPsimT{s}{d_1,r} \end{aligned} \right\}, \tag{S1}\label{prop:SRP:sim-inner:S1} \\
%%%
%& \snumRC{2}\cdot\prOd{s}{d_1}{d_2}{r} \geq \left\{ \begin{aligned} & \snumDX{(\!1\!)}{}\cdot\pSRPsimT{s}{d_1,r} + \rnumDX{(\!1\!)}{}\cdot\pSRPsimT{r}{d_1,d_2} \\ & + \left( \snumCX{3} + \snumCX{4} \right)\cdot\pSRPsimT{s}{d_1,r} \\ & + \snumCX{7}\cdot\pSRPsimT{s}{d_1,r} \end{aligned} \right\}, \tag{T1}\label{prop:SRP:sim-inner:T1} \\
%%%
%& \left( \snumUC{2} + \snumRC{2} \right) \prOb{s}{d_1}{d_2 r} \geq \left\{ \begin{aligned} & \left( \snumAM{1} + \snumDX{2}{}\right)\cdot\pSRPsimT{s}{d_2,r} \\ & + \left( \snumCX{1} + \snumCX{3} \right)\cdot\pSRPsimT{s}{d_2,r} \\ & + \snumCX{6}\cdot\pSRPsimT{s}{d_2,r} \end{aligned} \right\}, \tag{S2}\label{prop:SRP:sim-inner:S2} \\
%%%
%& \snumRC{1}\cdot\prOb{s}{d_1}{d_2 r} \geq \left\{ \begin{aligned} & \snumDX{(\!2\!)}{}\cdot\pSRPsimT{s}{d_2,r} + \rnumDX{(\!2\!)}{}\cdot\pSRPsimT{r}{d_1,d_2} \\ & + \left( \snumCX{2} + \snumCX{4} \right)\cdot\pSRPsimT{s}{d_2,r} \\ & + \snumCX{8}\cdot\pSRPsimT{s}{d_2,r} \end{aligned} \right\}, \tag{T2}\label{prop:SRP:sim-inner:T2}
%%%
%\end{align}

\noindent$\bullet$ Group~5, termed the {\em XOR condition}, has $3$ inequalities:
\begin{align}
& \sum_{l=1}^{4}\snumCX{l}\cdot\prOa{s}{d_1 d_2}{r} \geq \rnumOX{}\cdot\pSRPsimT{r}{d_1,d_2}, \tag{X0}\label{prop:SRP:sim-inner:X0}
\end{align}
and for each $(i,j)$ pair,
\begin{align}
& \snumAM{j}\!\cdot\!\pSRPsimT{s}{d_id_j,\overline{d_i}r} + \Big( \snumUC{i}\!+\!\snumRC{i}\!+\!\snumRC{j}\!+\!\sum_{l=1}^{4}\snumCX{l}\Big)\cdot\!\prOa{s}{d_i}{d_j r} \nN \\
& + \left( \snumCX{4+i}+\snumCX{6+i}+\snumDX{i}{}+\snumDX{(\!i\!)}{} \right)\cdot\pSRPsimT{s}{\overline{d_i}r} \nN \\
& + \left( \rnumUC{i} + \rnumRC + \rnumDX{(\!i\!)}{} + \rnumOX{}\right) \cdot\prOa{r}{d_i}{d_j} \nN \\
& \geq \left( \snumCX{7-i} + \snumCX{9-i} \right)\cdot\pSRPsimT{s}{d_i} + \left( \rnumCX{} \!+ \rnumDX{[i]}{} \right)\cdot\pSRPsimT{r}{d_i}. \tag{X}\label{prop:SRP:sim-inner:X}
\end{align}
%%-- Back up (longer version without symmetry)
%\begin{align}
%%%
%\begin{split}
%& \left\{ \begin{aligned} & \snumAM{2}\cdot\pSRPsimT{s}{d_1d_2,\overline{d_1}r} \\ & + \left( \snumUC{1}+\snumRC{1}+\snumRC{2}+\sum_{l=1}^{4}\snumCX{l}\right)\cdot\prOa{s}{d_1}{d_2 r} \\ & + \left( \snumCX{5}+\snumCX{7}+\snumDX{1}{}+\snumDX{(\!1\!)}{} \right)\cdot\pSRPsimT{s}{\overline{d_1}r} \\ & + \left( \rnumUC{1} + \rnumRC + \rnumDX{(\!1\!)}{} + \rnumOX{}\right) \cdot\prOa{r}{d_1}{d_2} \end{aligned} \right\} \\
%& \;\; \geq \left( \snumCX{6} + \snumCX{8} \right)\cdot\pSRPsimT{s}{d_1} + \left( \rnumCX{} + \rnumDX{[1]}{} \right)\cdot\pSRPsimT{r}{d_1},
%\end{split} \tag{X1}\label{prop:SRP:sim-inner:X1} \\
%\begin{split}
%& \left\{ \begin{aligned} & \snumAM{1}\cdot\pSRPsimT{s}{d_1d_2,\overline{d_2}r} \\ & + \left( \snumUC{2}+\snumRC{1}+\snumRC{2}+\sum_{l=1}^{4}\snumCX{l}\right)\cdot\prOc{s}{d_1}{d_2}{r} \\ & + \left( \snumCX{6}+\snumCX{8}+\snumDX{2}{}+\snumDX{(\!2\!)}{} \right)\cdot\pSRPsimT{s}{\overline{d_2}r} \\ & + \left( \rnumUC{2} + \rnumRC + \rnumDX{(\!2\!)}{} + \rnumOX{}\right) \cdot\prOb{r}{d_1}{d_2} \end{aligned} \right\} \\
%& \;\; \geq \left( \snumCX{5} + \snumCX{7} \right)\cdot\pSRPsimT{s}{d_2} + \left( \rnumCX{} + \rnumDX{[2]}{} \right)\cdot\pSRPsimT{r}{d_2},
%\end{split} \tag{X2}\label{prop:SRP:sim-inner:X2}
%%%
%\end{align}

\noindent$\bullet$ Group~6, termed the {\em decodability condition}, has $2$ inequalities: For each $(i,j)$ pair,
\begin{align}
& \!\!\!\Big( \snumUC{i} + \snumAM{j} + \!\!\!\!\!\!\sum_{\quad k\in\{1,2\}}\!\!\!\!\!\!\!\!\snumRC{k} + \sum_{l=1}^8 \snumCX{l} + \snumDX{i}{} + \snumDX{(\!i\!)}{} \Big) \cdot \pSRPsimT{s}{d_i} \,+ \nN \\
& \left(\rnumUC{i} + \rnumRC + \rnumOX{} + \rnumCX{} + \rnumDX{(\!i\!)}{} + \rnumDX{[i]}{}\right)\cdot\pSRPsimT{r}{d_i} \geq \RB{i}. \tag{D}\label{prop:SRP:sim-inner:D}
\end{align}

The intuition is as follows. \PropRef{prop:SRP:sim-inner} can be described based on packet movements in a queueing network, governed by the proposed LNC operations. Each $s$- and $r$-variable (except $t$-variables for time-sharing) is associated with a specific LNC operation performed by the source $s$ and the relay $r$, respectively. The inequalities \Ref{prop:SRP:sim-inner:E} to \Ref{prop:SRP:sim-inner:D} then describe the queueing process, where LHS and RHS of each inequality implies the packet insertion and removal condition of a queue. For the notational convenience, we define the following queue notations associated with these $14$ inequalities \Ref{prop:SRP:sim-inner:E} to \Ref{prop:SRP:sim-inner:D}:

%%-- Back up for the original writings
%The intuition is as follows. The proposed LNC inner bound is derived based on the ideas of describing the packet movements in a queueing network, where movements are governed by LNC operations. Each variable (except $t$-variables for time-sharing) in \PropRef{prop:SRP:sim-inner} is associated with a specific LNC operation. Note that $s$-variables are associated with LNC operations performed by the source $s$, while $r$-variables are associated with LNC operations performed by the relay $r$. The inequalities \Ref{prop:SRP:sim-inner:E} to \Ref{prop:SRP:sim-inner:D} then describe the queueing process for packet movements, where the LHS and the RHS of each inequality implies the packet insertion and removal conditions, respectively, of the corresponding queue by the related LNC operations. For notational convenience, we define the following queue notations associated with these $14$ inequalities \Ref{prop:SRP:sim-inner:E} to \Ref{prop:SRP:sim-inner:D}:

%ARXIV%\newpage

\setlength{\tabcolsep}{5pt}
\begin{table}[h]
    \caption{Queue denominations for the inequalities \Ref{prop:SRP:sim-inner:E} to \Ref{prop:SRP:sim-inner:D}}
    \centering
    {\footnotesize
        \renewcommand\arraystretch{1.75}
        \begin{tabular}{|l|l|l|l|l|}
        \hline
        (\ref{prop:SRP:sim-inner:E}1): $\SRPQe{1}$ & (\ref{prop:SRP:sim-inner:B}1): $\SRPQb{1}{2}$ & (\ref{prop:SRP:sim-inner:S}1): $\SRPQsx{1}{2}{}$ & \Ref{prop:SRP:sim-inner:X0}: $\SRPQstar$ \\
        (\ref{prop:SRP:sim-inner:E}2): $\SRPQe{2}$ & (\ref{prop:SRP:sim-inner:B}2): $\SRPQb{2}{1}$ & (\ref{prop:SRP:sim-inner:T}1): $\SRPQsxSP{1}{2}$ & (\ref{prop:SRP:sim-inner:X}1): $\SRPQx{1}{2}$ \\
        (\ref{prop:SRP:sim-inner:A}1): $\SRPQr{1}$ & \Ref{prop:SRP:sim-inner:M}: $\SRPQm$ & (\ref{prop:SRP:sim-inner:S}2): $\SRPQsx{2}{1}{}$ & (\ref{prop:SRP:sim-inner:X}2): $\SRPQx{2}{1}$ \\
        (\ref{prop:SRP:sim-inner:A}2): $\SRPQr{2}$ & & (\ref{prop:SRP:sim-inner:T}2): $\SRPQsxSP{2}{1}$ & (\ref{prop:SRP:sim-inner:D}1): $\SRPQd{1}$ \\
        & & & (\ref{prop:SRP:sim-inner:D}2): $\SRPQd{2}$ \\
        \hline
        \end{tabular}\label{tab:SRP:queue}
    }
\end{table}\setlength{\tabcolsep}{6pt}
\vspace*{-\baselineskip}\vspace*{+3pt}
\noindent where we use the index-after-reference to distinguish the session (i.e. flow) of focus of an inequality. For example, (\ref{prop:SRP:sim-inner:E}1) and (\ref{prop:SRP:sim-inner:E}2) are to denote the inequality \Ref{prop:SRP:sim-inner:E} when $(i,j)=(1,2)$ and $(i,j)=(2,1)$, respectively.

%%-- back up the original
%\setlength{\tabcolsep}{5pt}
%\begin{table}[h]
%    \caption{Queue denominations for the inequalities \Ref{prop:SRP:sim-inner:E1} to \Ref{prop:SRP:sim-inner:D2}}
%    \centering
%    {\footnotesize
%        \renewcommand\arraystretch{1.75}
%        \begin{tabular}{|l|l|l|l|l|}
%        \hline
%        \Ref{prop:SRP:sim-inner:E1}: $\SRPQe{1}$ & \Ref{prop:SRP:sim-inner:B1}: $\SRPQb{1}{2}$ & \Ref{prop:SRP:sim-inner:S1}: $\SRPQsx{1}{2}{}$ & \Ref{prop:SRP:sim-inner:X0}: $\SRPQstar$ \\
%        %%
%        \Ref{prop:SRP:sim-inner:E2}: $\SRPQe{2}$ & \Ref{prop:SRP:sim-inner:B2}: $\SRPQb{2}{1}$ & \Ref{prop:SRP:sim-inner:T1}: $\SRPQsxSP{1}{2}$ & \Ref{prop:SRP:sim-inner:X1}: $\SRPQx{1}{2}$ \\
%        %%
%        \Ref{prop:SRP:sim-inner:A1}: $\SRPQr{1}$ & \Ref{prop:SRP:sim-inner:M}: $\SRPQm$ & \Ref{prop:SRP:sim-inner:S2}: $\SRPQsx{2}{1}{}$ & \Ref{prop:SRP:sim-inner:X2}: $\SRPQx{2}{1}$ \\
%        %%
%        \Ref{prop:SRP:sim-inner:A2}: $\SRPQr{2}$ & & \Ref{prop:SRP:sim-inner:T2}: $\SRPQsxSP{2}{1}$ & \Ref{prop:SRP:sim-inner:D1}: $\SRPQd{1}$ \\
%        %%
%        & & & \Ref{prop:SRP:sim-inner:D2}: $\SRPQd{2}$ \\
%        \hline
%        \end{tabular}\label{tab:SRP:queue}
%    }
%\end{table}
%\setlength{\tabcolsep}{6pt}

For example, suppose that $\boldWB{1}=(X_1,\cdots,X_{n\RB{1}})$ packets and $\boldWB{2}=(Y_1,\cdots,Y_{n\RB{2}})$ packets are initially stored in queues $Q^{1}_{\phi}$ and $Q^{2}_{\phi}$, respectively, at source $s$. The superscript $k\in\{1,2\}$ indicates that the queue is for the packets intended to destination $d_k$. The subscript indicates that those packets have not been heard by any of $\{d_1,d_2,r\}$. The LNC operation corresponding to the variable $\snumUC{1}$ (resp. $\snumUC{2}$) is to send a session-$1$ packet $X_i$ (resp. a session-$2$ packet $Y_j$) uncodedly. Then the inequality (\ref{prop:SRP:sim-inner:E}1) (resp. (\ref{prop:SRP:sim-inner:E}2)) implies that whenever it is received by at least one of $\{d_1,d_2,r\}$, this packet is removed from the queue of $\SRPQe{1}$ (resp. $\SRPQe{2}$).

Depending on the reception status, a packet will either be moved to another queue or remain in the same queue. For example, the use of the $\snumUC{1}$-operation (sending $X_i\in\boldWB{1}$ uncodedly from source) will take $X_i$ from $\SRPQe{1}$ and insert it into $\SRPQd{1}$ as long as $\Zt{s}{d_1} \!=\! 1$ in the reception status $\boldZnt{s}$, i.e., when the intended destination $d_1$ correctly receives it. Similarly, when the reception status is $\Zt{s}{d_1} \!=\! \Zt{s}{d_2} \!=\! 0$ but $\Zt{s}{r}\!=\!1$, this packet will be inserted to the queue $\SRPQr{1}$ according to the packet movement rule of (\ref{prop:SRP:sim-inner:A}1); inserted to $\SRPQsx{1}{2}{}$ when $\Zt{s}{d_1} \!=\! \Zt{s}{r} \!=\! 0$ but $\Zt{s}{d_2}\!=\!1$ by (\ref{prop:SRP:sim-inner:S}1); and inserted to $\SRPQx{1}{2}$ when $\Zt{s}{d_1}\!=\!0$ but $\Zt{s}{d_2} \!=\! \Zt{s}{r}\!=\!1$ by (\ref{prop:SRP:sim-inner:X}1). Obviously when none of $\{d_1,d_2,r\}$ has received it, the packet $X_l$ simply remains in $\SRPQe{1}$.

%%-- Back up
%Depending on the reception status, a packet will either be moved to another queue or remain in the same queue. For example, the use of the $\snumUC{1}$-operation (sending $X_i\in\boldWB{1}$ uncodedly from source) will take $X_i$ from $\SRPQe{1}$ and insert it into $\SRPQd{1}$ when the reception status is $\pSRPsimT{s}{d_1}$, i.e., when the intended destination $d_1$ correctly receives it. Similarly, when the reception status is $\prOa{s}{d_1 d_2}{r}$, this packet will be inserted to the queue $\SRPQr{1}$ according to the packet movement rule of (\ref{prop:SRP:sim-inner:A}1); inserted to $\SRPQsx{1}{2}{}$ when $\prOd{s}{d_1}{d_2}{r}$ by (\ref{prop:SRP:sim-inner:S}1); and inserted to $\SRPQx{1}{2}$ when $\prOa{s}{d_1}{d_2 r}$ by (\ref{prop:SRP:sim-inner:X}1). Obviously when $\prO{s}{d_1 d_2 r}{}$, since any node in $\{d_1,d_2,r\}$ has not received at all, the packet $X_i$ simply remains in $\SRPQe{1}$.

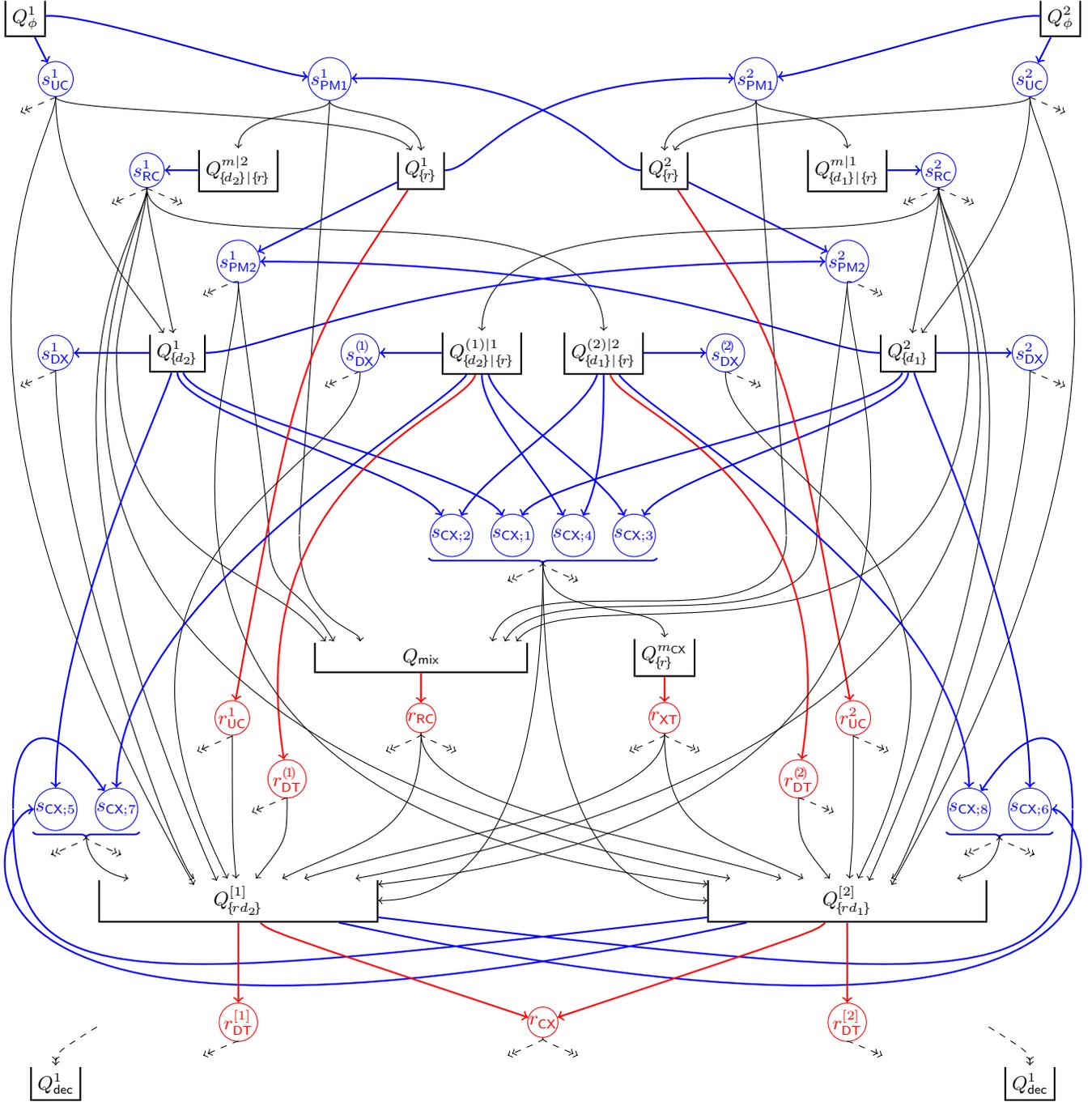
\begin{figure*}
    \hspace*{-1.9cm}
        \begin{tikzpicture}
        %-- Layer-0
        \node[] (L0) at (0.0,0) {};
        \node[rect={solid,thick,draw=black}{}{solid,thick,draw=black}{solid,thick,draw=black}] (E1) at ([shift={(-8.5,0)}]L0) {\small $\SRPQe{1}$};
        \node[rect={solid,thick,draw=black}{}{solid,thick,draw=black}{solid,thick,draw=black}] (E2) at ([shift={(+8.5,0)}]L0) {\small $\SRPQe{2}$};
        %-- Layer-1
        \node[] (L1) at ([shift={(0,-1.0)}]L0) {};
        \node[draw,blue,circle,inner sep=0pt, minimum size=12pt] (PM1) at ([shift={(-3.5,0)}]L1) {\small $\snumPM{1}$};
        \node[draw,blue,circle,inner sep=0pt, minimum size=12pt] (PM2) at ([shift={(+3.5,0)}]L1) {\small $\snumPM{2}$};
        \node[draw,blue,circle,inner sep=0pt, minimum size=12pt] (UC1) at ([shift={(-8,0)}]L1) {\small $\snumUC{1}$};
        \node[] (UC1dec) at ([shift={(-0.7,-0.6)}]UC1) {};  \draw[dashed,->>] (UC1) .. controls([shift={(0,-0.3)}]UC1) .. (UC1dec);
        \node[draw,blue,circle,inner sep=0pt, minimum size=12pt] (UC2) at ([shift={(+8,0)}]L1) {\small $\snumUC{2}$};
        \node[] (UC2dec) at ([shift={(+0.7,-0.6)}]UC2) {};  \draw[dashed,->>] (UC2) .. controls([shift={(0,-0.3)}]UC2) .. (UC2dec);
        %-- Layer-2
        \node[] (L2) at ([shift={(0,-1.5)}]L1) {};
        \node[rect={solid,thick,draw=black}{}{solid,thick,draw=black}{solid,thick,draw=black}] (B1) at ([shift={(-5,0)}]L2) {\small $\SRPQb{1}{2}$};
        \node[rect={solid,thick,draw=black}{}{solid,thick,draw=black}{solid,thick,draw=black}] (A1) at ([shift={(-2,0)}]L2) {\small $\SRPQr{1}$};
        \node[rect={solid,thick,draw=black}{}{solid,thick,draw=black}{solid,thick,draw=black}] (A2) at ([shift={(+2,0)}]L2) {\small $\SRPQr{2}$};
        \node[rect={solid,thick,draw=black}{}{solid,thick,draw=black}{solid,thick,draw=black}] (B2) at ([shift={(+5,0)}]L2) {\small $\SRPQb{2}{1}$};
        \node[draw,blue,circle,inner sep=0pt, minimum size=12pt] (RC1) at ([shift={(-1.5,0)}]B1) {\small $\snumRC{1}$};
        \node[] (RC1dec1) at ([shift={(-0.7,-0.6)}]RC1) {};  \draw[dashed,->>] (RC1) .. controls([shift={(0,-0.3)}]RC1) .. (RC1dec1);
        \node[] (RC1dec2) at ([shift={(+0.7,-0.6)}]RC1) {};  \draw[dashed,->>] (RC1) .. controls([shift={(0,-0.3)}]RC1) .. (RC1dec2);
        \node[draw,blue,circle,inner sep=0pt, minimum size=12pt] (RC2) at ([shift={(+1.5,0)}]B2) {\small $\snumRC{2}$};
        \node[] (RC2dec1) at ([shift={(-0.7,-0.6)}]RC2) {};  \draw[dashed,->>] (RC2) .. controls([shift={(0,-0.3)}]RC2) .. (RC2dec1);
        \node[] (RC2dec2) at ([shift={(+0.7,-0.6)}]RC2) {};  \draw[dashed,->>] (RC2) .. controls([shift={(0,-0.3)}]RC2) .. (RC2dec2);
        %-- Layer-{0,1,2} Lines
        \draw[blue,thick,->] (E1) .. controls([shift={(+1,0.1)}]E1) and ([shift={(-1,0.1)}]PM1) .. (PM1);
        \draw[blue,thick,->] (E2) .. controls([shift={(-1,0.1)}]E2) and ([shift={(+1,0.1)}]PM2) .. (PM2);
        \draw[blue,thick,->] (A2) .. controls([shift={(-1,0)}]A2) and ([shift={(+4,0.1)}]PM1) .. (PM1);
        \draw[blue,thick,->] (A1) .. controls([shift={(+1,0)}]A1) and ([shift={(-4,0.1)}]PM2) .. (PM2);
        \draw[->] (PM1) .. controls([shift={(0,-0.6)}]PM1) and ([shift={(0,+1)}]B1) .. (B1);
        \draw[->] (PM1) .. controls([shift={(0,-0.6)}]PM1) and ([shift={(0,+1)}]A1) .. (A1);
        \draw[->] (PM2) .. controls([shift={(0,-0.6)}]PM2) and ([shift={(0,+1)}]B2) .. (B2);
        \draw[->] (PM2) .. controls([shift={(0,-0.6)}]PM2) and ([shift={(0,+1)}]A2) .. (A2);
        \draw[blue,thick,->] (E1) -- (UC1);    \draw[blue,thick,->] (E2) -- (UC2);    \draw[blue,thick,->] (B1) -- (RC1);    \draw[blue,thick,->] (B2) -- (RC2);
        \draw[->] (UC1) .. controls([shift={(0,-0.6)}]UC1) and ([shift={(-0.5,+1)}]A1) .. (A1);
        \draw[->] (UC2) .. controls([shift={(0,-0.6)}]UC2) and ([shift={(+0.5,+1)}]A2) .. (A2);
        %-- Layer-3
        \node[] (L3) at ([shift={(0,-3.0)}]L2) {};
        \node[rect={solid,thick,draw=black}{}{solid,thick,draw=black}{solid,thick,draw=black}] (S1) at ([shift={(-6,0)}]L3) {\small $\SRPQsx{1}{2}{}$};
        \node[rect={solid,thick,draw=black}{}{solid,thick,draw=black}{solid,thick,draw=black}] (T1) at ([shift={(-1,0)}]L3) {\small $\SRPQsxSP{1}{2}$};
        \node[rect={solid,thick,draw=black}{}{solid,thick,draw=black}{solid,thick,draw=black}] (T2) at ([shift={(+1,0)}]L3) {\small $\SRPQsxSP{2}{1}$};
        \node[rect={solid,thick,draw=black}{}{solid,thick,draw=black}{solid,thick,draw=black}] (S2) at ([shift={(+6,0)}]L3) {\small $\SRPQsx{2}{1}{}$};
        \node[draw,blue,circle,inner sep=0pt, minimum size=12pt] (AM1) at ([shift={(+1.0,+1.5)}]S1) {\small $\snumAM{1}$};
        \node[] (AM1dec) at ([shift={(-0.7,-0.6)}]AM1) {};  \draw[dashed,->>] (AM1) .. controls([shift={(0,-0.3)}]AM1) .. (AM1dec);
        \node[draw,blue,circle,inner sep=0pt, minimum size=12pt] (AM2) at ([shift={(-1.0,+1.5)}]S2) {\small $\snumAM{2}$};
        \node[] (AM2dec) at ([shift={(+0.7,-0.6)}]AM2) {};  \draw[dashed,->>] (AM2) .. controls([shift={(0,-0.3)}]AM2) .. (AM2dec);
        \node[draw,blue,circle,inner sep=0pt, minimum size=12pt] (DX1) at ([shift={(-2.0,0)}]S1) {\small $\snumDX{1}{}$};
        \node[] (DX1dec) at ([shift={(-0.7,-0.6)}]DX1) {};  \draw[dashed,->>] (DX1) .. controls([shift={(0,-0.3)}]DX1) .. (DX1dec);
        \node[draw,blue,circle,inner sep=0pt, minimum size=12pt] (DX2) at ([shift={(+2.0,0)}]S2) {\small $\snumDX{2}{}$};
        \node[] (DX2dec) at ([shift={(+0.7,-0.6)}]DX2) {};  \draw[dashed,->>] (DX2) .. controls([shift={(0,-0.3)}]DX2) .. (DX2dec);
        \node[draw,blue,circle,inner sep=0pt, minimum size=12pt] (DX1*) at ([shift={(-2.0,0)}]T1) {\small $\snumDX{(\!1\!)}{}$};
        \node[] (DX1*dec) at ([shift={(-0.7,-0.6)}]DX1*) {};  \draw[dashed,->>] (DX1*) .. controls([shift={(0,-0.3)}]DX1*) .. (DX1*dec);
        \node[draw,blue,circle,inner sep=0pt, minimum size=12pt] (DX2*) at ([shift={(+2.0,0)}]T2) {\small $\snumDX{(\!2\!)}{}$};
        \node[] (DX2*dec) at ([shift={(+0.7,-0.6)}]DX2*) {};  \draw[dashed,->>] (DX2*) .. controls([shift={(0,-0.3)}]DX2*) .. (DX2*dec);
        %-- Layer-{2,3} Lines
        \draw[->] (RC1) .. controls([shift={(0,-0.6)}]RC1) .. (S1);
        \draw[->] (RC1) .. controls([shift={(0,-1.7)}]RC1) and ([shift={(0,+3)}]T2).. (T2);
        \draw[->] (RC2) .. controls([shift={(0,-0.6)}]RC2) .. (S2);
        \draw[->] (RC2) .. controls([shift={(0,-1.7)}]RC2) and ([shift={(0,+3)}]T1) .. (T1);
        \draw[blue,thick,->] (S2) .. controls([shift={(-1.5,0)}]S2) and ([shift={(+7.0,0)}]AM1) .. (AM1);
        \draw[blue,thick,->] (S1) .. controls([shift={(+1.5,0)}]S1) and ([shift={(-7.0,0)}]AM2) .. (AM2);
        \draw[blue,thick,->] (S1) -- (DX1);  \draw[blue,thick,->] (S2) -- (DX2);
        \draw[blue,thick,->] (T1) -- (DX1*); \draw[blue,thick,->] (T2) -- (DX2*);
        \draw[blue,thick,->] (A1) -- (AM1);
        \draw[blue,thick,->] (A2) -- (AM2);
        \draw[->] (UC1) .. controls([shift={(0,-2.0)}]UC1) and ([shift={(-0.7,+1)}]S1) .. (S1);
        \draw[->] (UC2) .. controls([shift={(0,-2.0)}]UC2) and ([shift={(+0.7,+1)}]S2) .. (S2);
        %-- Layer-4
        \node[] (L4) at ([shift={(0,-3.0)}]L3) {};
        \node[draw,blue,circle,inner sep=0pt, minimum size=12pt] (CX1) at ([shift={(-0.5,0)}]L4) {\small $\snumCX{1}$};
        \node[draw,blue,circle,inner sep=0pt, minimum size=12pt] (CX2) at ([shift={(-1.5,0)}]L4) {\small $\snumCX{2}$};
        \node[draw,blue,circle,inner sep=0pt, minimum size=12pt] (CX3) at ([shift={(+1.5,0)}]L4) {\small $\snumCX{3}$};
        \node[draw,blue,circle,inner sep=0pt, minimum size=12pt] (CX4) at ([shift={(+0.5,0)}]L4) {\small $\snumCX{4}$};
        \node[] (BraceStart) at ([shift={(-0.5,0)}]CX2) {};
        \node[] (BraceEnd) at ([shift={(+0.5,0)}]CX3) {};
        \draw[blue,thick,decoration={brace,mirror,raise=0.37cm},decorate] (BraceStart) -- (BraceEnd);
        \node[] (CX0) at ([shift={(0,-0.3)}]L4) {};
        \node[] (CX0dec1) at ([shift={(-0.7,-0.5)}]CX0) {};  \draw[dashed,->>] (CX0) .. controls([shift={(0,-0.2)}]CX0) .. (CX0dec1);
        \node[] (CX0dec2) at ([shift={(+0.7,-0.5)}]CX0) {};  \draw[dashed,->>] (CX0) .. controls([shift={(0,-0.2)}]CX0) .. (CX0dec2);
        \node[fill=none,draw=none,inner sep=0pt,minimum size=0pt] (PM1toM) at ([shift={(-2.5,0)}]CX2) {};
        \node[fill=none,draw=none,inner sep=0pt,minimum size=0pt] (PM2toM) at ([shift={(+2.5,0)}]CX3) {};
        \node[fill=none,draw=none,inner sep=0pt,minimum size=0pt] (AM1toM) at ([shift={(-3.0,0)}]CX2) {};
        \node[fill=none,draw=none,inner sep=0pt,minimum size=0pt] (AM2toM) at ([shift={(+3.0,0)}]CX3) {};
        \node[fill=none,draw=none,inner sep=0pt,minimum size=0pt] (RC1toM) at ([shift={(-1.0,0)}]S1) {};
        \node[fill=none,draw=none,inner sep=0pt,minimum size=0pt] (RC2toM) at ([shift={(+1.0,0)}]S2) {};
        %-- Layer-{3,4} Lines
        \draw[blue,thick,->] (S1) .. controls([shift={(+0.3,-1)}]S1) and ([shift={(-0.5,+1)}]CX1) .. (CX1);
        \draw[blue,thick,->] (S2) .. controls([shift={(-0.3,-1)}]S2) and ([shift={(+0.5,+1)}]CX1) .. (CX1);
        \draw[blue,thick,->] (S1) .. controls([shift={(   0,-1)}]S1) and ([shift={(-0.5,+1)}]CX2) .. (CX2);
        \draw[blue,thick,->] (T2) .. controls([shift={(-0.3,-1)}]T2) and ([shift={(+0.5,+1)}]CX2) .. (CX2);
        \draw[blue,thick,->] (T1) .. controls([shift={(+0.3,-1)}]T1) and ([shift={(-0.5,+1)}]CX3) .. (CX3);
        \draw[blue,thick,->] (S2) .. controls([shift={(   0,-1)}]S2) and ([shift={(+0.5,+1)}]CX3) .. (CX3);
        \draw[blue,thick,->] (T1) .. controls([shift={(   0,-1)}]T1) and ([shift={(-0.5,+1)}]CX4) .. (CX4);
        \draw[blue,thick,->] (T2) .. controls([shift={(   0,-1)}]T2) and ([shift={(+0.5,+1)}]CX4) .. (CX4);
        %-- Layer-5
        \node[] (L5) at ([shift={(0,-2.0)}]L4) {};
        \node[rect={solid,thick,draw=black}{}{solid,thick,draw=black}{solid,thick,draw=black}] (M) at ([shift={(-2,0)}]L5) {\small $~~~~~~~~~~~~\SRPQm~~~~~~~~~~~~$};
        \node[draw=none,rectangle,minimum width=1.5cm,minimum height=1.5cm] (Mleft) at ([shift={(-1.0,-0.5)}]M) {};
        \node[draw=none,rectangle,minimum width=1.5cm,minimum height=1.5cm] (Mright) at ([shift={(+1.0,-0.5)}]M) {};
        %\node[] (Mright) at ([shift={(+1.0,0)}]M) {};
        \node[rect={solid,thick,draw=black}{}{solid,thick,draw=black}{solid,thick,draw=black}] (X0) at ([shift={(+2,0)}]L5) {\small $\SRPQstar$};
        \node[] (X0left) at ([shift={(-1.0,0)}]X0) {};
        \node[] (X0right) at ([shift={(+1.0,0)}]X0) {};
        %-- Layer-{4,5} Lines
        \draw[->] (PM1) .. controls([shift={(0,-0.6)}]PM1) and ([shift={(0,+0.5)}]PM1toM) .. (PM1toM) .. controls([shift={(0,-1.5)}]PM1toM) and ([shift={(0,+1.0)}]Mleft) .. (Mleft);
        \draw[->] (AM1) .. controls([shift={(0,-0.6)}]AM1) and ([shift={(0,+0.5)}]AM1toM) .. (AM1toM) .. controls([shift={(0,-1.0)}]AM1toM) and ([shift={(-1.0,+1.75)}]Mleft) .. (Mleft);
        \draw[->] (RC1) .. controls([shift={(0,-0.6)}]RC1) and ([shift={(0,+0.5)}]RC1toM) .. (RC1toM) .. controls([shift={(0,-3.5)}]RC1toM) and ([shift={(-1.5,+2.0)}]Mleft) .. (Mleft);
        \draw[->] (PM2) .. controls([shift={(0,-0.6)}]PM2) and ([shift={(0,+0.5)}]PM2toM) .. (PM2toM) .. controls([shift={(0,-2.0)}]PM2toM) and ([shift={(+0.5,+2.2)}]Mright) .. (Mright);
        \draw[->] (AM2) .. controls([shift={(0,-0.6)}]AM2) and ([shift={(0,+0.5)}]AM2toM) .. (AM2toM) .. controls([shift={(0,-2.0)}]AM2toM) and ([shift={(+1.0,+2.0)}]Mright) .. (Mright);
        \draw[->] (RC2) .. controls([shift={(0,-0.6)}]RC2) and ([shift={(0,+0.5)}]RC2toM) .. (RC2toM) .. controls([shift={(0,-6.0)}]RC2toM) and ([shift={(+1.5,+2.0)}]Mright) .. (Mright);
        %\draw[->] (CX0) .. controls([shift={(0,-1.2)}]CX0) and ([shift={(0,+1.0)}]M) .. (M);
        \draw[->] (CX0) .. controls([shift={(0,-1.2)}]CX0) and ([shift={(0,+1.0)}]X0) .. (X0);
        %-- Layer-6
        \node[] (L6) at ([shift={(0,-1.0)}]L5) {};
        \node[draw,red,circle,inner sep=0pt, minimum size=12pt] (rUC1) at ([shift={(-5.1,0)}]L6) {\small $\rnumUC{1}$};
        \node[] (rUC1dec) at ([shift={(-0.7,-0.6)}]rUC1) {};  \draw[dashed,->>] (rUC1) .. controls([shift={(0,-0.3)}]rUC1) .. (rUC1dec);
        \node[draw,red,circle,inner sep=0pt, minimum size=12pt] (rUC2) at ([shift={(+5.1,0)}]L6) {\small $\rnumUC{2}$};
        \node[] (rUC2dec) at ([shift={(+0.7,-0.6)}]rUC2) {};  \draw[dashed,->>] (rUC2) .. controls([shift={(0,-0.3)}]rUC2) .. (rUC2dec);
        \node[draw,red,circle,inner sep=0pt, minimum size=12pt] (rRC) at ([shift={(-2,0)}]L6) {\small $\rnumRC$};
        \node[] (rRCdec1) at ([shift={(-0.7,-0.6)}]rRC) {};  \draw[dashed,->>] (rRC) .. controls([shift={(0,-0.3)}]rRC) .. (rRCdec1);
        \node[] (rRCdec2) at ([shift={(+0.7,-0.6)}]rRC) {};  \draw[dashed,->>] (rRC) .. controls([shift={(0,-0.3)}]rRC) .. (rRCdec2);
        \node[draw,red,circle,inner sep=0pt, minimum size=12pt] (rOX) at ([shift={(+2,0)}]L6) {\small $\rnumOX{}$};
        \node[] (rOXdec1) at ([shift={(-0.7,-0.6)}]rOX) {};  \draw[dashed,->>] (rOX) .. controls([shift={(0,-0.3)}]rOX) .. (rOXdec1);
        \node[] (rOXdec2) at ([shift={(+0.7,-0.6)}]rOX) {};  \draw[dashed,->>] (rOX) .. controls([shift={(0,-0.3)}]rOX) .. (rOXdec2);
        \node[draw,red,circle,inner sep=0pt, minimum size=12pt] (rDX1*) at ([shift={(-4.2,-1.0)}]L6) {\small $\rnumDX{(\!1\!)}{}$};
        \node[] (rDX1*dec) at ([shift={(-0.7,-0.6)}]rDX1*) {};  \draw[dashed,->>] (rDX1*) .. controls([shift={(0,-0.3)}]rDX1*) .. (rDX1*dec);
        \node[draw,red,circle,inner sep=0pt, minimum size=12pt] (rDX2*) at ([shift={(+4.2,-1.0)}]L6) {\small $\rnumDX{(\!2\!)}{}$};
        \node[] (rDX2*dec) at ([shift={(+0.7,-0.6)}]rDX2*) {};  \draw[dashed,->>] (rDX2*) .. controls([shift={(0,-0.3)}]rDX2*) .. (rDX2*dec);
        %-- Layer-{5,6} Lines
        \draw[red,thick,->] (M) -- (rRC); \draw[red,thick,->] (X0) -- (rOX);
        \draw[red,thick,->] (A1) .. controls([shift={(-1.0,0)}]DX1*) .. (rUC1);
        \draw[red,thick,->] (A2) .. controls([shift={(+1.0,0)}]DX2*) .. (rUC2);
        \draw[red,thick,->] (T1) .. controls([shift={(-0.3,-1)}]T1) and ([shift={(-3.5,1.0)}]CX2) .. (rDX1*);
        \draw[red,thick,->] (T2) .. controls([shift={(+0.3,-1)}]T2) and ([shift={(+3.5,1.0)}]CX3) .. (rDX2*);
        %-- Layer-7
        \node[] (L7) at ([shift={(0,-3.0)}]L6) {};
        \node[draw,blue,circle,inner sep=0pt, minimum size=12pt] (CX5) at ([shift={(-8,+1.5)}]L7) {\small $\snumCX{5}$};
        \node[draw,blue,circle,inner sep=0pt, minimum size=12pt] (CX7) at ([shift={(-7,+1.5)}]L7) {\small $\snumCX{7}$};
        \node[] (BraceStart) at ([shift={(-0.5,0)}]CX5) {};
        \node[] (BraceEnd) at ([shift={(+0.5,0)}]CX7) {};
        \draw[blue,thick,decoration={brace,mirror,raise=0.37cm},decorate] (BraceStart) -- (BraceEnd);
        \node[] (CX57) at ([shift={(-7.5,+1.2)}]L7) {};
        \node[] (CX57dec1) at ([shift={(-0.7,-0.5)}]CX57) {};  \draw[dashed,->>] (CX57) .. controls([shift={(0,-0.2)}]CX57) .. (CX57dec1);
        \node[] (CX57dec2) at ([shift={(+0.7,-0.5)}]CX57) {};  \draw[dashed,->>] (CX57) .. controls([shift={(0,-0.2)}]CX57) .. (CX57dec2);
        \node[draw,blue,circle,inner sep=0pt, minimum size=12pt] (CX6) at ([shift={(+8,+1.5)}]L7) {\small $\snumCX{6}$};
        \node[draw,blue,circle,inner sep=0pt, minimum size=12pt] (CX8) at ([shift={(+7,+1.5)}]L7) {\small $\snumCX{8}$};
        \node[] (BraceStart) at ([shift={(-0.5,0)}]CX8) {};
        \node[] (BraceEnd) at ([shift={(+0.5,0)}]CX6) {};
        \draw[blue,thick,decoration={brace,mirror,raise=0.37cm},decorate] (BraceStart) -- (BraceEnd);
        \node[] (CX68) at ([shift={(+7.5,+1.2)}]L7) {};
        \node[] (CX68dec1) at ([shift={(-0.7,-0.5)}]CX68) {};  \draw[dashed,->>] (CX68) .. controls([shift={(0,-0.2)}]CX68) .. (CX68dec1);
        \node[] (CX68dec2) at ([shift={(+0.7,-0.5)}]CX68) {};  \draw[dashed,->>] (CX68) .. controls([shift={(0,-0.2)}]CX68) .. (CX68dec2);
        \node[rect={solid,thick,draw=black}{}{solid,thick,draw=black}{solid,thick,draw=black}] (X1) at ([shift={(-5,0)}]L7) {\small $~~~~~~~~~~~~~~~~\SRPQx{1}{2}~~~~~~~~~~~~~~~~$};
        \node[draw=none,rectangle,minimum width=1.5cm,minimum height=1.5cm] (X1left) at ([shift={(-0.3,-0.5)}]X1) {};
        \node[] (X1forRC) at ([shift={(-0.3,+0.2)}]X1) {};
        \node[rect={solid,thick,draw=black}{}{solid,thick,draw=black}{solid,thick,draw=black}] (X2) at ([shift={(+5,0)}]L7) {\small $~~~~~~~~~~~~~~~~\SRPQx{2}{1}~~~~~~~~~~~~~~~~$};
        \node[draw=none,rectangle,minimum width=1.5cm,minimum height=1.5cm] (X2right) at ([shift={(+0.3,-0.5)}]X2) {};
        \node[] (X2forRC) at ([shift={(+0.3,+0.2)}]X2) {};
        %-- Layer-{6,7} Lines
        \draw[->] (rRC) .. controls([shift={(0,-2.0)}]rRC) and ([shift={(+2.0,+1.0)}]X1) .. (X1);
        \draw[->] (rRC) .. controls([shift={(0,-1.5)}]rRC) and ([shift={(-3.0,+1.0)}]X2) .. (X2);
        \draw[->] (rOX) .. controls([shift={(0,-1.5)}]rOX) and ([shift={(+3.0,+1.0)}]X1) .. (X1);
        \draw[->] (rOX) .. controls([shift={(0,-2.0)}]rOX) and ([shift={(-2.0,+1.0)}]X2) .. (X2);
        \draw[->] (UC1) .. controls([shift={(0,-1.0)}]UC1) and ([shift={(-2.7,0.0)}]DX1) .. (X1left);
        \draw[->] (UC2) .. controls([shift={(0,-1.0)}]UC2) and ([shift={(+2.7,0.0)}]DX2) .. (X2right);
        \draw[->] (RC1) .. controls([shift={(0,-1.0)}]RC1) and ([shift={(-2.0,-0.5)}]RC1toM) .. (X1forRC);
        \draw[->] (RC1) .. controls([shift={(0,-1.0)}]RC1) and ([shift={(-4.0,-6.0)}]RC1toM) .. (X2);
        \draw[->] (RC2) .. controls([shift={(0,-1.0)}]RC2) and ([shift={(+2.0,-0.5)}]RC2toM) .. (X2forRC);
        \draw[->] (RC2) .. controls([shift={(0,-1.0)}]RC2) and ([shift={(+4.0,-6.0)}]RC2toM) .. (X1);
        \draw[->] (AM1) .. controls([shift={(0,-2.0)}]AM1) and ([shift={(-3.0,+1.5)}]X1) .. (X2);
        \draw[->] (AM2) .. controls([shift={(0,-2.0)}]AM2) and ([shift={(+3.0,+1.5)}]X2) .. (X1);
        \draw[->] (CX0) .. controls([shift={(0,-4.0)}]CX0) and ([shift={(+4.0,0)}]X1) .. (X1);
        \draw[->] (CX0) .. controls([shift={(0,-4.0)}]CX0) and ([shift={(-4.0,0)}]X2) .. (X2);
        \draw[->] (DX1) .. controls([shift={(0,-3.0)}]DX1) and ([shift={(-1.0,0.5)}]X1) .. (X1);
        \draw[->] (DX2) .. controls([shift={(0,-3.0)}]DX2) and ([shift={(+1.0,0.5)}]X2) .. (X2);
        \draw[->] (DX1*) .. controls([shift={(0,-2.0)}]DX1*) and ([shift={(-3.5,0)}]AM1toM) .. (X1);
        \draw[->] (DX2*) .. controls([shift={(0,-2.0)}]DX2*) and ([shift={(+3.5,0)}]AM2toM) .. (X2);
        \draw[->] (rUC1) .. controls([shift={(0,-2.0)}]rUC1) .. (X1);
        \draw[->] (rUC2) .. controls([shift={(0,-2.0)}]rUC2) .. (X2);
        \draw[blue,thick,->] (S1) .. controls([shift={(-0.3,-1)}]S1) and ([shift={(0.0,+3.0)}]CX5) .. (CX5);
        \draw[blue,thick,->] (X2) .. controls([shift={(-6,-3.5)}]X1) and ([shift={(-1.0,0.0)}]CX5) .. (CX5);
        \draw[blue,thick,->] (T1) .. controls([shift={(-0.7,-1)}]T1) and ([shift={(0.0,+3.5)}]CX7) .. (CX7);
        \node[inner sep=0pt, minimum size=0pt] (CX5left) at ([shift={(-0.7,0)}]CX5) {};
        \draw[blue,thick,->] (X2) .. controls([shift={(-2.5,-1.5)}]X1) and ([shift={(0,-3.0)}]CX5left) .. (CX5left) .. controls([shift={(0,+1.5)}]CX5left) and ([shift={(-1.0,+1.5)}]CX7) .. (CX7);
        \draw[blue,thick,->] (S2) .. controls([shift={(+0.3,-1)}]S2) and ([shift={(0.0,+3.0)}]CX6) .. (CX6);
        \draw[blue,thick,->] (X1) .. controls([shift={(+6,-3.5)}]X2) and ([shift={(+1.0,0.0)}]CX6) .. (CX6);
        \draw[blue,thick,->] (T2) .. controls([shift={(+0.7,-1)}]T2) and ([shift={(0.0,+3.5)}]CX8) .. (CX8);
        \node[inner sep=0pt, minimum size=0pt] (CX6right) at ([shift={(+0.7,0)}]CX6) {};
        \draw[blue,thick,->] (X1) .. controls([shift={(+2.5,-1.5)}]X2) and ([shift={(0,-3.0)}]CX6right) .. (CX6right) .. controls([shift={(0,+1.5)}]CX6right) and ([shift={(+1.0,+1.5)}]CX8) .. (CX8);
        \draw[->] (CX57) .. controls([shift={(0,-0.2)}]CX57) and ([shift={(-2.5,0.5)}]X1) .. (X1);
        \draw[->] (CX68) .. controls([shift={(0,-0.2)}]CX68) and ([shift={(+2.5,0.5)}]X2) .. (X2);
        \draw[->] (rDX1*) .. controls([shift={(0,-1.0)}]rDX1*) .. (X1);
        \draw[->] (rDX2*) .. controls([shift={(0,-1.0)}]rDX2*) .. (X2);
        %-- Layer-8
        \node[] (L8) at ([shift={(0,-2.0)}]L7) {};
        \node[draw,red,circle,inner sep=0pt, minimum size=12pt] (rCX) at ([shift={(0,0)}]L8) {\small $\rnumCX{}$};
        \node[] (rCXdec1) at ([shift={(-0.7,-0.6)}]rCX) {};  \draw[dashed,->>] (rCX) .. controls([shift={(0,-0.3)}]rCX) .. (rCXdec1);
        \node[] (rCXdec2) at ([shift={(+0.7,-0.6)}]rCX) {};  \draw[dashed,->>] (rCX) .. controls([shift={(0,-0.3)}]rCX) .. (rCXdec2);
        \node[draw,red,circle,inner sep=0pt, minimum size=12pt] (rDX1**) at ([shift={(-5,0)}]L8) {\small $\rnumDX{[1]}{}$};
        \node[] (rDX1**dec) at ([shift={(-0.7,-0.6)}]rDX1**) {};  \draw[dashed,->>] (rDX1**) .. controls([shift={(0,-0.3)}]rDX1**) .. (rDX1**dec);
        \node[draw,red,circle,inner sep=0pt, minimum size=12pt] (rDX2**) at ([shift={(+5,0)}]L8) {\small $\rnumDX{[2]}{}$};
        \node[] (rDX2**dec) at ([shift={(+0.7,-0.6)}]rDX2**) {};  \draw[dashed,->>] (rDX2**) .. controls([shift={(0,-0.3)}]rDX2**) .. (rDX2**dec);
        \node[rect={solid,thick,draw=black}{}{solid,thick,draw=black}{solid,thick,draw=black}] (D1) at ([shift={(-8.0,-1.0)}]L8) {\small $\SRPQd{1}$};
        \node[] (D1dec) at ([shift={(+0.8,+1.0)}]D1) {};  \draw[dashed,->>] (D1dec) .. controls([shift={(0,+0.5)}]D1) .. (D1);
        \node[rect={solid,thick,draw=black}{}{solid,thick,draw=black}{solid,thick,draw=black}] (D2) at ([shift={(+8.0,-1.0)}]L8) {\small $\SRPQd{1}$};
        \node[] (D2dec) at ([shift={(-0.8,+1.0)}]D2) {};  \draw[dashed,->>] (D2dec) .. controls([shift={(0,+0.5)}]D2) .. (D2);
        %-- Layer-{7,8} Lines
        \draw[red,thick,->] (X1) .. controls([shift={(+0.5,-0.5)}]X1) .. (rCX);
        \draw[red,thick,->] (X2) .. controls([shift={(-0.5,-0.5)}]X2) .. (rCX);
        \draw[red,thick,->] (X1) -- (rDX1**);   \draw[red,thick,->] (X2) -- (rDX2**);
        \end{tikzpicture}
    \caption{Illustrations of The Queueing Network described by the inequalities (\ref{prop:SRP:sim-inner:E}1) to (\ref{prop:SRP:sim-inner:D}2) in \PropRef{prop:SRP:sim-inner}. The upper-side-open rectangle represents the queue, and the circle represents LNC encoding operation, where the blue means the encoding by the source $s$ and the red means the encoding by the relay $r$. The black outgoing arrows from a LNC operation (or from a set of LNC operations grouped by a curly brace) represent the packet movements process depending on the reception status, where the southwest and southeast dashed arrows are especially for into $\SRPQd{1}$ and into $\SRPQd{2}$, respectively.}% $\SRPQm$, $\SRPQsxSP{1}{2}$, $\SRPQx{1}{2}$, and $\SRPQstar$.}
    \label{fig:SRP:Qnetwork}
\end{figure*}

Fig.~\ref{fig:SRP:Qnetwork} illustrates the proposed queueing network and movement process represented by \PropRef{prop:SRP:sim-inner}. The full/detailed descriptions of the LNC operations and the corresponding packet movement process following the inequalities in \PropRef{prop:SRP:sim-inner} are relegated to \AppRef{app:SRP:queue_invariance}.

\subsection{The Properties of Queues and The Correctness Proof}\label{sec:SRP:Queue-Property}

Each queue in the queueing network, see Fig.~\ref{fig:SRP:Qnetwork}, is carefully designed to store packets in a specific format such that the queue itself can represent a specific scenario to be beneficial. In this subsection, we highlight the properties of the queues, which later will be used to prove the correctness of our achievability scheme of \PropRef{prop:SRP:sim-inner}.

%{\bf [Revise from HERE!!!]}
To that end, we first describe the properties of $\SRPQe{1}$, $\SRPQd{1}$, $\SRPQr{1}$, and $\SRPQsx{1}{2}{}$ since their purpose is clear in the sense that these queues collect pure session-$1$ packets (indicated by the superscript), but heard only by the nodes (in the subscript $\{\cdot\}$) or correctly decoded by the desired destination $d_1$ (by the subscript ${\small\mathsf{dec}})$. After that, we describe the properties of $\SRPQm$, and then explain $\SRPQb{1}{2}$, $\SRPQsxSP{1}{2}$, and $\SRPQx{1}{2}$ focusing on the queues of session-$1$. For example, $\SRPQb{1}{2}$ implies the queue that contains the packet mixtures (the superscript $m$), each  of session-$1$ and session-$2$, where such mixtures are known by $d_2$ and those session-$2$ packets used for mixtures related to a session-$1$ packet that is mixed with a session-$2$ packet, where such mixture is known by $d_2$ but the session-$2$ packet is known by $r$ as well. The properties of the queues related to the session-$2$ packets, i.e., $\SRPQe{2}$, $\SRPQd{2}$, $\SRPQr{2}$, $\SRPQsx{2}{1}{}$, $\SRPQb{2}{1}$, $\SRPQsxSP{2}{1}$, and $\SRPQx{2}{1}$, will be symmetrically explained by simultaneously swapping (a) session-$1$ and session-$2$ in the superscript; (b) $X$ and $Y$; (c) $i$ and $j$; and (d) $d_1$ and $d_2$, if applicable. The property of $\SRPQstar$ will be followed at last.

%{\color{red}For notational convenience, we denote $\SRPQd{1\cup2}$ as shorthand for $\SRPQd{1}\cup\SRPQd{2}$, the union of two decoding queues for each session.}
To help aid the explanations, we also define for each node in $\{d_1,d_2,r\}$, the {\em reception list} $\RL{d_1}$, $\RL{d_2}$, and $\RL{r}$, respectively, that records how the received packet is constituted. The reception list is a binary matrix of its column size fixed to $n(\RB{1}+\RB{2})$ but its row size being the number of received packets and thus variable (increasing) over the course of total time slots. For example, suppose that $d_1$ has received a pure session-$1$ packet $X_1$, a self-mixture $[X_1+X_2]$, and a cross-mixture $[X_3+Y_1]$. Then $\RL{d_1}$ will be \vspace*{-\baselineskip}

\begin{figure}[H]
    \begin{tikzpicture}
        \node[] (C1) at (0,0) {};
        \node[align=center, text width=1cm] at ([shift={(0,0)}]C1) {{\small $n\RB{1}$}};
        \node[] (BraceStart) at ([shift={(-1.5,0)}]C1) {};
        \node[] (BraceEnd) at ([shift={(+1.5,0)}]C1) {};
        \draw[black,thick,decoration={brace,raise=-0.3cm},decorate] (BraceStart) -- (BraceEnd);
        \node[] (C2) at ([shift={(3,0)}]C1) {};
        \node[align=center, text width=1cm] at ([shift={(0,0)}]C2) {{\small $n\RB{2}$}};
        \node[] (BraceStart) at ([shift={(-1.5,0)}]C2) {};
        \node[] (BraceEnd) at ([shift={(+1.5,0)}]C2) {};
        \draw[black,thick,decoration={brace,raise=-0.3cm},decorate] (BraceStart) -- (BraceEnd);
        \node[align=center, text width=3cm] at ([shift={(0,-0.5)}]C1) {{\small $1$ $0$ $\cdots$ $\cdots$ $\cdots$ $\cdots$}};
        \node[align=center, text width=3cm] at ([shift={(0,-0.9)}]C1) {{\small $1$ $1$ $0\;\cdot$ $\,\cdots$ $\cdots$ $\cdots$}};
        \node[align=center, text width=3cm] at ([shift={(0,-1.3)}]C1) {{\small $0$ $0$ $1$ $0$ $\cdots$ $\cdots$ $\cdots$}};
        %\node[align=center, text width=3cm] at ([shift={(0,-1.7)}]C1) {{\small $0$ $0$ $\cdots$ $\cdots$ $\cdots$ $\cdots$}};
        \node[align=center, text width=3cm] at ([shift={(0,-0.5)}]C2) {{\small $0$ $0$ $\cdots$ $\cdots$ $\cdots$ $\cdots$}};
        \node[align=center, text width=3cm] at ([shift={(0,-0.9)}]C2) {{\small $0$ $0$ $\cdots$ $\cdots$ $\cdots$ $\cdots$}};
        \node[align=center, text width=3cm] at ([shift={(0,-1.3)}]C2) {{\small $1$ $0$ $\cdots$ $\cdots$ $\cdots$ $\cdots$ }};
        %\node[align=center, text width=3cm] at ([shift={(0,-1.7)}]C2) {{\small $0$ $1$ $0\;\cdot$ $\,\cdots$ $\cdots$ $\cdots$ }};
    \end{tikzpicture}
\end{figure}
\vspace*{-\baselineskip}

\noindent such that the first row vector represents the pure $X_1$ received, the second row vector represents the mixture $[X_1+X_2]$ received, and the third row vector represents the mixture $[X_3+Y_1]$ received, all in a binary format. Namely, whenever a node receives a packet, whether such packet is pure or not, a new $n(\RB{1}+\RB{2})$-dimensional row vector is inserted into the reception list by marking the corresponding entries of $X_i$ or $Y_j$ as flagged (``1") or not flagged (``0") accordingly. From the previous example, $[X_1+X_2]$ in the reception list $\RL{d_1}$ means that the list contains a $n(\RB{1}+\RB{2})$-dimensional row vector of exactly $\{1,1,0,\cdots,0\}$. We then say that a pure packet is {\em not flagged} in the reception list, if the column of the corresponding entry contains all zeros. From the previous example, the pure session-$2$ packet $Y_2$ is not flagged in $\RL{d_1}$, meaning that $d_1$ has neither received $Y_2$ nor any mixture involving this $Y_2$. Note that ``not flagged" is a stronger definition than ``unknown". From the previous example, the pure session-$1$ packet $X_3$ is unknown to $d_1$ but still flagged in $\RL{d_1}$ as $d_1$ has received the mixture $[X_3+Y_1]$ involving this $X_3$. Another example is the pure $X_2$ that is flagged in $\RL{d_1}$ but $d_1$ knows this $X_2$ as it can use the received $X_1$ and the mixture $[X_1+X_2]$ to extract $X_2$. We sometimes abuse the reception list notation to denote the collective reception list by $\mathsf{RL}_{T}$ for some non-empty subset $T\subseteq\{d_1,d_2,r\}$. For example, $\RL{d_1,d_2,r}$ implies the vertical concatenation of all $\RL{d_1}$, $\RL{d_2}$, and $\RL{r}$.

We now describe the properties of the queues.

\noindent \makebox[1.6cm][l]{$\bullet$ $\SRPQe{1}$:} Every packet in this queue is {\em of a pure session-$1$} and {\em unknown} to any of $\{d_1,d_2,r\}$, even {\em not flagged} in $\RL{d_1,d_2,r}$. Initially, this queue contains all the session-$1$ packets $\boldWB{1}$, and will be empty in the end.

\noindent \makebox[1.6cm][l]{$\bullet$ $\SRPQd{1}$:} Every packet in this queue is {\em of a pure session-$1$} and {\em known} to $d_1$. Initially, this queue is empty but will contain all the session-$1$ packets $\boldWB{1}$ in the end.

\noindent \makebox[1.6cm][l]{$\bullet$ $\SRPQr{1}$:} Every packet in this queue is {\em of a pure session-$1$} and {\em known} by $r$ but {\em unknown} to any of $\{d_1,d_2\}$, even {\em not flagged} in $\RL{d_1,d_2}$.

\noindent \makebox[1.6cm][l]{$\bullet$ $\SRPQsx{1}{2}{}$:} Every packet in this queue is {\em of a pure session-$1$} and {\em known} by $d_2$ but {\em unknown} to any of $\{d_1,r\}$, even {\em not flagged} in $\RL{d_1,r}$.

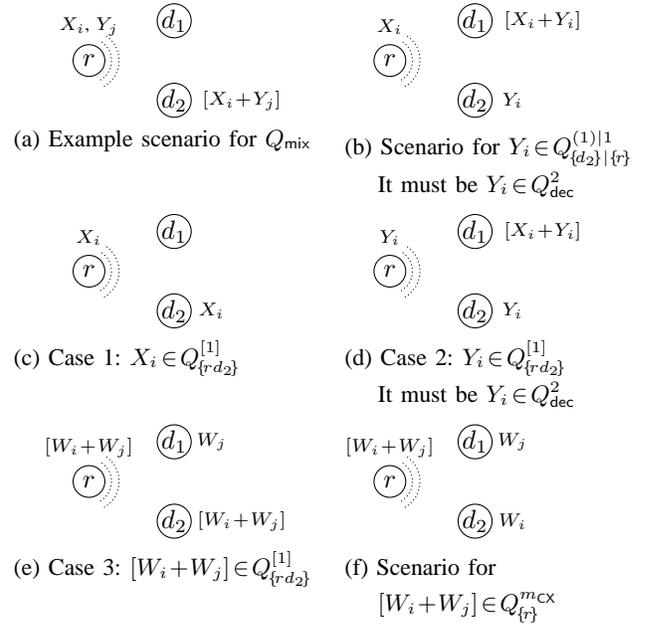
\begin{figure}
        \begin{tikzpicture}
        \node[draw,circle,inner sep=0pt,minimum size=12pt] (r) at (0,0) {$r$};
        \draw[densely dotted] ([shift={(-60:0.3)}]r) arc (-60:60:0.3);
        \draw[densely dotted] ([shift={(-60:0.4)}]r) arc (-60:60:0.4);
        \node[align=center] at ([shift={(0,0.45)}]r) {{\scriptsize $X_i$, $Y_j$}};
        \node[draw,circle,inner sep=0pt,minimum size=12pt] (d1) at ([shift={(25:1.25)}]r){$d_1$};
        \node[draw,circle,inner sep=0pt,minimum size=12pt] (d2) at ([shift={(-25:1.25)}]r){$d_2$};
        \node[align=flush left] at ([shift={(0.9,0)}]d2) {{\scriptsize $[X_i\!+\!Y_j]$}};
        \node[below=0.8cm, align=flush left, text width=5cm] at ([shift={(1.5,0)}]r) {{\small (a) Example scenario for $\SRPQm$}};
        \node[draw,circle,inner sep=0pt,minimum size=12pt] (r) at (4,0) {$r$};
        \draw[densely dotted] ([shift={(-60:0.3)}]r) arc (-60:60:0.3);
        \draw[densely dotted] ([shift={(-60:0.4)}]r) arc (-60:60:0.4);
        \node[align=center] at ([shift={(0,0.45)}]r) {{\scriptsize $X_i$}};
        \node[draw,circle,inner sep=0pt,minimum size=12pt] (d1) at ([shift={(25:1.25)}]r){$d_1$};
        \node[align=flush left] at ([shift={(0.9,0)}]d1) {{\scriptsize $[X_i\!+\!Y_i]$}};
        \node[draw,circle,inner sep=0pt,minimum size=12pt] (d2) at ([shift={(-25:1.25)}]r){$d_2$};
        \node[align=flush left] at ([shift={(0.5,0)}]d2) {{\scriptsize $Y_i$}};
        \node[below=0.8cm, align=flush left, text width=5cm] at ([shift={(1.9,0)}]r) {{\small (b) Scenario for $Y_i\!\in\!\SRPQsxSP{1}{2}$}};
        \node[below=1.35cm, align=flush left, text width=5cm] at ([shift={(1.9,0)}]r) {{\small $~~~$ It must be $Y_i\!\in\!\SRPQd{2}$}};
        \end{tikzpicture}
        \\[+3pt]
        \begin{tikzpicture}
        \node[draw,circle,inner sep=0pt,minimum size=12pt] (r) at (0,0) {$r$};
        \draw[densely dotted] ([shift={(-60:0.3)}]r) arc (-60:60:0.3);
        \draw[densely dotted] ([shift={(-60:0.4)}]r) arc (-60:60:0.4);
        \node[align=center] at ([shift={(0,0.45)}]r) {{\scriptsize $X_i$}};
        \node[draw,circle,inner sep=0pt,minimum size=12pt] (d1) at ([shift={(25:1.25)}]r){$d_1$};
        \node[draw,circle,inner sep=0pt,minimum size=12pt] (d2) at ([shift={(-25:1.25)}]r){$d_2$};
        \node[align=flush left] at ([shift={(0.5,0)}]d2) {{\scriptsize $X_i$}};
        \node[below=0.8cm, align=flush left, text width=5cm] at ([shift={(1.5,0)}]r) {{\small (c) Case~1: $X_i\!\in\!\SRPQx{1}{2}$}};
        \node[draw,circle,inner sep=0pt,minimum size=12pt] (r) at (4,0) {$r$};
        \draw[densely dotted] ([shift={(-60:0.3)}]r) arc (-60:60:0.3);
        \draw[densely dotted] ([shift={(-60:0.4)}]r) arc (-60:60:0.4);
        \node[align=center] at ([shift={(0,0.45)}]r) {{\scriptsize $Y_i$}};
        \node[draw,circle,inner sep=0pt,minimum size=12pt] (d1) at ([shift={(25:1.25)}]r){$d_1$};
        \node[align=flush left] at ([shift={(0.9,0)}]d1) {{\scriptsize $[X_i\!+\!Y_i]$}};
        \node[draw,circle,inner sep=0pt,minimum size=12pt] (d2) at ([shift={(-25:1.25)}]r){$d_2$};
        \node[align=flush left] at ([shift={(0.5,0)}]d2) {{\scriptsize $Y_i$}};
        \node[below=0.8cm, align=flush left, text width=5cm] at ([shift={(1.9,0)}]r) {{\small (d) Case~2: $Y_i\!\in\!\SRPQx{1}{2}$}};
        \node[below=1.35cm, align=flush left, text width=5cm] at ([shift={(1.9,0)}]r) {{\small $~~~$ It must be $Y_i\!\in\!\SRPQd{2}$}};
        \end{tikzpicture}
        \\[+3pt]
        \begin{tikzpicture}
        \node[draw,circle,inner sep=0pt,minimum size=12pt] (r) at (0,0) {$r$};
        \draw[densely dotted] ([shift={(-60:0.3)}]r) arc (-60:60:0.3);
        \draw[densely dotted] ([shift={(-60:0.4)}]r) arc (-60:60:0.4);
        \node[align=center] at ([shift={(0,0.45)}]r) {{\scriptsize $[W_i\!+\!W_j]$}};
        \node[draw,circle,inner sep=0pt,minimum size=12pt] (d1) at ([shift={(25:1.25)}]r){$d_1$};
        \node[align=flush left] at ([shift={(0.5,0)}]d1) {{\scriptsize $W_j$}};
        \node[draw,circle,inner sep=0pt,minimum size=12pt] (d2) at ([shift={(-25:1.25)}]r){$d_2$};
        \node[align=flush left] at ([shift={(0.9,0)}]d2) {{\scriptsize $[W_i\!+\!W_j]$}};
        \node[below=0.8cm, align=flush left, text width=5cm] at ([shift={(1.5,0)}]r) {{\small (e) Case~3: $[W_i\!+\!W_j]\!\in\!\SRPQx{1}{2}$}};
        \node[draw,circle,inner sep=0pt,minimum size=12pt] (r) at (4,0) {$r$};
        \draw[densely dotted] ([shift={(-60:0.3)}]r) arc (-60:60:0.3);
        \draw[densely dotted] ([shift={(-60:0.4)}]r) arc (-60:60:0.4);
        \node[align=center] at ([shift={(0,0.45)}]r) {{\scriptsize $[W_i\!+\!W_j]$}};
        \node[draw,circle,inner sep=0pt,minimum size=12pt] (d1) at ([shift={(25:1.25)}]r){$d_1$};
        \node[align=flush left] at ([shift={(0.5,0)}]d1) {{\scriptsize $W_j$}};
        \node[draw,circle,inner sep=0pt,minimum size=12pt] (d2) at ([shift={(-25:1.25)}]r){$d_2$};
        \node[align=flush left] at ([shift={(0.5,0)}]d2) {{\scriptsize $W_i$}};
        \node[below=0.9cm, align=flush left, text width=5cm] at ([shift={(1.9,0)}]r) {{\small (f) Scenario for}};
        \node[below=1.35cm, align=flush left, text width=5cm] at ([shift={(1.9,0)}]r) {{\small $~~~$ $[W_i\!+\!W_j]\!\in\!\SRPQstar$}};
        \end{tikzpicture}
        \caption{Illustrations of Scenarios of the Queues.}% $\SRPQm$, $\SRPQsxSP{1}{2}$, $\SRPQx{1}{2}$, and $\SRPQstar$.}
        \label{fig:SRP:queue_scenario}
\end{figure}

\noindent \makebox[1.6cm][l]{$\bullet$ $\SRPQm$:} Every packet in this queue is {\em of a linear sum $[X_i+Y_j]$} from a session-$1$ packet $X_i$ and a session-$2$ packet $Y_j$ such that at least one of the following conditions hold:
\begin{itemize}%\addtolength{\itemindent}{0.5cm}
\item[(a)] $[X_i+Y_j]$ is in $\RL{d_1}$; $X_i$ is {\em unknown} to $d_1$; and $Y_j$ is {\em known} by $r$ but {\em unknown} to $d_2$.
\item[(b)] $[X_i+Y_j]$ is in $\RL{d_2}$; $X_i$ is {\em known} by $r$ but {\em unknown} to $d_1$; and $Y_j$ is {\em unknown} to $d_2$.
\end{itemize}
The detailed clarifications are as follows. For a NC designer, one important consideration is to generate as many ``all-happy" scenarios as possible in an efficient manner so that single transmission benefits both destination simultaneously. One famous example is the {\em classic XOR} operation that a sender transmits a linear sum $[X_i+Y_j]$ when a session-$1$ packet $X_i$ is not yet delivered to $d_1$ but overheard by $d_2$ and a session-$2$ packet $Y_j$ is not yet delivered to $d_2$ but overheard by $d_1$. Namely, the source $s$ can perform such classic butterfly-style operation of sending the linear mixture $[X_i+Y_j]$ whenever such pair of $X_i$ and $Y_j$ is available. Similarly, $\SRPQm$ represents such an ``all-happy" scenario that the relay $r$ can benefit both destinations simultaneously by sending either $X_i$ or $Y_j$. For example, suppose that the source $s$ has transmitted a packet mixture $[X_i+Y_j]$ and it is received by $d_2$ only. And assume that $r$ already knows the individual $X_i$ and $Y_j$ but $X_i$ is unknown to $d_1$, see Fig.~\ref{fig:SRP:queue_scenario}(a). This example scenario falls into the second condition of $\SRPQm$ above. Then sending $X_i$ from the relay $r$ simultaneously enables $d_1$ to receive the desired $X_i$ and $d_2$ to decode the desired $Y_j$ by subtracting the received $X_i$ from the known $[X_i+Y_j]$. $\SRPQm$ collects such all-happy mixtures $[X_i+Y_j]$ that has been received by either $d_1$ or $d_2$ or both. In the same scenario, however, notice that $r$ cannot benefit both destinations simultaneously, if $r$ sends $Y_j$, instead of $X_i$. As a result, we use the notation $[X_i+Y_j]\!:\!W$ to denote the specific packet $W$ (known by $r$) that $r$ can send to benefit both destinations. In this second condition scenario of Fig.~\ref{fig:SRP:queue_scenario}(a), $\SRPQm$ is storing $[X_i+Y_j]\!:\!X_i$.

\noindent \makebox[1.6cm][l]{$\bullet$ $\SRPQb{1}{2}$:} Every packet in this queue is {\em of a linear sum $[X_i+Y_j]$} from a session-$1$ packet $X_i$ and a session-$2$ packet $Y_j$ such that they jointly satisfy the following conditions simultaneously.
\begin{itemize}%\addtolength{\itemindent}{0.5cm}
\item[(a)] $[X_i+Y_j]$ is in $\RL{d_2}$.
\item[(b)] $X_i$ is {\em unknown} to any of $\{d_1,d_2,r\}$, even {\em not flagged} in $\RL{d_1,r}$.
\item[(c)] $Y_j$ is {\em known} by $r$ but {\em unknown} to any of $\{d_1,d_2\}$, even {\em not flagged} in $\RL{d_1}$.
\end{itemize}

\noindent The scenario is the same as in Fig.~\ref{fig:SRP:queue_scenario}(a) when $r$ not having $X_i$. In this scenario, we have observed that $r$ cannot benefit both destinations by sending the known $Y_j$. $\SRPQb{1}{2}$ collects such unpromising $[X_i+Y_j]$ mixtures.

%%-- Back up for the original writings
%\noindent \makebox[1.6cm][l]{$\bullet$ $\SRPQb{1}{2}$:} Every packet in this queue is {\em of a linear sum $[X_i+Y_j]$} from a session-$1$ packet $X_i\!\not\in\!\SRPQd{1}$ and a session-$2$ packet $Y_j\!\not\in\!\SRPQd{2}$ such that it is {\em received by $d_2$ only} and {\em the pure $Y_j$ is only known by $r$.} Namely, {\em $[X_i+Y_j]$ is in $\RL{d_2}$} and {\em $Y_j$ is in $\RL{r}$} but {\em $X_i$ is not flagged in $\RL{d_1,r}$} and {\em $Y_j$ is not flagged in $\RL{d_1}$.} The scenario is the same as in Fig.~\ref{fig:SRP:queue_scenario}(a) when $r$ not having $X_i$. In this scenario, we have observed that $r$ cannot benefit both destinations by sending the known $Y_j$. $\SRPQb{1}{2}$ collects such unpromising $[X_i+Y_j]$ mixtures.

\noindent \makebox[1.6cm][l]{$\bullet$ $\SRPQsxSP{1}{2}$:} Every packet in this queue is {\em of a pure session-$2$} packet $Y_i$ such that there exists a pure session-$1$ packet $X_i$ that $Y_i$ is information equivalent to, and they jointly satisfy the following conditions simultaneously.
\begin{itemize}%\addtolength{\itemindent}{0.5cm}
\item[(a)] $[X_i+Y_i]$ is in $\RL{d_1}$.
\item[(b)] $X_i$ is {\em known} by $r$ but {\em unknown} to any of $\{d_1,d_2\}$.
\item[(c)] $Y_i$ is {\em known} by $d_2$ (i.e. already in $\SRPQd{2}$) but {\em unknown} to any of $\{d_1,r\}$, even not flagged in $\RL{r}$.
\end{itemize}

\noindent %Specifically, this queue represents the scenario that $d_1$ has received a mixture $[X_i+Y_i]$ and thus $X_i\!\not\in\!\SRPQd{1}$ can further be decoded whenever $d_1$ receives $Y_i$ later. And note that $Y_i$ is a session-$2$ packet but $Y_i$ is known by the desired $d_2$ already.
The concrete explanations are as follows. The main purpose of this queue is basically the same as $\SRPQsx{1}{2}{}$, i.e., to store session-$1$ packet overheard by $d_2$, so as to be used by the source $s$ for the classic XOR operation with the session-$2$ counterparts (e.g., any packet in $\SRPQsx{2}{1}{}$). Notice that any $X_i\!\in\!\SRPQsx{1}{2}{}$ is unknown to $r$ and thus $r$ cannot generate the corresponding linear mixture with the counterpart. However, because $X_i$ is unknown to the relay, $r$ cannot even naively deliver $X_i$ to the desired destination $d_1$. On the other hand, the queue $\SRPQsxSP{1}{2}$ here not only allows $s$ to perform the classic XOR operation but also admits naive delivery from $r$. To that end, consider the scenario in Fig.~\ref{fig:SRP:queue_scenario}(b). Here, $d_1$ has received a linear sum $[X_i+Y_i]$. Whenever $d_1$ receives $Y_i$ (session-$2$ packet), $d_1$ can use $Y_i$ and the known $[X_i+Y_i]$ to decode the desired $X_i$. This $Y_i$ is also known by $d_2$ (i.e., already in $\SRPQd{2}$), meaning that $Y_i$ is no more different than a session-$1$ packet overheard by $d_2$ but not yet delivered to $d_1$. Namely, such $Y_i$ can be treated as {\em information equivalent to $X_i$}. That is, using this session-$2$ packet $Y_i$ for the sake of session-$1$ does not incur any information duplicity because $Y_i$ is already received by the desired destination $d_2$.\footnote{This means that $d_2$ does not require $Y_i$ any more, and thus $s$ or $r$ can freely use this $Y_i$ in the network to represent not-yet-decoded $X_i$ instead.} For shorthand, we denote such $Y_i$ as $Y_i\equiv X_i$. As a result, the source $s$ can use this $Y_i$ as for session-$1$ when performing the classic XOR operation with a session-$2$ counterpart. Moreover, $r$ also knows the pure $X_i$ and thus relay can perform naive delivery for $d_1$ as well.

\noindent \makebox[1.6cm][l]{$\bullet$ $\SRPQx{1}{2}$:} Every packet in this queue is {\em of either a pure or a mixed} packet $\overline{W}$ satisfying the following conditions simultaneously.
\begin{itemize}%\addtolength{\itemindent}{0.5cm}
\item[(a)] $\overline{W}$ is {\em known} by both $r$ and $d_2$ but {\em unknown} to $d_1$.
\item[(b)] $d_1$ can extract a desired session-$1$ packet when $\overline{W}$ is further received.
\end{itemize}

\noindent Specifically, there are three possible cases based on how the packet $\overline{W}\!\in\!\SRPQx{1}{2}$ is constituted:

\begin{itemize}\addtolength{\itemindent}{0.6cm}
\item[Case~1:] $\overline{W}$ is a pure session-$1$ packet $X_i$. That is, $X_i$ is known by both $r$ and $d_2$ but unknown to $d_1$ as in Fig.~\ref{fig:SRP:queue_scenario}(c). Obviously, $d_1$ acquires this new $X_i$ when it is further delivered to $d_1$.
\item[Case~2:] $\overline{W}$ is a pure session-$2$ packet $Y_i\!\in\!\SRPQd{2}$. That is, $Y_i$ is already received by $d_2$ and known by $r$ as well but unknown to $d_1$. For such $Y_i$, as similar to the discussions of $\SRPQsxSP{1}{2}$, there exists a session-$1$ packet $X_i$ still unknown to $d_1$ where $X_i\equiv Y_i$, and their mixture $[X_i+Y_i]$ is in $\RL{d_1}$, see Fig.~\ref{fig:SRP:queue_scenario}(d). One can easily see that when $d_1$ further receives this $Y_i$, $d_1$ can use the received $Y_i$ and the known $[X_i+Y_i]$ to decode the desired $X_i$.
\item[Case~3:] $\overline{W}$ is a mixed packet of the form $[W_i+W_j]$ where $W_i$ and $W_j$ are pure but generic that can be either a session-$1$ or a session-$2$ packet. That is, the linear sum $[W_i+W_j]$ is known by both $r$ and $d_2$ but unknown to $d_1$. In this case, $W_i$ is still unknown to $d_1$ but $W_j$ is already received by $d_1$ so that whenever $[W_i+W_j]$ is delivered to $d_1$, $W_i$ can further be decoded. See Fig.~\ref{fig:SRP:queue_scenario}(e) for details. Specifically, there are two possible subcases depending on whether $W_i$ is of a pure session-$1$ or of a pure session-$2$:
    \begin{itemize}
    \item $W_i$ is a session-$1$ packet $X_i$. As discussed above, $X_i$ is unknown to $d_1$ and it is obvious that $d_1$ can decode the desired $X_i$ whenever $[W_i+W_j]$ is delivered to $d_1$.
    \item $W_i$ is a session-$2$ packet $Y_i\!\in\!\SRPQd{2}$. In this subcase, there exists a session-$1$ packet $X_i$ (other than $W_j$ in the above Case~3 discussions) still unknown to $d_1$ where $X_i\equiv Y_i$. Moreover, $[X_i+Y_i]$ is already in $\RL{d_1}$. As a result, $d_1$ can decode the desired $X_i$ whenever $[W_i+W_j]$ is delivered to $d_1$.
    \end{itemize}
\end{itemize}

\noindent The concrete explanations are as follows. The main purpose of this queue is basically the same as $\SRPQsxSP{1}{2}$ but the queue $\SRPQx{1}{2}$ here allows not only the source $s$ but also the relay $r$ to perform the classic XOR operation. As elaborated above, we have three possible cases depending on the form of the packet $\overline{W}\!\in\!\SRPQx{1}{2}$. Specifically, either a pure session-$1$ packet $X_i\!\not\in\!\SRPQd{1}$ (Case~1) or a pure session-$2$ packet $Y_i\!\in\!\SRPQd{2}$ (Case~2) or a mixture $[W_i+W_j]$ (Case~3) will be used when either $s$ or $r$ performs the classic XOR operation with a session-$2$ counterpart. For example, suppose that we have a packet $X\!\in\!\SRPQx{2}{1}$ (Case~2) as a session-$2$ counterpart. Symmetrically following the Case~2 scenario of $\SRPQx{1}{2}$ in Fig.~\ref{fig:SRP:queue_scenario}(d), we know that ${X}$ has been received by both $r$ and $d_1$. There also exists a session-$2$ packet $Y$ still unknown to $d_2$ where $Y\equiv X$, of which their mixture $[X+Y]$ is already in $\RL{d_2}$. For this session-$2$ counterpart $X$, consider any packet $\overline{W}$ in $\SRPQx{1}{2}$. Obviously, the relay $r$ knows both $\overline{W}$ and $X$ by assumption. As a result, either $s$ or $r$ can send their linear sum $[\overline{W}+X]$ as per the classic pairwise XOR operation. Since $d_1$ already knows $X$ by assumption, such mixture $[\overline{W}+X]$, when received by $d_1$, can be used to decode $\overline{W}$ and further decode a desired session-$1$ packet as discussed above. Moreover, if $d_2$ receives $[\overline{W}+X]$, then $d_2$ can use the known $\overline{W}$ to extract $X$ and further decode the desired $Y$ since $[X+Y]$ is already in $\RL{d_2}$ by assumption.

\noindent \makebox[1.6cm][l]{$\bullet$ $\SRPQstar$:} Every packet in this queue is {\em of a linear sum} $[W_i+W_j]$ that satisfies the following conditions simultaneously.

\begin{itemize}%\addtolength{\itemindent}{0.5cm}
\item[(a)] $[W_i+W_j]$ is in $\RL{r}$.
\item[(b)] $W_i$ is {\em known} by $d_2$ but {\em unknown} to any of $\{d_1,r\}$.
\item[(c)] $W_j$ is {\em known} by $d_1$ but {\em unknown} to any of $\{d_2,r\}$.
\end{itemize}

\noindent where $W_i$ and $W_j$ are pure but generic that can be either a session-$1$ or a session-$2$ packet. Specifically, there are four possible cases based on the types of $W_i$ and $W_j$ packets:

\begin{itemize}\addtolength{\itemindent}{0.6cm}
\item[Case~1:] $W_i$ is a pure session-$1$ packet $X_i$ and $W_j$ is a pure session-$2$ packet $Y_j$.
\item[Case~2:] $W_i$ is a pure session-$1$ packet $X_i$ and $W_j$ is a pure session-$1$ packet $X_j\!\in\!\SRPQd{1}$. For the latter $X_j$ packet, as similar to the discussions of $\SRPQsxSP{1}{2}$, there also exists a pure session-$2$ packet $Y_j$ still unknown to $d_2$ where $Y_j\equiv X_j$ and their mixture $[X_j+Y_j]$ is already in $\RL{d_2}$. As a result, later when $d_2$ decodes this $X_j$, $d_2$ can use $X_j$ and the known $[X_j+Y_j]$ to decode the desired $Y_j$.
\item[Case~3:] $W_i$ is a pure session-$2$ packet $Y_i\!\in\!\SRPQd{2}$ and $W_j$ is a pure session-$2$ packet $Y_j$. For the former $Y_i$ packet, there also exists a pure session-$1$ packet $X_i$ still unknown to $d_1$ where $X_i\equiv Y_i$ and $[X_i+Y_i]$ is already in $\RL{d_1}$. As a result, later when $d_1$ decodes this $Y_i$, $d_1$ can use $Y_i$ and the known $[X_i+Y_i]$ to decode the desired $X_i$.
\item[Case~3:] $W_i$ is a pure session-$2$ packet $Y_i\!\in\!\SRPQd{2}$ and $W_j$ is a pure session-$1$ packet $X_j\!\in\!\SRPQd{1}$. For the former $Y_i$ and the latter $X_j$ packets, the discussions follow the Case~3 and Case~2 above, respectively.
\end{itemize}

%\begin{itemize}
%\item[(i)] $[W_i+W_j]$ is generated by $\snumCX{1}$ and received only by $r$.
%\item[(i)] $[W_i+W_j]$ is generated by $\snumCX{2}$ and received only by $r$.
%\item[(i)] $[W_i+W_j]$ is generated by $\snumCX{3}$ and received only by $r$.
%\item[(i)] $[W_i+W_j]$ is generated by $\snumCX{4}$ and received only by $r$.
%\end{itemize}

%Note that both $W_i$ and $W_j$ are generic and thus can be either a session-$1$ or a session-$2$ packet. But we require that {\em before $W_i$ (resp. $W_j$) reached to $\SRPQstar$, $W_i$ was in either $\SRPQsx{1}{2}{}$ or $\SRPQsxSP{1}{2}$ (resp. either $\SRPQsx{2}{1}{}$ or $\SRPQsxSP{2}{1}$).}

\setlength{\tabcolsep}{3pt}
\begin{table}[h]
    \caption{Summary of the associated LNC operations \\ that moves packets into and takes packets out of.}
    \centering
    {\footnotesize
    \renewcommand\arraystretch{1.25}
    \begin{tabular}{|c|c|c|}
        \hline
        \multicolumn{1}{|c|}{\bf{LNC operations $\mapsto$}} & \multicolumn{1}{c|}{\bf{Queue}} & \multicolumn{1}{c|}{\bf{$\mapsto$ LNC operations}} \\
        \hline
        \hline
        & $\SRPQe{1}$ & {$\snumUC{1}$, $\snumPM{1}$} \\[+3pt]
        \hline
        {$\snumUC{1}$, $\snumPM{1}$} & $\SRPQr{1}$ & {$\snumPM{2}$, $\snumAM{1}$, $\rnumUC{1}$} \\[+3pt]
        \hline
        {$\snumPM{1}$} & $\SRPQb{1}{2}$ & {$\snumRC{1}$}  \\[+3pt]
        \hline
        \multirow{2}{*}{$\snumUC{1}$, $\snumRC{1}$} & \multirow{2}{*}{$\SRPQsx{1}{2}{}$} & {$\snumAM{2}$, $\snumDX{1}{}$} \\
        & & {$\snumCX{1}$, $\snumCX{2}$, $\snumCX{5}$} \\[+3pt]
        \hline
        \multirow{2}{*}{$\snumRC{2}$} & \multirow{2}{*}{$\SRPQsxSP{1}{2}$} & {$\snumDX{(\!1\!)}{}$, $\snumCX{3}$} \\
        & & {$\snumCX{4}$, $\snumCX{7}$, $\rnumDX{(\!1\!)}{}$} \\[+3pt]
        \hline
        {$\snumUC{1}$, $\snumAM{2}$, $\snumRC{1}$, $\snumDX{1}{}$} & \multirow{2}{*}{$\SRPQx{1}{2}$ (Case~1)} & \multirow{6}{*}{$\begin{aligned} \snumCX{6}&, \snumCX{8} \\ \rnumDX{[1]}{}&, \rnumCX{} \end{aligned}$} \\
        {$\snumCX{5}$, $\rnumUC{1}$, $\rnumDX{(\!1\!)}{}$, $\rnumRC$} & & {} \\[+3pt]
        \cline{1-2}
        {$\snumAM{2}$, $\snumRC{2}$, $\snumDX{(\!1\!)}{}$} & \multirow{2}{*}{$\SRPQx{1}{2}$ (Case~2)} & {} \\
        {$\snumCX{7}$, $\rnumRC$} & & {} \\[+3pt]
        \cline{1-2}
        {$\snumCX{1}$, $\snumCX{2}$} & \multirow{2}{*}{$\SRPQx{1}{2}$ (Case~3)} & {} \\
        {$\snumCX{3}$, $\snumCX{4}$, $\rnumOX{}$} & & {} \\[+3pt]
        \hline
        {$\snumUC{1}$, $\snumAM{1}$, $\snumRC{1}$, $\snumRC{2}$} & \multirow{4}{*}{$\SRPQd{1}$} & \\
        {$\snumDX{1}{}$, $\snumDX{(\!1\!)}{}$, $\{\snumCX{1}\,\text{to}\,\snumCX{8}\}$} & & \\
        {$\rnumUC{1}$, $\rnumDX{(\!1\!)}{}$, $\rnumDX{[1]}{}$} & & \\
        {$\rnumRC$, $\rnumOX{}$, $\rnumCX{}$} & & \\[+3pt]
        \hline
        \hline
        {$\snumPM{1}$, $\snumPM{2}$, $\snumAM{1}$, $\snumAM{2}$} & \multirow{2}{*}{$\SRPQm$} & \multirow{2}{*}{$\rnumRC$} \\
        {$\snumRC{1}$, $\snumRC{2}$} & & \\[+3pt]
        \hline
        {$\snumCX{1}$, $\snumCX{2}$, $\snumCX{3}$, $\snumCX{4}$} & $\SRPQstar$ & {$\rnumOX{}$} \\[+3pt]
        \hline
        \hline
        & $\SRPQe{2}$ & {$\snumUC{2}$, $\snumPM{2}$} \\[+3pt]
        \hline
        {$\snumUC{2}$, $\snumPM{2}$} & $\SRPQr{2}$ & {$\snumPM{1}$, $\snumAM{2}$, $\rnumUC{2}$} \\[+3pt]
        \hline
        {$\snumPM{2}$} & $\SRPQb{2}{1}$ & {$\snumRC{2}$} \\[+3pt]
        \hline
        \multirow{2}{*}{$\snumUC{2}$, $\snumRC{2}$} & \multirow{2}{*}{$\SRPQsx{2}{1}{}$} & {$\snumAM{1}$, $\snumDX{2}{}$} \\
        & & {$\snumCX{1}$, $\snumCX{3}$, $\snumCX{6}$} \\[+3pt]
        \hline
        \multirow{2}{*}{$\snumRC{1}$} & \multirow{2}{*}{$\SRPQsxSP{2}{1}$} & {$\snumDX{(\!2\!)}{}$, $\snumCX{2}$} \\
        & & {$\snumCX{4}$, $\snumCX{8}$, $\rnumDX{(\!2\!)}{}$} \\[+3pt]
        \hline
        {$\snumUC{2}$, $\snumAM{1}$, $\snumRC{2}$, $\snumDX{2}{}$} & \multirow{2}{*}{$\SRPQx{2}{1}$ (Case~1)} & \multirow{6}{*}{$\begin{aligned} \snumCX{5}&, \snumCX{7} \\ \rnumDX{[2]}{}&, \rnumCX{} \end{aligned}$} \\
        {$\snumCX{6}$, $\rnumUC{2}$, $\rnumDX{(\!2\!)}{}$, $\rnumRC$} & & {} \\[+3pt]
        \cline{1-2}
        {$\snumAM{1}$, $\snumRC{1}$, $\snumDX{(\!2\!)}{}$} & \multirow{2}{*}{$\SRPQx{2}{1}$ (Case~2)} & {} \\
        {$\snumCX{8}$, $\rnumRC$} & & {} \\[+3pt]
        \cline{1-2}
        {$\snumCX{1}$, $\snumCX{2}$} & \multirow{2}{*}{$\SRPQx{2}{1}$ (Case~3)} & {} \\
        {$\snumCX{3}$, $\snumCX{4}$, $\rnumOX{}$} & & {} \\[+3pt]
        \hline
        {$\snumUC{2}$, $\snumAM{2}$, $\snumRC{1}$, $\snumRC{2}$} & \multirow{4}{*}{$\SRPQd{2}$} & \\
        {$\snumDX{2}{}$, $\snumDX{(\!2\!)}{}$, $\{\snumCX{1}\,\text{to}\,\snumCX{8}\}$} & & \\
        {$\rnumUC{2}$, $\rnumDX{(\!2\!)}{}$, $\rnumDX{[2]}{}$} & & \\
        {$\rnumRC$, $\rnumOX{}$, $\rnumCX{}$} & & \\[+3pt]
        \hline
    \end{tabular}
    \label{tab:SRP:queue_in-out}
    }
\end{table}
\setlength{\tabcolsep}{6pt}

\noindent The concrete explanations are as follows. This queue represents the ``all-happy" scenario as similar to the butterfly-style operation by the relay $r$, i.e., sending a linear mixture $[W_i+W_j]$ using $W_i$ heard by $d_2$ and $W_j$ heard by $d_1$. Originally, $r$ must have known both individuals packets $W_i$ and $W_j$ to generate their linear sum. However, the sender in fact does not need to know both individuals to perform this classic XOR operation. The sender can still do the same operation even though it knows the linear sum $[W_i+W_j]$ only. This possibility only applies to the relay $r$ as all the messages including both individual packets are originated from the source $s$. As a result, this queue represents such scenario that the relay $r$ only knows the linear sum instead of individuals, as in Fig.~\ref{fig:SRP:queue_scenario}(f). More precisely, Cases~1 to~4 happen when the source $s$ performed one of four classic XOR operations $\snumCX{1}$ to $\snumCX{4}$, respectively, and the corresponding linear sum is received only by $r$, see \AppRef{app:SRP:queue_invariance} for details.

Based on the properties of queues, we now describe the correctness of \PropRef{prop:SRP:sim-inner}, our LNC inner bound. To that end, we first investigate all the LNC operations involved in \PropRef{prop:SRP:sim-inner} and prove the ``Queue Invariance", i.e., the queue properties explained above {\em remains invariant regardless of an LNC operation chosen}. Such long and tedious investigations are relegated to \AppRef{app:SRP:queue_invariance}. Then, the decodability condition \Ref{prop:SRP:sim-inner:D}, jointly with the Queue Invariance, imply that $\SRPQd{1}$ and $\SRPQd{2}$ will contain at least $n\RB{1}$ and $n\RB{2}$ number of pure session-$1$ and pure session-$2$ packets, respectively, in the end. This further means that, given a rate vector $(\RB{1},\RB{2})$, any $t$-, $s$-, and $r$-variables that satisfy the inequalities \Ref{prop:SRP:sim-inner:E} to \Ref{prop:SRP:sim-inner:D} in \PropRef{prop:SRP:sim-inner} will be achievable. The correctness proof of \PropRef{prop:SRP:sim-inner} is thus complete.

For readability, we also describe for each queue, the associated LNC operations that moves packet into and takes packets out of, see Table~\ref{tab:SRP:queue_in-out}.

\subsection{The General LNC Inner Bound}\label{sec:SRP:General}

The LNC inner bound in \PropRef{prop:SRP:sim-inner} has focused on the strong-relaying scenario and has considered mostly on cross-packets-mixing operations (i.e., mixing packets from different sessions when benefiting both destinations simultaneously). We now describe the general LNC inner bound that works in arbitrary $s$-PEC and $r$-PEC distributions, and also introduces self-packets-mixing operations (i.e., mixing packets from the same session for further benefits).

\begin{prop}\label{prop:SRP:sim-innerV2} A rate vector $(\RB{1},\RB{2})$ is LNC-achievable if there exist $2$ non-negative variables $t_s$ and $t_r$, $(6\times2 + 8 + 3\times2)$ non-negative $s$-variables:
\begin{align*}
& \big\{\snumUC{k},\; \snumPM{k},\; \snumAM{k},\; \snumRC{k},\; \snumDX{k}{}\;, \snumDX{(\!k\!)}{},\; \text{for all}\;k\in\{1,2\} \big\}, \\
& \big\{\snumCX{l} \,(l\!=\!1,\cdots\!,8) \big\}, \\
& \big\{\snumSX{k}{l} \,(l\!=\!1,2,3) \; \text{for all}\;k\in\{1,2\} \big\}.
\end{align*}

\noindent and $(2\times(3 \times 2 + 3))$ non-negative $w$-variables: For all $h\in\{s,r\}$,
\begin{align*}
& \big\{ \wnumUC{h}{k},\; \wnumDX{h}{(\!k\!)}{},\; \wnumDX{h}{[k]}{} \;:\; \text{for all}\;k\in\{1,2\} \big\}, \\
& \big\{ \wnumRC{h},\; \wnumOX{h}{},\; \wnumCX{h}{} \big\},
\end{align*}

\noindent such that jointly they satisfy the following five groups of linear conditions:
\end{prop}

\noindent$\bullet$ Group~1, termed the {\em time-sharing condition}, has $3$ inequalities:
\begin{align}
1 & \geq t_s + t_r, \label{prop:SRP:sim-innerV2:TS1} \\
t_s & \geq \sum_{k\in\{1,2\}} \!\! \left( \snumUC{k} + \snumPM{k} + \snumAM{k} + \snumRC{k} + \snumDX{k}{} + \snumDX{(\!k\!)}{} \right) \nN \\
& \qquad + \sum_{l=1}^{8} \snumCX{l} + \!\!\sum_{k\in\{1,2\}} \!\! \left( \snumSX{k}{1} + \snumSX{k}{2} + \snumSX{k}{3} \right) \nN \\
& \qquad + \sum_{k\in\{1,2\}} \!\! \left( \wnumUC{s}{k} + \wnumDX{s}{(\!k\!)}{} + \wnumDX{s}{[k]}{}\right) \nN \\
& \qquad + \wnumRC{s} + \wnumOX{s}{} + \wnumCX{s}{}, \label{prop:SRP:sim-innerV2:TS2} \\
t_r & \geq \!\sum_{k\in\{1,2\}} \!\!\! \left( \wnumUC{r}{k} \!+\! \wnumDX{r}{(\!k\!)}{} \!+\! \wnumDX{r}{[k]}{}\right) + \wnumRC{r} + \wnumOX{r}{} + \wnumCX{r}{}. \label{prop:SRP:sim-innerV2:TS3}
\end{align}

\noindent$\bullet$ Group~2, termed the {\em packets-originating condition}, has $2$ inequalities: Consider any $i,j\in\{1,2\}$
satisfying $i\neq j$. For each $(i,j)$ pair (out of the two choices $(1,2)$ and $(2,1)$),
\begin{align}
\RB{i} \geq \left(\snumUC{i} + \snumPM{i}\right)\cdot\pSRPsimT{s}{d_i,d_j,r}, \label{prop:SRP:sim-innerV2:E}
\end{align}
where \Ref{prop:SRP:sim-innerV2:E} is the same to \Ref{prop:SRP:sim-inner:E} in \PropRef{prop:SRP:sim-inner}.
%%-- Back up for original writings
%\begin{align}
%\RB{1} \geq \left( \snumUC{1} + \snumPM{1} \right) \cdot\pSRPsimT{s}{d_1,d_2,r}, \label{prop:SRP:sim-innerV2:E1} \\
%\RB{2} \geq \left( \snumUC{2} + \snumPM{2} \right) \cdot\pSRPsimT{s}{d_1,d_2,r}. \label{prop:SRP:sim-innerV2:E2}
%\end{align}

\noindent$\bullet$ Group~3, termed the {\em packets-mixing condition}, has $4$ inequalities: For each $(i,j)$ pair,
\begin{align}
\begin{split}
& \left(\snumUC{i}+\snumPM{i}\right)\cdot\prOa{s}{d_i d_j}{r} \geq (\snumPM{j} + \snumAM{i}) \cdot \pSRPsimT{s}{d_i,d_j} \\
& \quad + \left(\snumSX{i}{1}+\snumSX{i}{2}\right)\cdot\pSRPsimT{s}{d_i,d_j} + \!\!\sum_{h\in\{s,r\}}\!\!\wnumUC{h}{i} \cdot \pSRPsimT{h}{d_i,d_j},
\end{split} \label{prop:SRP:sim-innerV2:A} \\
& \snumPM{i}\cdot\prOd{s}{d_i}{d_j}{r} \geq \snumRC{i} \cdot \pSRPsimT{s}{d_i,d_j,r}, \label{prop:SRP:sim-innerV2:B}
\end{align}
%%-- Back up for original writings
%\begin{align}
%%%
%& \left(\snumUC{1}+\snumPM{1}\right)\cdot\prOa{s}{d_1 d_2}{r} \geq \left\{ \begin{aligned} & \left(\snumPM{2}+\snumAM{1}\right) \cdot \pSRPsimT{s}{d_1,d_2} \\ & + {\color{blue}\left(\snumSX{1}{1}+\snumSX{1}{2}\right)\cdot\pSRPsimT{s}{d_1,d_2}} \\ & + \!\!\sum_{h\in\{s,r\}}\!\!\wnumUC{h}{1} \cdot \pSRPsimT{h}{d_1,d_2} \end{aligned} \right\}, \label{prop:SRP:sim-innerV2:A1} \\
%%%
%& \left(\snumUC{2}+\snumPM{2}\right)\cdot\prOa{s}{d_1 d_2}{r} \geq \left\{ \begin{aligned} & \left(\snumPM{1}+\snumAM{2}\right) \cdot \pSRPsimT{s}{d_1,d_2} \\ & + {\color{blue}\left(\snumSX{2}{1}+\snumSX{2}{2}\right)\cdot\pSRPsimT{s}{d_1,d_2}} \\ & + \!\!\sum_{h\in\{s,r\}}\!\!\wnumUC{h}{2} \cdot \pSRPsimT{h}{d_1,d_2} \end{aligned} \right\}, \label{prop:SRP:sim-innerV2:A2} \\
%%%
%& \qquad\quad\;\;\; \snumPM{1}\cdot\prOd{s}{d_1}{d_2}{r} \geq \snumRC{1} \cdot \pSRPsimT{s}{d_1,d_2,r}, \label{prop:SRP:sim-innerV2:B1} \\
%%%
%& \qquad\quad\;\;\; \snumPM{2}\cdot\prOb{s}{d_1}{d_2 r} \geq \snumRC{2} \cdot \pSRPsimT{s}{d_1,d_2,r}, \label{prop:SRP:sim-innerV2:B2}
%\end{align}
and the following one inequality:
\begin{align}
& \snumPM{1}\!\cdot\!\pSRPsimT{s}{d_1,d_2 r} + \snumPM{2}\!\cdot\!\pSRPsimT{s}{d_2,d_1 r} + \snumAM{1}\!\cdot\!\pSRPsimT{s}{\overline{d_1}{d_2}} \,+ \nN \\
& \snumAM{2}\!\cdot\!\pSRPsimT{s}{d_1\overline{d_2}} \!+\! \left( \snumRC{1} \!+\! \snumRC{2} \right) \!\cdot\! \prOa{s}{d_1d_2}{r} \geq \!\!\!\!\sum_{h\in\{s,r\}}\!\!\!\!\wnumRC{h}\!\cdot\pSRPsimT{h}{d_1,d_2}. \label{prop:SRP:sim-innerV2:M}
\end{align}
%%-- Back up for original writings
%\begin{align}
%\begin{split}
%& \left\{ \begin{aligned} & \snumPM{1}\cdot\pSRPsimT{s}{d_1,d_2 r} + \snumPM{2}\cdot\pSRPsimT{s}{d_2,d_1 r} \\ & + \snumAM{1}\cdot\pSRPsimT{s}{\overline{d_1}d_2} + \snumAM{2}\cdot\pSRPsimT{s}{d_1\overline{d_2}} \\ & + \left( \snumRC{1} + \snumRC{2} \right) \cdot \prOa{s}{d_1d_2}{r} \end{aligned}\right\} \\
%& \qquad\qquad\qquad\qquad\quad\;\;\; \geq \!\!\sum_{h\in\{s,r\}}\!\!\wnumRC{h}\cdot\pSRPsimT{h}{d_1,d_2}, \label{prop:SRP:sim-innerV2:M}
%\end{split}
%\end{align}
where \Ref{prop:SRP:sim-innerV2:B} is the same to \Ref{prop:SRP:sim-inner:B} in \PropRef{prop:SRP:sim-inner}.

\noindent$\bullet$ Group~4, termed the {\em classic XOR condition by source only}, has $4$ inequalities:
\begin{align}
& \left( \snumUC{i} + \snumRC{i} \right) \prOd{s}{d_i}{d_j}{r} \geq \left(\snumAM{j} + \snumDX{i}{}\right)\cdot\pSRPsimT{s}{d_i,r} \, + \nN \\
& \quad \left( \snumCX{1} + \snumCX{1+i} \right)\cdot\,\pSRPsimT{s}{d_i,r} + \snumCX{4+i}\cdot\pSRPsimT{s}{d_i,r} \, + \nN \\
& \quad \left(\snumSX{i}{1}+\snumSX{i}{3}\right)\cdot\pSRPsimT{s}{d_i,r}, \label{prop:SRP:sim-innerV2:S} \\
& \snumRC{j}\cdot\prOd{s}{d_i}{d_j}{r} + \snumSX{i}{1}\cdot\prOb{s}{d_i}{d_j r} \geq \snumDX{(\!i\!)}{}\cdot\pSRPsimT{s}{d_i,r}\, + \nN \\
& \; \!\!\sum_{h\in\{s,r\}}\!\!\wnumDX{h}{(\!i\!)}{}\cdot\pSRPsimT{h}{d_i,d_j}\,+ \left( \snumCX{1+j} + \snumCX{4} \right)\cdot \, \pSRPsimT{s}{d_i,r}\, + \nN \\
& \; \snumCX{6+i}\cdot\pSRPsimT{s}{d_i,r} + \snumSX{i}{2}\cdot\pSRPsimT{s}{d_i d_j,r} + \snumSX{i}{3}\cdot\pSRPsimT{s}{d_i r,d_j}. \label{prop:SRP:sim-innerV2:T}
\end{align}

\noindent$\bullet$ Group~5, termed the {\em XOR condition}, has $3$ inequalities:

\begin{align}
& \sum_{l=1}^{4}\snumCX{l}\cdot\prOa{s}{d_1 d_2}{r} \geq \!\!\sum_{h\in\{s,r\}}\!\!\wnumOX{h}{}\cdot\pSRPsimT{h}{d_1,d_2}, \label{prop:SRP:sim-innerV2:X0}
\end{align}
and for each $(i,j)$ pair,
\begin{align}
& \snumAM{j}\!\cdot\!\pSRPsimT{s}{d_id_j,\overline{d_i}r} + \Big( \snumUC{i}\!+\!\snumRC{i}\!+\!\snumRC{j}\!+\!\sum_{l=1}^{4}\snumCX{l}\Big)\cdot\!\prOa{s}{d_i}{d_j r} \nN \\
& + \left( \snumCX{4+i}+\snumCX{6+i}+\snumDX{i}{}+\snumDX{(\!i\!)}{} \right)\cdot\pSRPsimT{s}{\overline{d_i}r} \nN \\
& + \left(\snumSX{i}{1}+\snumSX{i}{2}+\snumSX{i}{3}\right)\cdot\left(\pSRPsimT{s}{d_j}+\pSRPsimT{s}{r}-\p{s}{d_id_jr}\right) \nN \\
& + \!\!\sum_{h\in\{s,r\}}\!\!\left( \wnumUC{h}{i} + \wnumRC{h} + \wnumDX{h}{(\!i\!)}{} + \wnumOX{h}{}\right) \cdot\pSRPsimT{h}{\overline{d_i}d_j} \nN \\
& \geq \left( \snumCX{7-i} + \snumCX{9-i} \right)\cdot\pSRPsimT{s}{d_i} \nN \\
& \quad + \sum_{h\in\{s,r\}}\!\!\left( \wnumCX{h} \!+ \wnumDX{h}{[i]}{} \right)\cdot\pSRPsimT{h}{d_i}. \label{prop:SRP:sim-innerV2:X}
\end{align}

\noindent$\bullet$ Group~6, termed the {\em decodability condition}, has $2$ inequalities: For each $(i,j)$ pair,
\begin{align}
& \!\!\!\Big( \snumUC{i} + \snumAM{j} + \!\!\!\!\!\!\sum_{\quad k\in\{1,2\}}\!\!\!\!\!\!\!\!\snumRC{k} + \sum_{l=1}^8 \snumCX{l} + \snumDX{i}{} + \snumDX{(\!i\!)}{} \Big) \cdot \pSRPsimT{s}{d_i} \nN \\
& \;\; + \left(\snumSX{i}{1} + \snumSX{i}{2} + \snumSX{i}{3}\right)\cdot\pSRPsimT{s}{d_i} \nN \\
& \;\; + \!\!\!\sum_{h\in\{s,r\}}\!\!\!\left(\wnumUC{h}{i} + \wnumRC{h} + \wnumOX{h}{} + \wnumCX{h}{}\right)\cdot\pSRPsimT{h}{d_i} \nN \\
& \;\; + \!\!\!\sum_{h\in\{s,r\}}\!\!\!\left(\wnumDX{h}{(\!i\!)}{} + \wnumDX{h}{[i]}{}\right)\cdot\pSRPsimT{h}{d_i} \geq \RB{i}, \label{prop:SRP:sim-innerV2:D}
\end{align}
%%-- Back up for original writings
%\begin{align}
%%%
%\begin{split}
%& \left( \snumUC{1} + \snumAM{1} + \!\!\!\!\!\!\sum_{\quad k\in\{1,2\}}\!\!\!\!\!\!\!\!\snumRC{k} + \sum_{l=1}^8 \snumCX{l} + \snumDX{1}{} + \snumDX{(\!1\!)}{} \right) \cdot \pSRPsimT{s}{d_1} \\
%& \;\; + {\color{blue}\left(\snumSX{1}{1} + \snumSX{1}{2} + \snumSX{1}{3}\right)\cdot\pSRPsimT{s}{d_1}} \\
%& \;\; + \!\!\!\sum_{h\in\{s,r\}}\!\!\!\left(\wnumUC{h}{1} + \wnumRC{h} + \wnumOX{h}{} + \wnumCX{h}{}\right)\cdot\pSRPsimT{h}{d_1} \\
%& \;\; + \!\!\!\sum_{h\in\{s,r\}}\!\!\!\left(\wnumDX{h}{(\!1\!)}{} + \wnumDX{h}{[1]}{}\right)\cdot\pSRPsimT{h}{d_1} \geq \RB{1},
%\end{split} \label{prop:SRP:sim-innerV2:D1} \\
%%%
%\begin{split}
%& \left( \snumUC{2} + \snumAM{2} + \!\!\!\!\!\!\sum_{\quad k\in\{1,2\}}\!\!\!\!\!\!\!\!\snumRC{k} + \sum_{l=1}^8 \snumCX{l} + \snumDX{2}{} + \snumDX{(\!2\!)}{} \right) \cdot \pSRPsimT{s}{d_2} \\
%& \;\; + {\color{blue}\left(\snumSX{2}{1} + \snumSX{2}{2} + \snumSX{2}{3}\right)\cdot\pSRPsimT{s}{d_2}} \\
%& \;\; + \!\!\!\sum_{h\in\{s,r\}}\!\!\!\left(\wnumUC{h}{2} + \wnumRC{h} + \wnumOX{h}{} + \wnumCX{h}{}\right)\cdot\pSRPsimT{h}{d_2} \\
%& \;\; + \!\!\!\sum_{h\in\{s,r\}}\!\!\!\left(\wnumDX{h}{(\!2\!)}{} + \wnumDX{h}{[2]}{}\right)\cdot\pSRPsimT{h}{d_2} \geq \RB{2}.
%\end{split} \label{prop:SRP:sim-innerV2:D2}
%%%
%\end{align}

The main difference to \PropRef{prop:SRP:sim-inner} (for the strong-relaying scenario) can be summarized as follows. Recall that all the messages $\boldWBtot=(\boldWB{1},\boldWB{2})$ are originated from the source $s$ and the knowledge space of the relay $r$ at time $t$, i.e., $\SPt{r}$ always satisfies $\Mtot\supseteq\SPt{r}$. As a result, $s$ can always mimic any LNC encoding operation that $r$ can perform regardless of any time $t\in\{1,\cdots,n\}$. Therefore, we allow $s$ to mimic the same encoding operations that $r$ does and thus the $r$-variables in \PropRef{prop:SRP:sim-inner} is now replaced by the $w$-variables associated with both $s$ and $r$, where the performer is distinguished by the superscript $(h)$, $h\in\{s,r\}$. For that, the conditions \Ref{prop:SRP:sim-inner:A}, \Ref{prop:SRP:sim-inner:T}, \Ref{prop:SRP:sim-inner:X0}, and \Ref{prop:SRP:sim-inner:X} that are associated with $r$-variables has changed to \Ref{prop:SRP:sim-innerV2:A}, \Ref{prop:SRP:sim-innerV2:T}, \Ref{prop:SRP:sim-innerV2:X}, and \Ref{prop:SRP:sim-innerV2:X}, respectively, by replacing $r$-variables into $w$-variables with the superscript $(h)$, $h\in\{s,r\}$. The $r$-PEC probabilities are also replaced by a generic notation $\pSRPsimT{h}{\cdot}$, $h\in\{s,r\}$. On the other hand, the other conditions that are associated only with $s$-variables, i.e., \Ref{prop:SRP:sim-inner:E}, \Ref{prop:SRP:sim-inner:B}, \Ref{prop:SRP:sim-inner:M}, and \Ref{prop:SRP:sim-inner:S} remain the same as before by \Ref{prop:SRP:sim-innerV2:E}, \Ref{prop:SRP:sim-innerV2:B}, \Ref{prop:SRP:sim-innerV2:M}, and \Ref{prop:SRP:sim-innerV2:S}, respectively. In addition to the above systematic changes, we also consider the more advanced LNC encoding operations that the source $s$ can do, i.e., self-packets-mixing operations $\{\snumSX{k}{l}\,(l\!=\!1,2,3) : \text{for all } k\!\in\!\{1,2\}\}$. By these newly added $6$ $s$-variables, \Ref{prop:SRP:sim-innerV2:A}, \Ref{prop:SRP:sim-innerV2:S} to \Ref{prop:SRP:sim-innerV2:T}, and \Ref{prop:SRP:sim-innerV2:X} to \Ref{prop:SRP:sim-innerV2:D} are updated accordingly.

The queueing network described in \SecRef{sec:SRP:Queue-Property} remains the same as before, but we have additional self-packets-mixing operations $\{\snumSX{k}{l}\,(l\!=\!1,2,3) : \text{for all } k\!\in\!\{1,2\}\}$ for the general LNC inner bound. The LNC encoding operations and the packet movement process of the newly added $s$-variables $\snumSX{k}{l}$ can also be found in \AppRef{app:SRP:queue_invariance:V2}.

\section{Numerical Evaluation}\label{sec:SRP:Eval}

\begin{figure}
    \includegraphics[width=9cm,height=4.5cm]{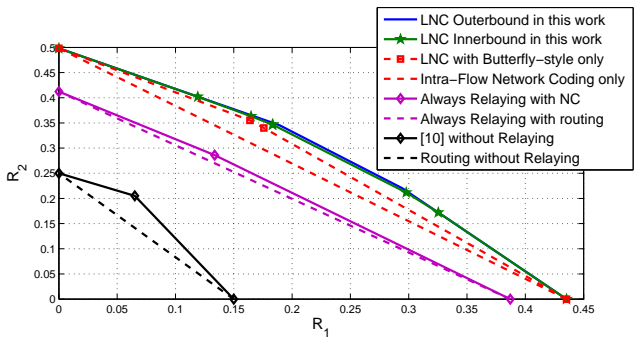}
    \caption{Comparison of LNC regions with different achievable rates\label{fig:SRP:comp}}
\end{figure}

Consider a smart repeater network with marginal channel success probabilities: (a) $s$-PEC: $\pSRPsimT{s}{d_1}=0.15$, $\pSRPsimT{s}{d_2}=0.25$, and $\pSRPsimT{s}{r}=0.8$; and (b) $r$-PEC: $\pSRPsimT{r}{d_1}=0.75$ and $\pSRPsimT{r}{d_2}=0.85$. And we assume that all the erasure events are independent. We will use the results in \PropsRef{prop:SRP:LP-outer}{prop:SRP:sim-inner} to find the largest $(\RB{1},\RB{2})$ value for this example scenario.

Fig.~\ref{fig:SRP:comp} compares the LNC capacity outer bound (\PropRef{prop:SRP:LP-outer}) and the LNC inner bound (\PropRef{prop:SRP:sim-inner}) with different achievability schemes. The smallest rate region is achieved by simply performing uncoded direct transmission without using the relay $r$. The second achievability scheme is the $2$-receiver broadcast channel LNC from the source $s$ in \cite{GeorgiadisTassiulas:NetCod09} while still not exploiting $r$ at all. The third and fourth schemes always use $r$ for any packet delivery. Namely, both schemes do not allow $2$-hop delivery from $s$. Then $r$ in the third scheme uses pure routing while $r$ performs the $2$-user broadcast channel LNC in the fourth scheme. The fifth scheme performs the time-shared transmission between $s$ and $r$, while allowing only intra-flow network coding. The sixth scheme is derived from using only the classic butterfly-style LNCs corresponding to $\snumCX{l}\,(l\!\!=\!\!1,\cdots\!,8)$, $\rnumCX{}$, and $\rnumOX{}$. That is, we do not allow $s$ to perform fancy operations such as $\snumPM{k}$, $\snumAM{k}$, $\snumRC{k}$, and $\rnumRC$. One can see that the result is strictly suboptimal.

In summary, one can see that our proposed LNC inner bound closely approaches to the LNC capacity outer bound in all angles. This shows that the newly-identified LNC operations other than the classic butterfly-style LNCs are critical in approaching the LNC capacity. The detailed rate region description of each sub-optimal achievability scheme can be found in \AppRef{app:SRP:Schemes}.

%%-- Back up for original writings
%The sixth and seventh schemes are derived from our achievability scheme with some restrictions. For the sixth scheme, we do not allow $s$ to perform fancy LNC operations such as $\snumPM{k}$, $\snumAM{k}$, and $\snumSX{k}{l}$. By (\ref{prop:SRP:sim-inner:B}1) to \Ref{prop:SRP:sim-inner:M}, the corresponding follow-up LNC operations such as $\snumRC{k}$ and $\rnumRC$ cannot be utilized. Namely, we only allow $s$ and $r$ to perform the classic butterfly-style XOR at most when benefiting both destinations simultaneously. One can see that the result is strictly suboptimal. However, the seventh scheme only shuts down the Self-XOR operations $\snumSX{k}{l}$, thus allowing the advanced premixing-$(\snumPM{k},\snumAM{k})$ and reactive-$(\snumRC{k},\wnumRC{h})$ styles in addition to the butterfly-style. It shows that the proposed fancy LNC operations other than the butterfly-style XORs are critical in approaching the LNC capacity.

\begin{figure}
    \includegraphics[width=9cm,height=4.5cm]{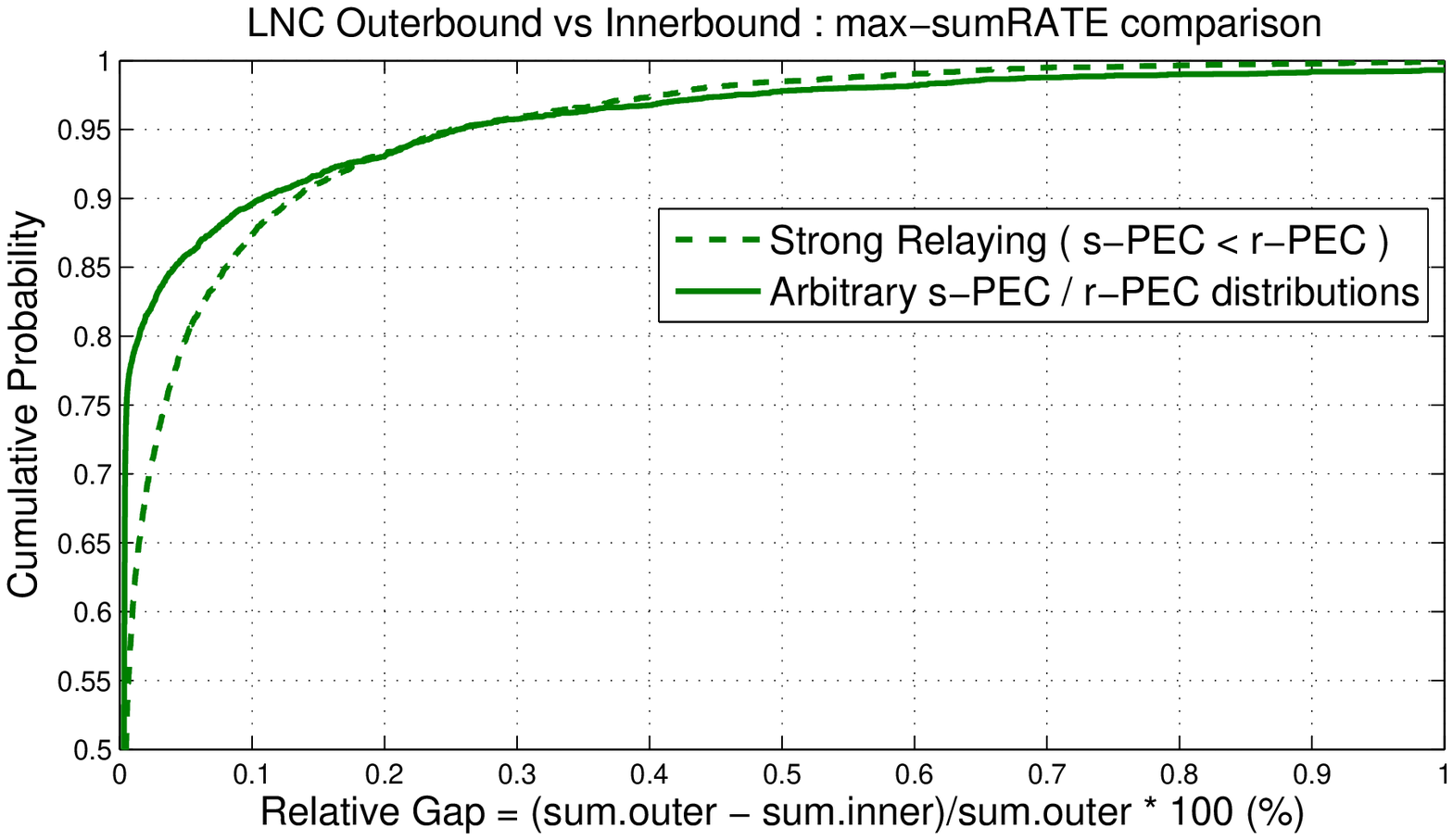}
    \caption{The cumulative distribution of the relative gap between the outer and the inner bounds. The LNC outer bound is described in \PropRef{prop:SRP:LP-outer}, and the inner bounds are described in \PropsRef{prop:SRP:sim-inner}{prop:SRP:sim-innerV2}, respectively.}
    \label{fig:SRP:CDF1}
\end{figure}

Fig.~\ref{fig:SRP:CDF1} examines the relative gaps between the outer bound and two inner bounds by choosing the channel parameters $\pSRPsimT{s}{\cdot}$ and $\pSRPsimT{r}{\cdot}$ uniformly randomly while obeying (a) the strong-relaying condition in \DefRef{def:SRP:strong-relaying} when using \PropRef{prop:SRP:sim-inner}; and (b) the arbitrary $s$-PEC and $r$-PEC distributions when using \PropRef{prop:SRP:sim-innerV2}. For any chosen parameter instance, we use a linear programming solver to find the largest sum rate $\Rsum$ of the LNC outer bound in \PropRef{prop:SRP:LP-outer}, which is denoted by $\RB{\mathsf{sum.outer}}$. Similarly, we find the largest sum rate $\Rsum$ that satisfies the LNC inner bound in \PropRef{prop:SRP:sim-inner} (resp. \PropRef{prop:SRP:sim-innerV2}) and denote it by $\RB{\mathsf{sum.inner}}$. We then compute the relative gap per each experiment, $(\RB{\mathsf{sum.outer}} - \RB{\mathsf{sum.inner}})/\RB{\mathsf{sum.outer}}$, and then repeat the experiment $10000$ times, and plot the cumulative distribution function (cdf) in unit of percentage. We can see that with more than $85\%$ of the experiments, the relative gap between the outer and inner bound is smaller than $0.08\%$ for Case~(a) and $0.04\%$ for Case~(b).

\section{Conclusion}\label{sec:SRP:Conclusion}

This work studies the LNC capacity of the smart repeater packet erasure network for two unicast flows. The capacity region has been effectively characterized by the proposed linear-subspace-based outer bound, and the capacity-approaching LNC scheme with newly identified LNC operations other than the previously well-known classic butterfly-style operations.

%\newpage
\appendices
\renewcommand{\thesubsectiondis}{\thesection-\arabic{subsection}.}
\renewcommand{\thesubsection}{\thesection-\arabic{subsection}}

\section{List of Coding Types for $\NETori$ and $\NEToriR$}\label{app:SRP:NETori}

We enumerate the $154$ {\em Feasible Types} (FTs) defined in \Ref{def:SRP:TYPE} that the source $s$ can transmit in the following way:
\begin{align}
\NETori \triangleq & \{ \msf{00000},\;\msf{00010},\;\msf{00020},\;\msf{00030},\;\msf{00070},\;\msf{00110}, \nonumber \\
&\msf{00130},\;\msf{00170},\;\msf{00220},\;\msf{00230},\;\msf{00270},\;\msf{00330},\nonumber\\
&\msf{00370},\;\msf{00570},\;\msf{00770},\;\msf{00A70},\;\msf{00B70},\;\msf{00F70},\nonumber\\
&\msf{00F71},\;\msf{01010},\;\msf{01030},\;\msf{01070},\;\msf{01110},\;\msf{01130},\nonumber\\
&\msf{01170},\;\msf{01230},\;\msf{01270},\;\msf{01330},\;\msf{01370},\;\msf{01570},\nonumber\\
&\msf{01770},\;\msf{01A70},\;\msf{01B70},\;\msf{01F70},\;\msf{01F71},\;\msf{02020},\nonumber\\
&\msf{02030},\;\msf{02070},\;\msf{02130},\;\msf{02170},\;\msf{02220},\;\msf{02230},\nonumber\\
&\msf{02270},\;\msf{02330},\;\msf{02370},\;\msf{02570},\;\msf{02770},\;\msf{02A70},\nonumber\\
&\msf{02B70},\;\msf{02F70},\;\msf{02F71},\;\msf{03030},\;\msf{03070},\;\msf{03130},\nonumber\\
&\msf{03170},\;\msf{03230},\;\msf{03270},\;\msf{03330},\;\msf{03370},\;\msf{03570},\nonumber\\
&\msf{03770},\;\msf{03A70},\;\msf{03B70},\;\msf{03F70},\;\msf{03F71},\;\msf{07070},\nonumber\\
&\msf{07170},\;\msf{07270},\;\msf{07370},\;\msf{07570},\;\msf{07770},\;\msf{07A70},\nonumber\\
&\msf{07B70},\;\msf{07F70},\;\msf{07F71},\;\msf{11110},\;\msf{11130},\;\msf{11170},\nonumber\\
&\msf{11330},\;\msf{11370},\;\msf{11570},\;\msf{11770},\;\msf{11B70},\;\msf{11F70},\nonumber\\
&\msf{11F71},\;\msf{13130},\;\msf{13170},\;\msf{13330},\;\msf{13370},\;\msf{13570},\nonumber\\
&\msf{13770},\;\msf{13B70},\;\msf{13F70},\;\msf{13F71},\;\msf{17170},\;\msf{17370},\nonumber\\
&\msf{17570},\;\msf{17770},\;\msf{17B70},\;\msf{17F70},\;\msf{17F71},\;\msf{22220},\nonumber\\
&\msf{22230},\;\msf{22270},\;\msf{22330},\;\msf{22370},\;\msf{22770},\;\msf{22A70},\nonumber\\
&\msf{22B70},\;\msf{22F70},\;\msf{22F71},\;\msf{23230},\;\msf{23270},\;\msf{23330},\nonumber\\
&\msf{23370},\;\msf{23770},\;\msf{23A70},\;\msf{23B70},\;\msf{23F70},\;\msf{23F71},\nonumber\\
&\msf{27270},\;\msf{27370},\;\msf{27770},\;\msf{27A70},\;\msf{27B70},\;\msf{27F70},\nonumber\\
&\msf{27F71},\;\msf{33330},\;\msf{33370},\;\msf{33770},\;\msf{33B70},\;\msf{33F70},\nonumber\\
&\msf{33F71},\;\msf{37370},\;\msf{37770},\;\msf{37B70},\;\msf{37F70},\;\msf{37F71},\nonumber\\
&\msf{57570},\;\msf{57770},\;\msf{57F70},\;\msf{57F71},\;\msf{77770},\;\msf{77F70},\nonumber\\
&\msf{77F71},\;\msf{A7A70},\;\msf{A7B70},\;\msf{A7F70},\;\msf{A7F71},\;\msf{B7B70},\nonumber\\
&\msf{B7F70},\;\msf{B7F71},\;\msf{F7F70},\;\msf{F7F71} \}, \nonumber %\label{def:SRP:TYPEs2}
\end{align}
where each $5$-digit index $\DIGIT{1}\DIGIT{2}\DIGIT{3}\DIGIT{4}\DIGIT{5}$ represent a $15$-bitstring $\b$ of which $\DIGIT{1}$ is a hexadecimal of first four bits, $\DIGIT{2}$ is a octal of the next three bits, $\DIGIT{3}$ is a hexadecimal of the next four bits, $\DIGIT{4}$ is a octal of the next three bits, and $\DIGIT{5}$ is binary of the last bit. The subset of $\NETori$ that the relay $r$ can transmit, i.e., $\NEToriR$ are listed separately in the following:
\begin{align}
\NEToriR \triangleq & \{ \msf{00F71},\;\msf{01F71},\;\msf{02F71},\;\msf{03F71},\;\msf{07F71},\;\msf{11F71},\nonumber\\
&\msf{13F71},\;\msf{17F71},\;\msf{22F71},\;\msf{23F71},\;\msf{27F71},\;\msf{33F71},\nonumber\\
&\msf{37F71},\;\msf{57F71},\;\msf{77F71},\;\msf{A7F71},\;\msf{B7F71},\;\msf{F7F71} \}, \nonumber %\label{def:SRP:TYPEr2}
\end{align}

Recall that the $b_{15}$ of a $15$-bitstring $\b$ represents whether the coding subset belongs to $\SRPAt{15}$ or not, and $\SRPAt{15} \TReq \SPtPREV{r}$ by definition \Ref{def:SRP:AsetR2}. As a result, any coding type with $b_{15}=1$ implies that it  lies in the knowledge space of the relay $r$. The enumerated $\NEToriR$ in the above is thus a collection of such coding subsets in $\NETori$ with $\DIGIT{5}=1$.

\section{LNC encoding operations, Packet Movement Process, and Queue Invariance in \PropRef{prop:SRP:sim-inner}}\label{app:SRP:queue_invariance}

In the following, we will describe all the LNC encoding operations and the corresponding packet movement process of \PropRef{prop:SRP:sim-inner} one by one, and then prove that the Queue Invariance explained in \SecRef{sec:SRP:Queue-Property} always holds.

To simplify the analysis, we will ignore the null reception, i.e., none of $\{d_1,d_2,r\}$ receives a transmitted packet, because nothing will happen in the queueing network. Moreover, we exploit the following symmetry: For those variables whose superscript indicates the session information $k\!\in\!\{1,2\}$ (either session-$1$ or session-$2$), here we describe session-$1$ ($k\!=\!1$) only. Those variables with $k\!=\!2$ in the superscript will be symmetrically explained by simultaneously swapping (a) session-$1$ and session-$2$ in the superscript; (b) $X$ and $Y$; (c) $i$ and $j$; and (d) $d_1$ and $d_2$, if applicable.

\noindent \makebox[1.2cm][l]{$\bullet$ $\snumUC{1}$:} The source $s$ transmits $X_i\!\in\!\SRPQe{1}$. Depending on the reception status, the packet movement process following the inequalities in \PropRef{prop:SRP:sim-inner} is summarized as follows.
\tableSTARTreduce
\begin{table}[H]
    \centering
    \subfloat[]{
        \centering
        {\footnotesize
        \begin{tabular}{|c|c|c|}
        \hline
        \multicolumn{1}{|c}{\bf{Departure}} & \multicolumn{1}{|m{1.2cm}|}{\centering \bf{Reception} \\ \bf{Status}} & \multicolumn{1}{c|}{\bf{Insertion}} \\
        \hline
        \hline
        \multirow{7}{*}{$\SRPQe{1}\!\xrightarrow{X_i}$} & $\overline{d_1d_2}r$ & $\xrightarrow{X_i}\SRPQr{1}$ \\[+3pt]
        \cline{2-3}
        & $\overline{d_1}d_2\overline{r}$ & $\xrightarrow{X_i}\SRPQsx{1}{2}{}$ \\[+3pt]
        \cline{2-3}
        & $d_1\overline{d_2r}$ & {$\xrightarrow{X_i}\SRPQd{1}$} \\
        \cline{2-3}
        & $\overline{d_1}d_2r$ & $\xrightarrow[\text{Case~1}]{X_i}\SRPQx{1}{2}$ \\[+3pt]
        \cline{2-3}
        & $d_1\overline{d_2}r$ & $\xrightarrow{X_i}\SRPQd{1}$ \\
        \cline{2-3}
        & $d_1d_2\overline{r}$ & $\xrightarrow{X_i}\SRPQd{1}$ \\
        \cline{2-3}
        & $d_1d_2r$ & $\xrightarrow{X_i}\SRPQd{1}$ \\
        \hline
        \end{tabular}
        }\label{tab:SRP:sUC1}
    }
\end{table}
\tableENDreduce
\begin{itemize}
\item[-] {\bf Departure}: One property for $X_i\!\in\!\SRPQe{1}$ is that $X_i$ must be unknown to any of $\{d_1,d_2,r\}$. As a result, whenever $X_i$ is received by any of them, $X_i$ must be removed from $\SRPQe{1}$ for the Queue Invariance.
    %Whenever $X_i$ is received by any of $\{r,d_1,d_2\}$, it is removed from $\SRPQe{1}$ and thus the property of $\SRPQe{1}$ is still satisfied.
\item[-] {\bf Insertion}: One can easily verify that the queue properties for $\SRPQr{1}$, $\SRPQsx{1}{2}{}$, $\SRPQd{1}$, and $\SRPQx{1}{2}$ hold for the corresponding insertions.
%Whenever $d_1$ receives it, we insert $X_i$ to $\SRPQd{1}$ and still satisfy its property because now $d_1$ knows this pure $X_i$. When only $r$ receives it, we insert $X_i$ to $\SRPQr{1}$ and still satisfy its property because now only $r$ knows this pure $X_i$. When only $d_2$ receives it, we insert $X_i$ to $\SRPQsx{1}{2}{}$ and still satisfy its property because now only $d_2$ knows this pure $X_i$. Similarly, when both $\{r,d_1\}$ receive it, we insert $X_i$ to $\SRPQx{1}{2}$ and still satisfy its property because
\end{itemize}

\noindent \makebox[1.2cm][l]{$\bullet$ $\snumUC{2}$:} $s$ transmits $Y_j\!\in\!\SRPQe{2}$. The movement process is symmetric to $\snumUC{1}$.

\noindent \makebox[1.2cm][l]{$\bullet$ $\snumPM{1}$:} $s$ transmits a mixture $[X_i+Y_j]$ from $X_i\!\in\!\SRPQe{1}$ and $Y_j\!\in\!\SRPQr{2}$. The movement process is as follows.
\tableSTARTreduce
\begin{table}[H]
    \centering
    \subfloat[]{
        \centering
        {\footnotesize
        \begin{tabular}{|c|c|c|}
        \hline
        $\SRPQe{1}\!\xrightarrow{X_i}$ & $\overline{d_1d_2}r$ & $\xrightarrow{X_i}\SRPQr{1}$ \\[+3pt]
        \hline
        \multirow{6}{*}{$\SRPQe{1}\!\xrightarrow{X_i}$,  $\SRPQr{2}\!\!\xrightarrow{Y_j}$} & $\overline{d_1}d_2\overline{r}$ & $\xrightarrow{[X_i+Y_j]}\SRPQb{1}{2}$ \\[+3pt]
        \cline{2-3}
        & $d_1\overline{d_2r}$ & $\xrightarrow{[X_i+Y_j]:Y_j}\SRPQm$ \\[+3pt]
        \cline{2-3}
        & $\overline{d_1}d_2r$ & $\xrightarrow{[X_i+Y_j]:X_i}\SRPQm$ \\[+3pt]
        \cline{2-3}
        \rule{0pt}{2.6ex} & $d_1\overline{d_2}r$ & \multirow{2}{*}{$\xrightarrow{[X_i+Y_j]:Y_j}\SRPQm$} \\
        \cline{2-2}
        & $d_1d_2\overline{r}$ &   \\
        \cline{2-3}
        & $d_1d_2r$ & $\xrightarrow{[X_i+Y_j]:\text{ either }X_i\text{ or }Y_j}\SRPQm$ \\[+3pt]
        \hline
        \end{tabular}
        }\label{tab:SRP:sPM1}
    }
\end{table}
\tableENDreduce
\begin{itemize}
\item[-] {\bf Departure}: The property for $X_i\!\in\!\SRPQe{1}$ is that $X_i$ must be unknown to any of $\{d_1,d_2,r\}$, even not flagged in $\RL{d_1,d_2,r}$. As a result, whenever the mixture $[X_i+Y_j]$ is received by any of $\{d_1,d_2,r\}$, $X_i$ must be removed from $\SRPQe{1}$. Similarly, the property for $Y_j\!\in\!\SRPQr{2}$ is that $Y_j$ must be unknown to any of $\{d_1,d_2\}$, even not flagged in $\RL{d_1,d_2}$. Therefore, whenever the mixture is received by any of $\{d_1,d_2\}$, $Y_j$ must be removed from $\SRPQr{2}$.
\item[-] {\bf Insertion}: When only $r$ receives the mixture, $r$ can use the known $Y_j$ and the received $[X_i+Y_j]$ to extract the pure $X_i$. As a result, we can insert $X_i$ to $\SRPQr{1}$ as it is not flagged in $\RL{d_1,d_2}$. The case when only $d_2$ receives the mixture satisfies the properties of $\SRPQb{1}{2}$ as $r$ knows the pure $Y_j$ only while $d_2$ knows the mixture $[X_i+Y_j]$ only. As a result, we can insert $[X_i+Y_j]$ to $\SRPQb{1}{2}$. The remaining reception cases fall into at least one of two conditions of $\SRPQm$. For example when only $d_1$ receives the mixture, now $[X_i+Y_j]$ is in $\RL{d_1}$ while $Y_j$ is still known by $r$ only. This corresponds to the first condition of $\SRPQm$. One can easily verify that other cases satisfy either one of or both properties of $\SRPQm$. Following the packet format for $\SRPQm$, we insert $[X_i+Y_j]:W$ into $\SRPQm$ where $W$ denotes the packet in $r$ that can benefit both destinations when transmitted. From the previous example when only $d_1$ receives the mixture, we insert $[X_i+Y_j]:Y_j$ into $\SRPQm$ as sending the known $Y_j$ from $r$ simultaneously enables $d_2$ to receive the desired $Y_j$ and $d_1$ to decode the desired $X_i$ by subtracting $Y_j$ from the received $[X_i+Y_j]$.
\end{itemize}

\noindent \makebox[1.2cm][l]{$\bullet$ $\snumPM{2}$:} $s$ transmits a mixture $[X_i\!+\!Y_j]$ from $X_i\!\in\!\SRPQr{1}$ and $Y_j\!\in\!\SRPQe{2}$. The movement process is symmetric to $\snumPM{1}$.

\noindent \makebox[1.2cm][l]{$\bullet$ $\snumAM{1}$:} $s$ transmits a mixture $[X_i+Y_j]$ from $X_i\!\in\!\SRPQr{1}$ and $Y_j\!\in\!\SRPQsx{2}{1}{}$. The movement process is as follows.
\tableSTARTreduce
\begin{table}[H]
    \centering
    \subfloat[]{
        \centering
        {\footnotesize
        \begin{tabular}{|c|c|c|}
        \hline
        $\SRPQsx{2}{1}{}\!\!\!\xrightarrow{Y_j}$ & $\overline{d_1d_2}r$ & $\xrightarrow[\text{Case~1}]{Y_j}\SRPQx{2}{1}$ \\[+3pt]
        \hline
        %--back up : $\begin{aligned}& \SRPQr{1}\!\!\!\xrightarrow{X_i}, \\ & \SRPQsx{2}{1}{}\!\!\!\xrightarrow{Y_j}\end{aligned}$
        $\SRPQr{1}\!\!\!\xrightarrow{X_i}$, $\SRPQsx{2}{1}{}\!\!\!\xrightarrow{Y_j}$ & $\overline{d_1}d_2\overline{r}$ & {$\xrightarrow{[X_i+Y_j]:X_i}\SRPQm$}\\[+3pt]
        \hline
        $\SRPQr{1}\!\!\!\xrightarrow{X_i}$ & $d_1\overline{d_2r}$ & {$\xrightarrow{X_i}\SRPQd{1}$} \\[+3pt]
        \hline
        \multirow{4}{*}{$\SRPQr{1}\!\!\!\xrightarrow{X_i}$, $\SRPQsx{2}{1}{}\!\!\!\xrightarrow{Y_j}$} & $\overline{d_1}d_2r$ & {$\xrightarrow{[X_i+Y_j]:X_i}\SRPQm$} \\
        \cline{2-3}
        & $d_1\overline{d_2}r$ & {$\xrightarrow{X_i}\SRPQd{1}$, $\xrightarrow[\text{Case~1}]{Y_j}\SRPQx{2}{1}$} \\[+3pt]
        \cline{2-3}
        & $d_1d_2\overline{r}$ & {$\xrightarrow{X_i}\SRPQd{1}$, $\xrightarrow[\text{Case~2}]{X_i(\equiv Y_j)}\SRPQx{2}{1}$} \\[+3pt]
        \cline{2-3}
        & $d_1d_2r$ & {$\xrightarrow{X_i}\SRPQd{1}$, $\xrightarrow[\text{Case~1}]{Y_j}\SRPQx{2}{1}$} \\[+3pt]
        \hline
        \end{tabular}
        }\label{tab:SRP:sAM1}
    }
\end{table}
\tableENDreduce
\begin{itemize}
\item[-] {\bf Departure}: The property for $X_i\!\in\!\SRPQr{1}$ is that $X_i$ must be unknown to any of $\{d_1,d_2\}$, even not flagged in $\RL{d_1,d_2}$. As a result, whenever the mixture $[X_i+Y_j]$ is received by any of $\{d_1,d_2\}$, $X_i$ must be removed from $\SRPQr{1}$. Similarly, the property for $Y_j\!\in\!\SRPQsx{2}{1}{}$ is that $Y_j$ must be unknown to any of $\{d_2,r\}$, even not flagged in $\RL{d_2,r}$. Therefore, whenever the mixture is received by any of $\{d_2,r\}$, $Y_j$ must be removed from $\SRPQsx{2}{1}{}$.
\item[-] {\bf Insertion}: Whenever $d_1$ receives the mixture, $d_1$ can use the known $Y_j$ and the received $[X_i+Y_j]$ to extract the pure/desired $X_i$. As a result, we can insert $X_i$ into $\SRPQd{1}$ whenever $d_1$ receives. The cases when $d_2$ receives but $d_1$ does not fall into the second condition of $\SRPQm$ as $[X_i+Y_j]$ is in $\RL{d_2}$ and $X_i$ is known by $r$ only. Namely, $r$ can benefit both destinations simultaneously by sending the known $X_i$. For those two reception status $\overline{d_1}d_2\overline{r}$ and $\overline{d_1}d_2 r$, we can thus insert this mixture into $\SRPQm$ as $[X_i+Y_j]\!:\!X_i$. Whenever $r$ receives the mixture, $r$ can use the known $X_i$ and the received $[X_i+Y_j]$ to extract the pure $Y_j$. Now $Y_j$ is known by both $r$ and $d_1$ but still unknown to $d_2$ even if $d_2$ receives this mixture $[X_i+Y_j]$ as well. As a result, $Y_j$ can be moved to $\SRPQx{2}{1}$ as the Case~1 insertion. But for the reception status of $\overline{d_1}d_2 r$, note from the previous discussion that we can insert the mixture into $\SRPQm$ since $d_2$ receives the mixture but $d_1$ does not. In this case, we chose to use more efficient $\SRPQm$ that can handle both sessions simultaneously. Finally when the reception status is $d_1 d_2\overline{r}$, we have that $X_i$ is known by both $r$ and $d_1$ while the mixture $[X_i+Y_j]$ is received by $d_2$. Namely, $X_i$ is still unknown to $d_2$ but when it is delivered, $d_2$ can use $X_i$ and the received $[X_i+Y_j]$ to extract a desired session-$2$ packet $Y_j$. Moreover, $X_i$ is already in $\SRPQd{1}$ and thus can be used as an information-equivalent packet for $Y_j$. This scenario is exactly the same as the Case~2 of $\SRPQx{2}{1}$ and thus we can move $X_i$ into $\SRPQx{2}{1}$ as the Case~2 insertion.
\end{itemize}

\noindent \makebox[1.2cm][l]{$\bullet$ $\snumAM{2}$:} $s$ transmits a mixture $[X_i+Y_j]$ from $X_i\!\in\!\SRPQsx{1}{2}{}$ and $Y_j\!\in\!\SRPQr{2}$. The movement process is symmetric to $\snumAM{1}$.

\noindent \makebox[1.2cm][l]{$\bullet$ $\snumRC{1}$:} $s$ transmits $X_i$ of the mixture $[X_i+Y_j]$ in $\SRPQb{1}{2}$. The movement process is as follows.
\tableSTARTreduce
\begin{table}[H]
    \centering
    \subfloat[]{
        \centering
        {\footnotesize
        \begin{tabular}{|c|c|c|}
        \hline
        \multirow{7}{*}{$\SRPQb{1}{2}\!\xrightarrow{[X_i+Y_j]}$} & $\overline{d_1d_2}r$ & $\xrightarrow{[X_i+Y_j]:X_i}\SRPQm$ \\[+3pt]
        \cline{2-3}
        & $\overline{d_1}d_2\overline{r}$ & {$\xrightarrow{X_i}\SRPQsx{1}{2}{}$, $\xrightarrow{Y_j}\SRPQd{2}$} \\[+3pt]
        \cline{2-3}
        & $d_1\overline{d_2r}$ & {$\xrightarrow{X_i}\SRPQd{1}$, $\xrightarrow{X_i}\SRPQsxSP{2}{1}$} \\[+3pt]
        \cline{2-3}
        & $\overline{d_1}d_2r$ & {$\xrightarrow[\text{Case~1}]{X_i}\SRPQx{1}{2}$, $\xrightarrow{Y_j}\SRPQd{2}$} \\[+3pt]
        \cline{2-3}
        & $d_1\overline{d_2}r$ & {$\xrightarrow{X_i}\SRPQd{1}$, $\xrightarrow[\text{Case~2}]{X_i(\equiv Y_j)}\SRPQx{2}{1}$} \\[+3pt]
        \cline{2-3}
        & $d_1d_2\overline{r}$ & \multirow{2}{*}{$\xrightarrow{X_i}\SRPQd{1}$, $\xrightarrow{Y_j}\SRPQd{2}$} \\
        \cline{2-2}
        & $d_1d_2r$ & \\
        \hline
        \end{tabular}
        }\label{tab:SRP:sRC1}
    }
\end{table}
\tableENDreduce
\begin{itemize}
\item[-] {\bf Departure}: One condition for $[X_i+Y_j]\!\in\!\SRPQb{1}{2}$ is that $X_i$ is unknown to any of $\{d_1,d_2,r\}$. As a result, whenever $X_i$ is received by any of $\{d_1,d_2,r\}$, the mixture $[X_i+Y_j]$ must be removed from $\SRPQb{1}{2}$.
\item[-] {\bf Insertion}: From the conditions of $\SRPQb{1}{2}$, we know that $X_i$ is unknown to $d_1$ and $Y_j$ is known only by $r$. As a result, whenever $d_1$ receives $X_i$, $d_1$ receives the new session-$1$ packet and thus we can insert $X_i$ into $\SRPQd{1}$. Whenever $d_2$ receives $X_i$, $d_2$ can use the known $[X_i+Y_j]$ and the received $X_i$ to subtract the pure $Y_j$. We can thus insert $Y_j$ into $\SRPQd{2}$. The case when only $r$ receives $X_i$ falls into the first condition of $\SRPQm$ as $[X_i+Y_j]$ is in $\RL{d_2}$ and $X_i$ is known by $r$ only. In this case, $r$ can benefit both destinations simultaneously by sending the received $X_i$. For this reception status of $\overline{d_1 d_2}r$, we thus insert the mixture into $\SRPQm$ as $[X_i+Y_j]\!:\!X_i$. The remaining reception status to consider are $\overline{d_1}d_2\overline{r}$, $d_1\overline{d_2 r}$, $\overline{d_1}d_2 r$, and $d_1\overline{d_2}r$. The first when only $d_2$ receives $X_i$ falls into the property of $\SRPQsx{1}{2}{}$ as $X_i$ is known only by $d_2$ and not flagged in $\RL{d_1,r}$. Thus we can insert $X_i$ into $\SRPQsx{1}{2}{}$. Obviously, $d_2$ can decode $Y_j$ from the previous discussion. For the second when only $d_1$ receives $X_i$, we first have $X_i\!\in\!\SRPQd{1}$ while $X_i$ is unknown to any of $\{d_2,r\}$. Moreover, $Y_j$ is known by $r$ only and $[X_i+Y_j]$ is in $\RL{d_2}$. This scenario falls exactly into $\SRPQsx{2}{1}{}$ and thus we can insert $X_i$ into $\SRPQsx{2}{1}{}$. The third case when both $d_2$ and $r$ receive $X_i$ falls exactly into Case~1 of $\SRPQx{1}{2}$ as $X_i$ is now known by both $d_2$ and $r$ but still unknown to $d_1$. And obviously, $d_2$ can decode $Y_j$ from the previous discussion. For the fourth case when both $d_1$ and $r$ receive $X_i$, we now have that $r$ contains $\{X_i,Y_j\}$; $d_1$ contains $X_i$; and $d_2$ contains $[X_i+Y_j]$. That is, $X_i$ is already in $\SRPQd{1}$ and known by $r$ as well but still unknown to $d_2$. Moreover, $d_2$ can decode the desired session-$2$ packet $Y_j$ when it receives $X_i$ further. As a result, $X_i$ can be used as an information-equivalent packet for $Y_j$ and can be moved into $\SRPQx{2}{1}$ as the Case~2 insertion.
\end{itemize}

\noindent \makebox[1.2cm][l]{$\bullet$ $\snumRC{2}$:} $s$ transmits $Y_j$ of $[X_i+Y_j]\in\SRPQb{2}{1}$. The movement process is symmetric to $\snumRC{1}$.

\noindent \makebox[1.2cm][l]{$\bullet$ $\snumDX{1}{}$:} $s$ transmits $X_i\in\SRPQsx{1}{2}{}$. The movement process is as follows.
\tableSTARTreduce
\begin{table}[H]
    \centering
    \subfloat[]{
        \centering
        {\footnotesize
        \begin{tabular}{|c|c|c|}
        \hline
        $\SRPQsx{1}{2}{}\!\!\!\xrightarrow{X_i}$ & $\overline{d_1d_2}r$ & $\xrightarrow[\text{Case~1}]{X_i}\SRPQx{1}{2}$ \\[+3pt]
        \hline
        \rule{0pt}{2.6ex} {do nothing} & $\overline{d_1}d_2\overline{r}$ & {do nothing} \\
        \hline
        \multirow{5}{*}{$\SRPQsx{1}{2}{}\!\!\!\xrightarrow{X_i}$} & $d_1\overline{d_2r}$ & {$\xrightarrow{X_i}\SRPQd{1}$} \\
        \cline{2-3}
        & $\overline{d_1}d_2r$ & $\xrightarrow[\text{Case~1}]{X_i}\SRPQx{1}{2}$ \\[+3pt]
        \cline{2-3}
        \rule{0pt}{2.6ex} & $d_1\overline{d_2}r$ & \multirow{3}{*}{$\xrightarrow{X_i}\SRPQd{1}$} \\
        \cline{2-2}
        & $d_1d_2\overline{r}$ & \\
        \cline{2-2}
        & $d_1d_2r$ & \\
        \hline
        \end{tabular}
        }\label{tab:SRP:sDX1}
    }
\end{table}
\tableENDreduce
\begin{itemize}
\item[-] {\bf Departure}: One condition for $X_i\!\in\!\SRPQsx{1}{2}{}$ is that $X_i$ must be unknown to any of $\{d_1,r\}$. As a result, $X_i$ must be removed from $\SRPQsx{1}{2}{}$ whenever it is received by any of $\{d_1,r\}$.
\item[-] {\bf Insertion}: Whenever $d_1$ receives $X_i$, it receives a new session-$1$ packet and thus we can insert $X_i$ into $\SRPQd{1}$. If $X_i$ is received by $r$ but not by $d_1$, then $X_i$ will be known by both $d_2$ and $r$ (since $d_2$ already knows $X_i$) but still unknown to $d_1$. This falls exactly into the first-case scenario of $\SRPQx{1}{2}$ and thus we can move $X_i$ into $\SRPQx{1}{2}$ as the Case~1 insertion.
\end{itemize}

\noindent \makebox[1.2cm][l]{$\bullet$ $\snumDX{2}{}$:} $s$ transmits $Y_j\in\SRPQsx{2}{1}{}$. The movement process is symmetric to $\snumDX{1}{}$.

\noindent \makebox[1.2cm][l]{$\bullet$ $\snumDX{(1)}{}$:} $s$ transmits $Y_i\in\SRPQsxSP{1}{2}$. The movement process is as follows.
\tableSTARTreduce
\begin{table}[H]
    \centering
    \subfloat[]{
        \centering
        {\footnotesize
        \begin{tabular}{|c|c|c|}
        \hline
        $\SRPQsxSP{1}{2}\!\!\!\xrightarrow{Y_i}$ & $\overline{d_1d_2}r$ & $\xrightarrow[\text{Case~2}]{Y_i}\SRPQx{1}{2}$ \\[+3pt]
        \hline
        \rule{0pt}{2.6ex} {do nothing} & $\overline{d_1}d_2\overline{r}$ & {do nothing} \\
        \hline
        \multirow{5}{*}{$\SRPQsxSP{1}{2}\!\!\!\xrightarrow{Y_i}$} & $d_1\overline{d_2r}$ & {$\xrightarrow{X_i(\equiv Y_i)}\SRPQd{1}$} \\
        \cline{2-3}
        & $\overline{d_1}d_2r$ & $\xrightarrow[\text{Case~2}]{Y_i}\SRPQx{1}{2}$ \\[+3pt]
        \cline{2-3}
        \rule{0pt}{2.6ex} & $d_1\overline{d_2}r$ & \multirow{3}{*}{$\xrightarrow{X_i(\equiv Y_i)}\SRPQd{1}$} \\
        \cline{2-2}
        & $d_1d_2\overline{r}$ & \\
        \cline{2-2}
        & $d_1d_2r$ & \\
        \hline
        \end{tabular}
        }\label{tab:SRP:sDX1*}
    }
\end{table}
\tableENDreduce
\begin{itemize}
\item[-] {\bf Departure}: One property for $Y_i\!\in\!\SRPQsxSP{1}{2}$ is that $Y_i$ must be unknown to any of $\{d_1,r\}$. As a result, whenever $Y_i$ is received by any of $\{d_1,r\}$, $Y_i$ must be removed from $\SRPQsxSP{1}{2}$.
\item[-] {\bf Insertion}: From the property of $Y_i\!\in\!\SRPQsxSP{1}{2}$, we know that $Y_i\!\in\!\SRPQd{2}$; there exists a session-$1$ packet $X_i$ still unknown to $d_1$ where $X_i\equiv Y_i$; and $[X_i+Y_i]$ is in $\RL{d_1}$. As a result, whenever $d_1$ receives $Y_i$, $d_1$ can use the received $Y_i$ and the known $[X_i+Y_i]$ to extract $X_i$ and thus we can insert $X_i$ into $\SRPQd{1}$. If $Y_i$ is received by $r$ but not by $d_1$, then $Y_i$ will be known by both $d_2$ and $r$ but unknown to $d_1$, where $[X_i+Y_i]$ is in $\RL{d_1}$. Thus when $d_1$ receives $Y_i$, $d_1$ can further decode the desired $X_i$. Moreover, $Y_i$ is already in $\SRPQd{2}$. As a result, we can move $Y_i$ into $\SRPQx{1}{2}$ as the Case~2 insertion.
\end{itemize}

\noindent \makebox[1.2cm][l]{$\bullet$ $\snumDX{(2)}{}$:} $s$ transmits $X_j\in\SRPQsxSP{2}{1}$. The movement process is symmetric to $\snumDX{(1)}{}$.

\noindent \makebox[1.2cm][l]{$\bullet$ $\snumCX{1}$:} $s$ transmits $[X_i+Y_j]$ from $X_i\in\SRPQsx{1}{2}{}$ and $Y_j\in\SRPQsx{2}{1}{}$. The movement process is as follows.
\tableSTARTreduce
\begin{table}[H]
    \centering
    \subfloat[]{
        \centering
        {\footnotesize
        \begin{tabular}{|c|c|c|}
        \hline
        $\begin{aligned}& \SRPQsx{1}{2}{}\!\xrightarrow{X_i}, \\ & \SRPQsx{2}{1}{}\!\xrightarrow{Y_j}\end{aligned}$ & $\overline{d_1d_2}r$ & $\xrightarrow{[X_i+Y_j]}\SRPQstar$ \\[+3pt]
        \hline
        $\SRPQsx{2}{1}{}\!\xrightarrow{Y_j}$ & $\overline{d_1}d_2\overline{r}$ & {$\xrightarrow{Y_j}\SRPQd{2}$} \\[+3pt]
        \hline
        $\SRPQsx{1}{2}{}\!\xrightarrow{X_i}$ & $d_1\overline{d_2r}$ & {$\xrightarrow{X_i}\SRPQd{1}$} \\[+3pt]
        \hline
        \multirow{4}{*}{$\begin{aligned}& \SRPQsx{1}{2}{}\!\xrightarrow{X_i}, \\ & \SRPQsx{2}{1}{}\!\xrightarrow{Y_j}\end{aligned}$} & $\overline{d_1}d_2r$ & {$\xrightarrow[\text{Case~3}]{[X_i+Y_j]}\SRPQx{1}{2}$, $\xrightarrow{Y_j}\SRPQd{2}$} \\[+3pt]
        \cline{2-3}
        & $d_1\overline{d_2}r$ & {$\xrightarrow{X_i}\SRPQd{1}$, $\xrightarrow[\text{Case~3}]{[X_i+Y_j]}\SRPQx{2}{1}$} \\[+3pt]
        \cline{2-3}
        & $d_1d_2\overline{r}$ & \multirow{2}{*}{$\xrightarrow{X_i}\SRPQd{1}$, $\xrightarrow{Y_j}\SRPQd{2}$} \\
        \cline{2-2}
        & $d_1d_2r$ & \\
        \hline
        \end{tabular}
        }\label{tab:SRP:sCX1}
    }
\end{table}
\tableENDreduce
\begin{itemize}
\item[-] {\bf Departure}: One condition for $X_i\!\in\!\SRPQsx{1}{2}{}$ is that $X_i$ must be unknown to any of $\{d_1,r\}$, even not flagged in $\RL{d_1,r}$. As a result, whenever the mixture is received by any of $\{d_1,r\}$, $X_i$ must be removed from $\SRPQsx{1}{2}{}$. Symmetrically for $Y_j\!\in\!\SRPQsx{2}{1}{}$, whenever the mixture is received by any of $\{d_2,r\}$, $Y_j$ must be removed from $\SRPQsx{2}{1}{}$.
\item[-] {\bf Insertion}: Whenever $d_1$ receives the mixture $[X_i+Y_j]$, $d_1$ can use the known $Y_j\!\in\!\SRPQsx{2}{1}{}$ and the received $[X_i+Y_j]$ to extract the desired $X_i$ and thus we can insert $X_i$ into $\SRPQd{1}$. Similarly, whenever $d_2$ receives this mixture, $d_2$ can use the known $X_i\!\in\!\SRPQsx{1}{2}{}$ and the received $[X_i+Y_j]$ to extract the desired $Y_j$ and thus we can insert $Y_j$ into $\SRPQd{2}$. The remaining reception status are $\overline{d_1 d_2}r$, $\overline{d_1}d_2 r$, and $d_2\overline{d_2}r$. The first when only $r$ receives the mixture exactly falls into the first-case scenario of $\SRPQstar$ as $[X_i+Y_j]$ is in $\RL{r}$; $X_i\!\in\!\SRPQsx{1}{2}{}$ is known by $d_2$ only; and $Y_j\!\in\!\SRPQsx{2}{1}{}$ is known by $d_1$ only. As a result, $r$ can then send this mixture $[X_i+Y_j]$ to benefit both destinations. The second case when both $d_2$ and $r$ receive the mixture, jointly with the assumption $Y_j\!\in\!\SRPQsx{2}{1}{}$, falls exactly into the third-case scenario of $\SRPQx{1}{2}$ where $W_i$ is a pure session-$1$ packet. As a result, we can move $[X_i+Y_j]$ into $\SRPQx{1}{2}$ as the Case~3 insertion. (And obviously, $d_2$ can decode $Y_j$ from the previous discussion.) The third case when both $d_1$ and $r$ receive the mixture follows symmetrically to the second case of $\overline{d_1}d_2 r$ and thus we can insert $[X_i+Y_j]$ into $\SRPQx{2}{1}$ as the Case~3 insertion.
\end{itemize}

\noindent \makebox[1.2cm][l]{$\bullet$ $\snumCX{2}$:} $s$ transmits $[X_i\!+\!X_j]$ from $X_i\in\SRPQsx{1}{2}{}$ and $X_j\in\SRPQsxSP{2}{1}$. The movement process is as follows.
\tableSTARTreduce
\begin{table}[H]
    \centering
    \subfloat[]{
        \centering
        {\footnotesize
        \begin{tabular}{|c|c|c|}
        \hline
        $\begin{aligned}& \SRPQsx{1}{2}{}\!\xrightarrow{X_i}, \\ & \SRPQsxSP{2}{1}\!\xrightarrow{X_j}\end{aligned}$ & $\overline{d_1d_2}r$ & $\xrightarrow{[X_i+X_j]}\SRPQstar$ \\[+3pt]
        \hline
        $\SRPQsxSP{2}{1}\!\xrightarrow{X_j}$ & $\overline{d_1}d_2\overline{r}$ & {$\xrightarrow{Y_j(\equiv X_j)}\SRPQd{2}$} \\[+3pt]
        \hline
        $\SRPQsx{1}{2}{}\!\xrightarrow{X_i}$ & $d_1\overline{d_2r}$ & {$\xrightarrow{X_i}\SRPQd{1}$} \\[+3pt]
        \hline
        \multirow{4}{*}{$\begin{aligned}& \SRPQsx{1}{2}{}\!\xrightarrow{X_i}, \\ & \SRPQsxSP{2}{1}\!\xrightarrow{X_j}\end{aligned}$} & $\overline{d_1}d_2r$ & {$\xrightarrow[\text{Case~3}]{[X_i+X_j]}\SRPQx{1}{2}$, $\xrightarrow{Y_j(\equiv X_j)}\SRPQd{2}$} \\[+3pt]
        \cline{2-3}
        & $d_1\overline{d_2}r$ & {$\xrightarrow{X_i}\SRPQd{1}$, $\xrightarrow[\text{Case~3}]{[X_i+X_j]}\SRPQx{2}{1}$} \\[+3pt]
        \cline{2-3}
        & $d_1d_2\overline{r}$ & \multirow{2}{*}{$\xrightarrow{X_i}\SRPQd{1}$, $\xrightarrow{Y_j(\equiv X_j)}\SRPQd{2}$} \\
        \cline{2-2}
        & $d_1d_2r$ & \\
        \hline
        \end{tabular}
        }\label{tab:SRP:sCX2}
    }
\end{table}
\tableENDreduce
\begin{itemize}
\item[-] {\bf Departure}: One condition for $X_i\!\in\!\SRPQsx{1}{2}{}$ is that $X_i$ must be unknown to any of $\{d_1,r\}$, even not flagged in $\RL{d_1,r}$. As a result, whenever the mixture $[X_i+X_j]$ is received by any of $\{d_1,r\}$, $X_i$ must be removed from $\SRPQsx{1}{2}{}$. From the property for $X_j\!\in\!\SRPQsxSP{2}{1}$, we know that $X_j$ is unknown to any of $\{d_2,r\}$, even not flagged in $\RL{r}$. As a result, whenever $r$ receives the mixture $[X_i+X_j]$, $X_j$ must be removed from $\SRPQsxSP{2}{1}$. Moreover, whenever $d_2$ receives this mixture, $d_2$ can use the known $X_i\!\in\!\SRPQsx{1}{2}{}$ and the received $[X_i+X_j]$ to decode $X_j$ and thus $X_j$ must be removed from $\SRPQsxSP{2}{1}$.
\item[-] {\bf Insertion}: From the properties of $X_i\!\in\!\SRPQsx{1}{2}{}$ and $X_j\!\in\!\SRPQsxSP{2}{1}$, we know that $r$ contains $Y_j$ (still unknown to $d_2$ and $Y_j\equiv X_j$); $d_1$ contains $X_j$; and $d_2$ contains $\{X_i,[Y_j+X_j]\}$ already. Therefore, whenever $d_1$ receives the mixture $[X_i+X_j]$, $d_1$ can use the known $X_j$ and the received $[X_i+X_j]$ to extract the desired $X_i$ and thus we can insert $X_i$ into $\SRPQd{1}$. Similarly, whenever $d_2$ receives this mixture, $d_2$ can use the known $\{X_i,[Y_j+X_j]\}$ and the received $[X_i+X_j]$ to extract the desired $Y_j$, and thus we can insert $Y_j$ into $\SRPQd{2}$. The remaining reception status are $\overline{d_1 d_2}r$, $\overline{d_1}d_2 r$, and $d_2\overline{d_2}r$. One can see that the case when only $r$ receives the mixture exactly falls into the Case~2 scenario of $\SRPQstar$. For the second case when both $d_2$ and $r$ receive the mixture, now $r$ contains $\{Y_j,[X_i+X_j]\}$; $d_1$ contained $X_j$ before; and $d_2$ contains $\{X_i,[Y_j+X_j],[X_i+X_j]\}$. This falls exactly into the third-case scenario of $\SRPQx{1}{2}$ where $W_i$ is a pure session-$1$ packet $X_i$. As a result, we can move $[X_i+X_j]$ into $\SRPQx{1}{2}$ as the Case~3 insertion. (And obviously, $d_2$ can decode the desired $Y_j$ from the previous discussion.) For the third case when both $d_1$ and $r$ receive the mixture, now $r$ contains $\{Y_j,[X_i+X_j]\}$; $d_1$ contains $\{X_j,[X_i+X_j]\}$; and $d_2$ contained $\{X_i,[Y_j+X_j]\}$ before, where we now have $X_i\!\in\!\SRPQd{1}$ from the previous discussion. This falls exactly into the third-case scenario of $\SRPQx{2}{1}$ where $W_j$ is a pure session-$1$ packet $X_j\!\in\!\SRPQd{1}$. Note that delivering $[X_i+X_j]$ will enable $d_2$ to further decode the desired $Y_j$. Thus we can move $[X_i+X_j]$ into $\SRPQx{2}{1}$ as the Case~3 insertion.
\end{itemize}

\noindent \makebox[1.2cm][l]{$\bullet$ $\snumCX{3}$:} $s$ transmits $[Y_i+Y_j]$ from $Y_i\in\SRPQsxSP{1}{2}$ and $Y_j\in\SRPQsx{2}{1}{}$. The movement process is as follows.
\tableSTARTreduce
\begin{table}[H]
    \centering
    \subfloat[]{
        \centering
        {\footnotesize
        \begin{tabular}{|c|c|c|}
        \hline
        $\begin{aligned}& \SRPQsxSP{1}{2}\!\xrightarrow{Y_i}, \\ & \SRPQsx{2}{1}{}\!\xrightarrow{Y_j}\end{aligned}$ & $\overline{d_1d_2}r$ & $\xrightarrow{[Y_i+Y_j]}\SRPQstar$ \\[+3pt]
        \hline
        $\SRPQsx{2}{1}{}\!\xrightarrow{Y_j}$ & $\overline{d_1}d_2\overline{r}$ & {$\xrightarrow{Y_j}\SRPQd{2}$} \\[+3pt]
        \hline
        $\SRPQsxSP{1}{2}\!\xrightarrow{Y_i}$ & $d_1\overline{d_2r}$ & {$\xrightarrow{X_i(\equiv Y_i)}\SRPQd{1}$} \\[+3pt]
        \hline
        \multirow{4}{*}{$\begin{aligned}& \SRPQsxSP{1}{2}\!\xrightarrow{Y_i}, \\ & \SRPQsx{2}{1}{}\!\xrightarrow{Y_j}\end{aligned}$} & $\overline{d_1}d_2r$ & {$\xrightarrow[\text{Case~3}]{[Y_i+Y_j]}\SRPQx{1}{2}$, $\xrightarrow{Y_j}\SRPQd{2}$} \\[+3pt]
        \cline{2-3}
        & $d_1\overline{d_2}r$ & {$\xrightarrow{X_i(\equiv Y_i)}\SRPQd{1}$, $\xrightarrow[\text{Case~3}]{[Y_i+Y_j]}\SRPQx{2}{1}$} \\[+3pt]
        \cline{2-3}
        & $d_1d_2\overline{r}$ & \multirow{2}{*}{$\xrightarrow{X_i(\equiv Y_i)}\SRPQd{1}$, $\xrightarrow{Y_j}\SRPQd{2}$} \\
        \cline{2-2}
        & $d_1d_2r$ & \\
        \hline
        \end{tabular}
        }\label{tab:SRP:sCX3}
    }
\end{table}
\tableENDreduce
\begin{itemize}
\item[-] {\bf Departure}: From the property for $Y_i\!\in\!\SRPQsxSP{1}{2}$, we know that $Y_i$ is unknown to any of $\{d_1,r\}$, even not flagged in $\RL{r}$. As a result, whenever $r$ receives the mixture $[Y_i+Y_j]$, $Y_i$ must be removed from $\SRPQsxSP{1}{2}$. Moreover, whenever $d_1$ receives this mixture, $d_1$ can use the known $Y_j\!\in\!\SRPQsx{2}{1}{}$ and the received $[Y_i+Y_j]$ to decode $Y_i$ and thus $Y_i$ must be removed from $\SRPQsxSP{1}{2}$. One condition for $Y_j\!\in\!\SRPQsx{2}{1}{}$ is that $Y_j$ must be unknown to any of $\{d_2,r\}$, even not flagged in $\RL{d_2,r}$. As a result, whenever the mixture $[Y_i+Y_j]$ is received by any of $\{d_2,r\}$, $Y_j$ must be removed from $\SRPQsx{2}{1}{}$.
\item[-] {\bf Insertion}: From the properties of $Y_i\!\in\!\SRPQsxSP{1}{2}$ and $Y_j\!\in\!\SRPQsx{2}{1}{}$, we know that $r$ contains $X_i$ (still unknown to $d_1$ and $X_i\equiv Y_i$); $d_1$ contains $\{Y_j,[X_i+Y_i]\}$; and $d_2$ contains $Y_i$ already. Therefore, whenever $d_1$ receives the mixture $[Y_i+Y_j]$, $d_1$ can use the known $\{Y_j,[X_i+Y_i]\}$ and the received $[Y_i+Y_j]$ to extract the desired $X_i$ and thus we can insert $X_i$ into $\SRPQd{1}$. Similarly, whenever $d_2$ receives this mixture, $d_2$ can use the known $Y_i$ and the received $[Y_i+Y_j]$ to extract the desired $Y_j$, and thus we can insert $Y_j$ into $\SRPQd{2}$. The remaining reception status are $\overline{d_1 d_2}r$, $\overline{d_1}d_2 r$, and $d_2\overline{d_2}r$. One can see that the first case when only $r$ receives the mixture exactly falls into the Case~3 scenario of $\SRPQstar$. For the second case when both $d_2$ and $r$ receive the mixture, now $r$ contains $\{X_i,[Y_i+Y_j]\}$; $d_1$ contained $\{Y_j,[X_i+Y_i]\}$ before; and $d_2$ contains $\{Y_i,[Y_i+Y_j]\}$, where we now have $Y_j\!\in\!\SRPQd{2}$ from the previous discussion. This falls exactly into the third-case scenario of $\SRPQx{1}{2}$ where $W_i$ is a pure session-$2$ packet $Y_i$. Note that delivering $[Y_i+Y_j]$ will enable $d_1$ to further decode the desired $X_i$. Thus we can move $[Y_i+Y_j]$ into $\SRPQx{1}{2}$ as the Case~3 insertion. For the third case when both $d_1$ and $r$ receive the mixture, now $r$ contains $\{X_i,[Y_i+Y_j]\}$; $d_1$ contains $\{Y_j,[X_i+Y_i],[Y_i+Y_j]\}$; and $d_2$ contained $Y_i$ before. This falls exactly into the third-case scenario of $\SRPQx{2}{1}$ where $W_j$ is a pure session-$2$ packet $Y_j$. As a result, we can move $[Y_i+Y_j]$ into $\SRPQx{2}{1}$ as the Case~3 insertion. (And obviously, $d_1$ can decode the desired $X_i$ from the previous discussion.)
\end{itemize}

\noindent \makebox[1.2cm][l]{$\bullet$ $\snumCX{4}$:} $s$ transmits $[Y_i+X_j]$ from $Y_i\in\SRPQsxSP{1}{2}$ and $X_j\in\SRPQsxSP{2}{1}$. The movement process is as follows.
\tableSTARTreduce
\begin{table}[H]
    \centering
    \subfloat[]{
        \centering
        {\footnotesize
        \begin{tabular}{|c|c|c|}
        \hline
        $\begin{aligned}& \SRPQsxSP{1}{2}\!\xrightarrow{Y_i}, \\ & \SRPQsxSP{2}{1}\!\xrightarrow{X_j}\end{aligned}$ & $\overline{d_1d_2}r$ & $\xrightarrow{[Y_i+X_j]}\SRPQstar$ \\[+3pt]
        \hline
        $\SRPQsxSP{2}{1}\!\xrightarrow{X_j}$ & $\overline{d_1}d_2\overline{r}$ & {$\xrightarrow{Y_j(\equiv X_j)}\SRPQd{2}$} \\[+3pt]
        \hline
        $\SRPQsxSP{1}{2}\!\xrightarrow{Y_i}$ & $d_1\overline{d_2r}$ & {$\xrightarrow{X_i(\equiv Y_i)}\SRPQd{1}$} \\[+3pt]
        \hline
        \multirow{4}{*}{$\begin{aligned}& \SRPQsxSP{1}{2}\!\xrightarrow{Y_i}, \\ & \SRPQsxSP{2}{1}\!\xrightarrow{X_j}\end{aligned}$} & $\overline{d_1}d_2r$ & {$\xrightarrow[\text{Case~3}]{[Y_i+X_j]}\SRPQx{1}{2}$, $\xrightarrow{Y_j(\equiv X_j)}\SRPQd{2}$} \\[+3pt]
        \cline{2-3}
        & $d_1\overline{d_2}r$ & {$\xrightarrow{X_i(\equiv Y_i)}\SRPQd{1}$, $\xrightarrow[\text{Case~3}]{[Y_i+X_j]}\SRPQx{2}{1}$} \\[+3pt]
        \cline{2-3}
        & $d_1d_2\overline{r}$ & \multirow{2}{*}{$\xrightarrow{X_i(\equiv Y_i)}\SRPQd{1}$, $\xrightarrow{Y_j(\equiv X_j)}\SRPQd{2}$} \\
        \cline{2-2}
        & $d_1d_2r$ & \\
        \hline
        \end{tabular}
        }\label{tab:SRP:sCX4}
    }
\end{table}
\tableENDreduce
\begin{itemize}
\item[-] {\bf Departure}: From the property for $Y_i\!\in\!\SRPQsxSP{1}{2}$, we know that $Y_i$ is unknown to any of $\{d_1,r\}$, even not flagged in $\RL{r}$. As a result, whenever $r$ receives the mixture $[Y_i+X_j]$, $Y_i$ must be removed from $\SRPQsxSP{1}{2}$. Moreover, $X_j\!\in\!\SRPQsxSP{2}{1}$ is known by $d_1$. As a result, whenever $d_1$ receives the mixture, $d_1$ can use the known $X_j$ and the received $[Y_i+X_j]$ to decode $Y_i$ and thus $Y_i$ must be removed from $\SRPQsxSP{1}{2}$. Symmetrically for $X_j\!\in\!\SRPQsxSP{2}{1}$, whenever the mixture is received by any of $\{d_2,r\}$, $X_j$ must be removed from $\SRPQsxSP{2}{1}$.
\item[-] {\bf Insertion}: From the properties of $Y_i\!\in\!\SRPQsxSP{1}{2}$ and $X_j\!\in\!\SRPQsxSP{2}{1}$, we know that $r$ contains $\{X_i,Y_j\}$ where $X_i$ (resp. $Y_j$) is still unknown to $d_1$ (resp. $d_2)$ and $X_i\equiv Y_i$ (resp. $Y_j\equiv X_j$); $d_1$ contains $\{[X_i+Y_i],X_j\}$; and $d_2$ contains $\{Y_i,[Y_j+X_j]\}$ already. Therefore, whenever $d_1$ receives the mixture $[Y_i+X_j]$, $d_1$ can use the known $\{[X_i+Y_i],X_j\}$ and the received $[Y_i+X_j]$ to extract the desired $X_i$ and thus we can insert $X_i$ into $\SRPQd{1}$. Similarly, whenever $d_2$ receives this mixture, $d_2$ can use the known $\{Y_i,[Y_j+X_j]\}$ and the received $[Y_i+X_j]$ to extract the desired $Y_j$, and thus we can insert $Y_j$ into $\SRPQd{2}$. The remaining reception status are $\overline{d_1 d_2}r$, $\overline{d_1}d_2 r$, and $d_2\overline{d_2}r$. One can see that the first case when only $r$ receives the mixture exactly falls into the Case~4 scenario of $\SRPQstar$. For the second case when both $d_2$ and $r$ receive the mixture, now $r$ contains $\{X_i,Y_j,[Y_i+X_j]\}$; $d_1$ contained $\{[X_i+Y_i],X_j\}$ before; and $d_2$ contains $\{Y_i,[Y_j+X_j],[Y_i+X_j]\}$ where we now have $X_j\!\in\!\SRPQd{1}$ from the previous discussion. This falls exactly into the third-case scenario of $\SRPQx{1}{2}$ where $W_i$ is a pure session-$2$ packet $Y_i$. Note that delivering $[Y_i+X_j]$ will enable $d_1$ to further decode the desired $X_i$. Thus we can move $[Y_i+X_j]$ into $\SRPQx{1}{2}$ as the Case~3 insertion. For the third case when both $d_1$ and $r$ receive the mixture, now $r$ contains $\{X_i,Y_j,[Y_i+X_j]\}$; $d_1$ contains $\{[X_i+Y_i],X_j,[Y_i+X_j]\}$; and $d_2$ contained $\{Y_i,[Y_j+X_j]\}$ before, where we now have $Y_i\!\in\!\SRPQd{2}$ from the previous discussion. This falls exactly into the third-case scenario of $\SRPQx{2}{1}$ where $W_j$ is a pure session-$2$ packet $X_j$. Note that delivering $[Y_i+X_j]$ will enable $d_2$ to further decode the desired $Y_j$. Thus we can move $[Y_i+X_j]$ into $\SRPQx{2}{1}$ as the Case~3 insertion.
\end{itemize}

\noindent \makebox[1.2cm][l]{$\bullet$ $\snumCX{5}$:} $s$ transmits $[X_i+\overline{W}_j]$ from $X_i\in\SRPQsx{1}{2}{}$ and $\overline{W}_j\in\SRPQx{2}{1}$. The movement process is as follows.
\tableSTARTreduce
\begin{table}[H]
    \centering
    \subfloat[]{
        \centering
        {\footnotesize
        \begin{tabular}{|c|c|c|}
        \hline
        $\SRPQsx{1}{2}{}\!\!\!\xrightarrow{X_i}$ & $\overline{d_1d_2}r$ & $\xrightarrow[\text{Case~1}]{X_i}\SRPQx{1}{2}$ \\[+3pt]
        \hline
        $\SRPQx{2}{1}\!\!\!\xrightarrow{\overline{W}_j}$ & $\overline{d_1}d_2\overline{r}$ & {$\xrightarrow{Y_j(\equiv \overline{W}_j)}\SRPQd{2}$}\\[+3pt]
        \hline
        $\SRPQsx{1}{2}{}\!\!\!\xrightarrow{X_i}$ & $d_1\overline{d_2r}$ & {$\xrightarrow{X_i}\SRPQd{1}$} \\[+3pt]
        \hline
        $\begin{aligned}& \SRPQsx{1}{2}{}\!\!\!\xrightarrow{X_i}, \\ & \SRPQx{2}{1}\!\!\!\xrightarrow{\overline{W}_j}\end{aligned}$ & $\overline{d_1}d_2r$ & {$\xrightarrow[\text{Case~1}]{X_i}\SRPQx{1}{2}$, $\xrightarrow{Y_j(\equiv \overline{W}_j)}\SRPQd{2}$} \\[+3pt]
        \hline
        $\SRPQsx{1}{2}{}\!\!\!\xrightarrow{X_i}$ & $d_1\overline{d_2}r$ & {$\xrightarrow{X_i}\SRPQd{1}$} \\[+3pt]
        \hline
        \multirow{4}{*}{$\begin{aligned} & \SRPQsx{1}{2}{}\!\!\!\xrightarrow{X_i}, \\ & \SRPQx{2}{1}\!\!\!\xrightarrow{\overline{W}_j} \end{aligned}$} & \multirow{2}{*}{$d_1d_2\overline{r}$} & \multirow{4}{*}{$\xrightarrow{X_i}\SRPQd{1}$, $\xrightarrow{Y_j(\equiv \overline{W}_j)}\SRPQd{2}$} \\
        & & \\
        \cline{2-2}
        & \multirow{2}{*}{$d_1d_2r$} & \\
        & & \\
        \hline
        \end{tabular}
        }\label{tab:SRP:sCX5}
    }
\end{table}
\tableENDreduce
\begin{itemize}
\item[-] {\bf Departure}: The property for $X_i\!\in\!\SRPQsx{1}{2}{}$ is that $X_i$ must be unknown to any of $\{d_1,r\}$, even not flagged in $\RL{d_1,r}$. As a result, whenever the mixture $[X_i+\overline{W}_j]$ is received by any of $\{d_1,r\}$, $X_i$ must be removed from $\SRPQsx{1}{2}{}$. Similarly, one condition for $\overline{W}_j\!\in\!\SRPQx{2}{1}$ is that $\overline{W}_j$ must be unknown to $d_2$. However when $d_2$ receives the mixture, $d_2$ can use the known $X_i\!\in\!\SRPQsx{1}{2}{}$ and the received $[X_i+\overline{W}_j]$ to decode $\overline{W}_j$. Thus $\overline{W}_j$ must be removed from $\SRPQx{2}{1}$ whenever $d_2$ receives.
\item[-] {\bf Insertion}: From the properties of $X_i\!\in\!\SRPQsx{1}{2}{}$ and $\overline{W}_j\!\in\!\SRPQx{2}{1}$, we know that $r$ contains $\overline{W}_j$; $d_1$ contains $\overline{W}_j$; and $d_2$ contains $X_i$ already. Therefore, whenever $d_1$ receives this mixture, $d_1$ can use the known $\overline{W}_j$ and the received $[X_i+\overline{W}_j]$ to extract the desired $X_i$ and thus we can insert $X_i$ into $\SRPQd{1}$. Similarly, whenever $d_2$ receives this mixture, $d_2$ can use the known $X_i$ and the received $[X_i+\overline{W}_j]$ to extract $\overline{W}_j$. We now need to consider case by case when $\overline{W}_j$ was inserted into $\SRPQx{2}{1}$. If it was the Case~1 insertion, then $\overline{W}_j$ is a pure session-$2$ packet $Y_j$ and thus we can simply insert $Y_j$ into $\SRPQd{2}$. If it was the Case~2 insertion, then $\overline{W}_j$ is a pure session-$2$ packet $X_j\!\in\!\SRPQd{1}$ and there exists a session-$2$ packet $Y_j$ still unknown to $d_2$ where $Y_j\equiv X_j$. Moreover, $d_2$ has received $[Y_j+X_j]$. As a result, $d_2$ can further decode $Y_j$ and thus we can insert $Y_j$ into $\SRPQd{2}$. If it was the Case~3 insertion, then $\overline{W}_j$ is a mixed form of $[W_i+W_j]$ where $W_i$ is already known by $d_2$ but $W_j$ is not. As a result, $d_2$ can decode $W_j$ upon receiving $\overline{W}_j=[W_i+W_j]$. Note that $W_j$ in the Case~3 insertion $\overline{W}_j=[W_i+W_j]\!\in\!\SRPQx{2}{1}$ comes from either $\SRPQsx{2}{1}{}$ or $\SRPQsxSP{2}{1}$. If $W_j$ was coming from $\SRPQsx{2}{1}{}$, then $W_j$ is a session-$2$ packet $Y_j$ and we can simply insert $Y_j$ into $\SRPQd{2}$. If $W_j$ was coming from $\SRPQsxSP{2}{1}$, then $W_j$ is a session-$1$ packet $X_j$ and there also exists a session-$2$ packet $Y_j$ still unknown to $d_2$ where $Y_j\equiv X_j$. Moreover, $d_2$ has received $[Y_j+X_j]$. As a result, $d_2$ can further use the known $[Y_j+X_j]$ and the extracted $X_j$ to decode $Y_j$ and thus we can insert $Y_j$ into $\SRPQd{2}$. In a nutshell, whenever $d_2$ receives the mixture $[X_i+\overline{W}_j]$, a session-$2$ packet $Y_j$ that was unknown to $d_2$ can be newly decoded. The remaining reception status are $\overline{d_1 d_2}r$ and $\overline{d_1}d_2 r$. For both cases when $r$ receives the mixture but $d_1$ does not, $r$ can use the known $\overline{W}_j$ and the received $[X_i+\overline{W}_j]$ to extract $X_i$. Since $X_i$ is now known by both $r$ and $d_2$ but unknown to $d_1$, we can thus move $X_i$ into $\SRPQx{1}{2}$ as the Case~1 insertion.
\end{itemize}

\noindent \makebox[1.2cm][l]{$\bullet$ $\snumCX{6}$:} $s$ transmits $[\overline{W}_i+Y_j]$ from $\overline{W}_i\in\SRPQx{1}{2}$ and $Y_j\in\SRPQsx{2}{1}{}$. The movement process is symmetric to $\snumCX{5}$.

\noindent \makebox[1.2cm][l]{$\bullet$ $\snumCX{7}$:} $s$ transmits $[Y_i+\overline{W}_j]$ from $Y_i\in\SRPQsxSP{1}{2}$ and $\overline{W}_j\in\SRPQx{2}{1}$. The movement process is as follows.
\tableSTARTreduce
\begin{table}[H]
    \centering
    \subfloat[]{
        \centering
        {\footnotesize
        \begin{tabular}{|c|c|c|}
        \hline
        $\SRPQsxSP{1}{2}\!\!\!\xrightarrow{Y_i}$ & $\overline{d_1d_2}r$ & $\xrightarrow[\text{Case~2}]{Y_i}\SRPQx{1}{2}$ \\[+3pt]
        \hline
        $\SRPQx{2}{1}\!\!\!\xrightarrow{\overline{W}_j}$ & $\overline{d_1}d_2\overline{r}$ & {$\xrightarrow{Y_j(\equiv \overline{W}_j)}\SRPQd{2}$}\\[+3pt]
        \hline
        $\SRPQsxSP{1}{2}\!\!\!\xrightarrow{Y_i}$ & $d_1\overline{d_2r}$ & {$\xrightarrow{X_i(\equiv Y_i)}\SRPQd{1}$} \\[+3pt]
        \hline
        $\begin{aligned}& \SRPQsxSP{1}{2}\!\!\!\xrightarrow{Y_i}, \\ & \SRPQx{2}{1}\!\!\!\xrightarrow{\overline{W}_j}\end{aligned}$ & $\overline{d_1}d_2r$ & {$\xrightarrow[\text{Case~2}]{Y_i}\SRPQx{1}{2}$, $\xrightarrow{Y_j(\equiv \overline{W}_j)}\SRPQd{2}$} \\[+3pt]
        \hline
        $\SRPQsxSP{1}{2}\!\!\!\xrightarrow{Y_i}$ & $d_1\overline{d_2}r$ & {$\xrightarrow{X_i(\equiv Y_i)}\SRPQd{1}$} \\[+3pt]
        \hline
        \multirow{4}{*}{$\begin{aligned} & \SRPQsxSP{1}{2}\!\!\!\xrightarrow{Y_i}, \\ & \SRPQx{2}{1}\!\!\!\xrightarrow{\overline{W}_j} \end{aligned}$} & \multirow{2}{*}{$d_1d_2\overline{r}$} & \multirow{2}{*}{$\xrightarrow{X_i(\equiv Y_i)}\SRPQd{1}$,} \\
        & & \\
        \cline{2-2}
        & \multirow{2}{*}{$d_1d_2r$} & \multirow{2}{*}{$\xrightarrow{Y_j(\equiv \overline{W}_j)}\SRPQd{2}$} \\
        & & \\
        \hline
        \end{tabular}
        }\label{tab:SRP:sCX7}
    }
\end{table}
\tableENDreduce
\begin{itemize}
\item[-] {\bf Departure}:
    From the property for $Y_i\!\in\!\SRPQsxSP{1}{2}$, we know that $Y_i$ is unknown to any of $\{d_1,r\}$, even not flagged in $\RL{r}$. As a result, whenever $r$ receives the mixture $[Y_i+\overline{W}_j]$, $Y_i$ must be removed from $\SRPQsxSP{1}{2}$. Moreover, $\overline{W}_j\!\in\!\SRPQx{2}{1}$ is known by $d_1$. As a result, whenever $d_1$ receives the mixture, $d_1$ can use the known $\overline{W}_j$ and the received $[Y_i+\overline{W}_j]$ to decode $Y_i$ and thus $Y_i$ must be removed from $\SRPQsxSP{1}{2}$. Similarly, one condition for $\overline{W}_j\!\in\!\SRPQx{2}{1}$ is that $\overline{W}_j$ must be unknown to $d_2$. However when $d_2$ receives the mixture, $d_2$ can use the known $Y_i\!\in\!\SRPQsxSP{1}{2}$ and the received $[Y_i+\overline{W}_j]$ to decode $\overline{W}_j$. Thus $\overline{W}_j$ must be removed from $\SRPQx{2}{1}$ whenever $d_2$ receives.
\item[-] {\bf Insertion}: From the properties of $Y_i\!\in\!\SRPQsxSP{1}{2}$ and $\overline{W}_j\!\in\!\SRPQx{2}{1}$, we know that $r$ contains $\{X_i,\overline{W}_j\}$; $d_1$ contains $\{[X_i+Y_i],\overline{W}_j\}$; and $d_2$ contains $Y_i$ already. Therefore, whenever $d_1$ receives this mixture, $d_1$ can use the known $\{[X_i+Y_i],\overline{W}_j\}$ and the received $[Y_i+\overline{W}_j]$ to extract the desired $X_i$ and thus we can insert $X_i$ into $\SRPQd{1}$. Similarly, whenever $d_2$ receives this mixture, $d_2$ can use the known $Y_i$ and the received $[Y_i+\overline{W}_j]$ to extract $\overline{W}_j$. We now need to consider case by case when $\overline{W}_j$ was inserted into $\SRPQx{2}{1}$. If it was the Case~1 insertion, then $\overline{W}_j$ is a pure session-$2$ packet $Y_j$ and thus we can simply insert $Y_j$ into $\SRPQd{2}$. If it was the Case~2 insertion, then $\overline{W}_j$ is a pure session-$1$ packet $X_j\!\in\!\SRPQd{1}$ and there exists a session-$2$ packet $Y_j$ still unknown to $d_2$ where $Y_j\equiv X_j$. Moreover, $d_2$ has received $[Y_j+X_j]$. As a result, $d_2$ can further decode $Y_j$ and thus we can insert $Y_j$ into $\SRPQd{2}$. If it was the Case~3 insertion, then $\overline{W}_j$ is a mixed form of $[W_i+W_j]$ where $W_i$ is already known by $d_2$ but $W_j$ is not. As a result, $d_2$ can decode $W_j$ upon receiving $\overline{W}_j=[W_i+W_j]$. Note that $W_j$ in the Case~3 insertion $\overline{W}_j=[W_i+W_j]\!\in\!\SRPQx{2}{1}$ comes from either $\SRPQsx{2}{1}{}$ or $\SRPQsxSP{2}{1}$. If $W_j$ was coming from $\SRPQsx{2}{1}{}$, then $W_j$ is a session-$2$ packet $Y_j$ and we can simply insert $Y_j$ into $\SRPQd{2}$. If $W_j$ was coming from $\SRPQsxSP{2}{1}$, then $W_j$ is a session-$1$ packet $X_j$ and there also exists a session-$2$ packet $Y_j$ still unknown to $d_2$ where $Y_j\equiv X_j$. Moreover, $d_2$ has received $[Y_j+X_j]$. As a result, $d_2$ can further use the known $[Y_j+X_j]$ and the extracted $X_j$ to decode $Y_j$ and thus we can insert $Y_j$ into $\SRPQd{2}$. In a nutshell, whenever $d_2$ receives the mixture $[Y_i+\overline{W}_j]$, a session-$2$ packet $Y_j$ that was unknown to $d_2$ can be newly decoded. The remaining reception status are $\overline{d_1 d_2}r$ and $\overline{d_1}d_2 r$. For both cases when $r$ receives the mixture but $d_1$ does not, $r$ can use the known $\overline{W}_j$ and the received $[Y_i+\overline{W}_j]$ to extract $Y_i$. Since $Y_i$ is now known by both $r$ and $d_2$ but $[X_i+Y_i]$ is in $\RL{d_1}$, we can thus move $Y_i$ into $\SRPQx{1}{2}$ as the Case~2 insertion.
\end{itemize}

\noindent \makebox[1.2cm][l]{$\bullet$ $\snumCX{8}$:} $s$ transmits $[\overline{W}_i+X_j]$ from $\overline{W}_i\in\SRPQx{1}{2}$ and $X_j\in\SRPQsxSP{2}{1}$. The movement process is symmetric to $\snumCX{7}$.

\noindent \makebox[1.2cm][l]{$\bullet$ $\rnumUC{1}$:} $r$ transmits $X_i$ from $X_i\in\SRPQr{1}$. The movement process is as follows.
\tableSTARTreduce
\begin{table}[H]
    \centering
    \subfloat[]{
        \centering
        {\footnotesize
        \begin{tabular}{|c|c|c|}
        \hline
        \multirow{3}{*}{$\SRPQr{1}\!\xrightarrow{X_i}$} & $\overline{d_1}d_2$ & $\xrightarrow[\text{Case~1}]{X_i}\SRPQx{1}{2}$ \\[+3pt]
        \cline{2-3}
        \rule{0pt}{2.6ex} & $d_1\overline{d_2}$ & \multirow{2}{*}{$\xrightarrow{X_i}\SRPQd{1}$} \\
        \cline{2-2}
        & $d_1 d_2$ & \\[+3pt]
        \hline
        \end{tabular}
        }\label{tab:SRP:rUC1}
    }
\end{table}
\tableENDreduce
\begin{itemize}
\item[-] {\bf Departure}: One condition for $X_i\!\in\!\SRPQr{1}$ is that $X_i$ must be unknown to any of $\{d_1,d_2\}$. As a result, whenever $X_i$ is received by any of $\{d_1,d_2\}$, $X_i$ must be removed from $\SRPQr{1}$.
\item[-] {\bf Insertion}: From the above discussion, we know that $X_i$ is unknown to $d_1$. As a result, whenever $X_i$ is received by $d_1$, we can insert $X_i$ to $\SRPQd{1}$. If $X_i$ is received by $d_2$ but not by $d_1$, then $X_i$ is now known by both $d_2$ and $r$ but still unknown to $d_1$. This exactly falls into the first-case scenario of $\SRPQx{1}{2}$ and thus we can move $X_i$ into $\SRPQx{1}{2}$ as the Case~1 insertion.
\end{itemize}

\noindent \makebox[1.2cm][l]{$\bullet$ $\rnumUC{2}$:} $r$ transmits $Y_j$ from $Y_j\in\SRPQr{2}$. The movement process is symmetric to $\rnumUC{1}$.

\noindent \makebox[1.2cm][l]{$\bullet$ $\rnumDX{(1)}{}$:} $r$ transmits $X_i$ that is known by $r$ only and information equivalent from $Y_i\!\in\!\SRPQsxSP{1}{2}$. The movement process is as follows.
\tableSTARTreduce
\begin{table}[H]
    \centering
    \subfloat[]{
        \centering
        {\footnotesize
        \begin{tabular}{|c|c|c|}
        \hline
        \multirow{3}{*}{$\SRPQsxSP{1}{2}\!\!\!\xrightarrow{Y_i}$} & $\overline{d_1}d_2$ & {$\xrightarrow[\text{Case~1}]{X_i}\SRPQx{1}{2}$} \\[+3pt]
        \cline{2-3}
        \rule{0pt}{2.6ex} & $d_1\overline{d_2}$ & \multirow{2}{*}{$\xrightarrow{X_i(\equiv Y_i)}\SRPQd{1}$} \\
        \cline{2-2}
        & $d_1 d_2$ & \\
        \hline
        \end{tabular}
        }\label{tab:SRP:rDX1*}
    }
\end{table}
\tableENDreduce
\begin{itemize}
\item[-] {\bf Departure}: From the property for $Y_i\!\in\!\SRPQsxSP{1}{2}$, we know that there exists an information-equivalent session-$1$ packet $X_i$ that is known by $r$ but unknown to any of $\{d_1,d_2\}$. As a result, whenever $X_i$ is received by any of $\{d_1,d_2\}$, $Y_i$ must be removed from $\SRPQsxSP{1}{2}$.
\item[-] {\bf Insertion}: From the above discussion, we know that $X_i$ is unknown to $d_1$ and thus we can insert $X_i$ to $\SRPQd{1}$ whenever $X_i$ is received by $d_1$. If $X_i$ is received by $d_2$ but not by $d_1$, then $X_i$ is now known by both $d_2$ and $r$ but still unknown to $d_1$. This exactly falls into the first-case scenario of $\SRPQx{1}{2}$ and thus we can move $X_i$ into $\SRPQx{1}{2}$ as the Case~1 insertion.
\end{itemize}

\noindent \makebox[1.2cm][l]{$\bullet$ $\rnumDX{(2)}{}$:} $r$ transmits $Y_j$ that is known by $r$ only and information equivalent from $X_j\!\in\!\SRPQsxSP{2}{1}$. The movement process is symmetric to $\rnumDX{(1)}{}$.

\noindent \makebox[1.2cm][l]{$\bullet$ $\rnumRC$:} $r$ transmits $W$ known by $r$ for the packet of the form $[X_i+Y_j]:W\!\in\!\SRPQm$. The movement process is as follows.
\tableSTARTreduce
\begin{table}[H]
    \centering
    \subfloat[]{
        \centering
        {\footnotesize
        \begin{tabular}{|c|c|c|}
        \hline
        \multirow{3}{*}{$\SRPQm\!\xrightarrow{[X_i+Y_j]:W}$} & $\overline{d_1}d_2$ & {$\begin{aligned} & \text{either} \xrightarrow[\text{Case~1}]{X_i}\SRPQx{1}{2} \text{ or} \xrightarrow[\text{Case~2}]{Y_j}\SRPQx{1}{2}, \\ & \xrightarrow{Y_j}\SRPQd{2} \end{aligned}$} \\[+3pt]
        \cline{2-3}
        & $d_1\overline{d_2}$ & {$\begin{aligned} & \xrightarrow{X_i}\SRPQd{1}, \\ & \text{either} \xrightarrow[\text{Case~1}]{Y_j}\SRPQx{2}{1} \text{ or} \xrightarrow[\text{Case~2}]{X_i}\SRPQx{2}{1} \end{aligned}$} \\[+3pt]
        \cline{2-3}
        & $d_1 d_2$ & {$\xrightarrow{X_i}\SRPQd{1}$, $\xrightarrow{Y_j}\SRPQd{2}$} \\[+3pt]
        \hline
        \end{tabular}
        }\label{tab:SRP:rRC}
    }
\end{table}
\tableENDreduce
\begin{itemize}
\item[-] {\bf Departure}: From the conditions of $[X_i+Y_j]:W\!\in\!\SRPQm$, we know that $\SRPQm$ is designed to benefit both destinations simultaneously when $r$ transmits $W$. That is, whenever $d_1$ (resp. $d_2$) receives $W$, $d_1$ (resp. $d_2$) can decode the desired $X_i$ (resp. $Y_j$), regardless whether the packet $W$ is of a session-$1$ or of a session-$2$. However from the conditions of $\SRPQm$, we know that $X_i$ is unknown to $d_1$ and $Y_j$ is unknown to $d_2$. Therefore, whenever $W$ is received by any of $\{d_1,d_2\}$, $[X_i+Y_j]:W$ must be removed from $\SRPQx{1}{2}$.
\item[-] {\bf Insertion}: From the above discussions, we know that $d_1$ (resp. $d_2$) can decode the desired $X_i$ (resp. $Y_j$) when $W$ is received by $d_1$ (resp. $d_2$). As a result, we can insert $X_i$ into $\SRPQd{1}$ (resp. $Y_j$ into $\SRPQd{2}$) when $d_1$ (resp. $d_2$) receives $W$. We now consider two reception status $\overline{d_1}d_2$ and $d_1\overline{d_2}$. From the conditions of $\SRPQm$, note that $W$ is always known by $r$ and can be either $X_i$ or $Y_j$. Moreover, $X_i$ (resp. $Y_j$) is unknown to $d_1$ (resp. $d_2$). For the first reception case $\overline{d_1}d_2$, if $X_i$ was chosen as $W$ to benefit both destinations, then $X_i$ is now known by both $d_2$ and $r$ but still unknown to $d_1$. This exactly falls into the first-case scenario of $\SRPQx{1}{2}$ and thus we move $X_i$ into $\SRPQx{1}{2}$ as the Case~1 insertion. On the other hand, if $Y_j$ was chosen as $W$ to benefit both destinations, then we know that $Y_j$ is now known by both $d_2$ and $r$, and that $[X_i+Y_j]$ is already in $\RL{d_1}$. This exactly falls into the second-case scenario of $\SRPQx{1}{2}$ and thus we can move $Y_j\!\in\!\SRPQd{2}$ into $\SRPQx{1}{2}$ as the Case~2 insertion. The second reception case $d_1\overline{d_2}$ will follow the the previous arguments symmetrically.
\end{itemize}

\noindent \makebox[1.2cm][l]{$\bullet$ $\rnumOX{}$:} $r$ transmits $[W_i+W_j]\!\in\!\SRPQstar$. The movement process is as follows.
\tableSTARTreduce
\begin{table}[H]
    \centering
    \subfloat[]{
        \centering
        {\footnotesize
        \begin{tabular}{|c|c|c|}
        \hline
        \multirow{6}{*}{$\SRPQstar\!\!\xrightarrow{[W_i+W_j]}$} & \multirow{2}{*}{$\overline{d_1}d_2$} & $\xrightarrow[\text{Case~3}]{[W_i+W_j]}\SRPQx{1}{2}$, \\
        & & $\xrightarrow{Y_j(\equiv W_j)}\SRPQd{2}$ \\[+3pt]
        \cline{2-3}
        & \multirow{2}{*}{$d_1\overline{d_2}$} & $\xrightarrow{X_i(\equiv W_i)}\SRPQd{1}$, \\
        & & $\xrightarrow[\text{Case~3}]{[W_i+W_j]}\SRPQx{2}{1}$ \\[+3pt]
        \cline{2-3}
        & \multirow{2}{*}{$d_1 d_2$} & $\xrightarrow{X_i(\equiv W_i)}\SRPQd{1}$, \\
        & & $\xrightarrow{Y_j(\equiv W_j)}\SRPQd{2}$ \\[+3pt]
        \hline
        \end{tabular}
        }\label{tab:SRP:rOX**}
    }
\end{table}
\tableENDreduce
\begin{itemize}
\item[-] {\bf Departure}: From the property for $[W_i+W_j]\!\in\!\SRPQstar$, we know that $W_i$ is known only by $d_2$ and that $W_j$ is only known by $d_1$. As a result, whenever $d_1$ receives this mixture, $d_1$ can use the known $W_j$ and the received $[W_i+W_j]$ to extract $W_i$ and thus the mixture must be removed from $\SRPQstar$. Similarly, whenever $d_2$ receives this mixture, $d_2$ can use the known $W_i$ and the received $[W_i+W_j]$ to extract $W_j$ and thus the mixture must be removed from $\SRPQstar$.
\item[-] {\bf Insertion}: From the above discussions, we have observed that whenever $d_1$ (resp. $d_2$) receives the mixture, $d_1$ (resp. $d_2$) can extract $W_i$ (resp. $W_j$). From the four cases study of $\SRPQstar$, we know that $d_1$ (resp. $d_2$) can decode a desired session-$1$ packet $X_i$ (resp. session-$2$ packet $Y_j$) whenever $d_1$ (resp. $d_2$) receives the mixture, and thus we can insert $X_i$ (resp. $Y_j$) into $\SRPQd{1}$ (resp. $\SRPQd{2}$). We now consider the reception status $\overline{d_1}d_2$ and $d_1\overline{d_2}$. If $d_2$ receives the mixture but $d_1$ does not, then $d_1$ contained $W_j$ and $d_2$ now contains $[W_i+W_j]$. Moreover, $[W_i+W_j]$ was transmitted from $r$. This falls exactly into the third-case scenario of $\SRPQx{1}{2}$. As a result, we can move $[W_i+W_j]$ into $\SRPQx{1}{2}$ as the Case~3 insertion. The case when the reception status is $d_1\overline{d_2}$ can be symmetrically followed such that we can move $[W_i+W_j]$ into $\SRPQx{2}{1}$ as the Case~3 insertion.
\end{itemize}

\noindent \makebox[1.2cm][l]{$\bullet$ $\rnumDX{[1]}{}$:} $r$ transmits $\overline{W}_i\!\in\!\SRPQx{1}{2}$. The movement process is as follows.
\tableSTARTreduce
\begin{table}[H]
    \centering
    \subfloat[]{
        \centering
        {\footnotesize
        \begin{tabular}{|c|c|c|}
        \hline
        \rule{0pt}{2.6ex} {do nothing} & $\overline{d_1}d_2$ & {do nothing} \\
        \hline
        \rule{0pt}{2.6ex} \multirow{2}{*}{$\SRPQx{1}{2}\!\!\!\xrightarrow{\overline{W}_i}$} & $d_1\overline{d_2}$ & \multirow{2}{*}{$\xrightarrow{X_i(\equiv \overline{W}_i)}\SRPQd{1}$} \\
        \cline{2-2}
        & $d_1 d_2$ & \\
        \hline
        \end{tabular}
        }\label{tab:SRP:rDX1**}
    }
\end{table}
\tableENDreduce
\begin{itemize}
\item[-] {\bf Departure}: One condition for $\overline{W}_i\!\in\!\SRPQx{1}{2}$ is that $\overline{W}_i$ is known by $d_2$ unknown to $d_1$. As a result, whenever $d_1$ receives, $\overline{W}_i$ must be removed from $\SRPQx{1}{2}$. Since $\overline{W}_i\!\in\!\SRPQx{1}{2}$ is already known by $d_2$, nothing happens if it is received by $d_2$.
\item[-] {\bf Insertion}: From the previous observation, we only need to consider the reception status when $d_1$ receives $\overline{W}_i$. For those $d_1\overline{d_2}$ and $d_1 d_2$, we need to consider case by case when $\overline{W}_i$ was inserted into $\SRPQx{1}{2}$. If it was the Case~1 insertion, then $\overline{W}_i$ is a pure session-$1$ packet $X_i$ and thus we can simply insert $X_i$ into $\SRPQd{1}$. If it was the Case~2 insertion, then $\overline{W}_i$ is a pure session-$2$ packet $Y_i\!\in\!\SRPQd{2}$ and there exists a session-$1$ packet $X_i$ still unknown to $d_1$ where $X_i\equiv Y_i$. Moreover, $d_1$ has received $[X_i+Y_i]$. As a result, $d_1$ can further decode $X_i$ and thus we can insert $X_i$ into $\SRPQd{1}$. If it was the Case~3 insertion, then $\overline{W}_i$ is a mixed form of $[W_i+W_j]$ where $W_j$ is already known by $d_1$ but $W_i$ is not. As a result, $d_1$ can decode $W_i$ upon receiving $\overline{W}_i=[W_i+W_j]$. Note that $W_i$ in the Case~3 insertion $\overline{W}_i=[W_i+W_j]\!\in\!\SRPQx{1}{2}$ comes from either $\SRPQsx{1}{2}{}$ or $\SRPQsxSP{1}{2}$. If $W_i$ was coming from $\SRPQsx{1}{2}{}$, then $W_i$ is a session-$1$ packet $X_i$ and we can simply insert $X_i$ into $\SRPQd{1}$. If $W_i$ was coming from $\SRPQsxSP{1}{2}$, then $W_i$ is a session-$2$ packet $Y_i$ and there also exists a session-$1$ packet $X_i$ still unknown to $d_1$ where $X_i\equiv Y_i$. Moreover, $d_1$ has received $[X_i+Y_i]$. As a result, $d_1$ can further use the known $[X_i+Y_i]$ and the extracted $Y_i$ to decode $X_i$ and thus we can insert $X_i$ into $\SRPQd{1}$. In a nutshell, whenever $d_1$ receives $\overline{W}_i$, a session-$1$ packet $X_i$ that was unknown to $d_1$ can be newly decoded.
\end{itemize}

\noindent \makebox[1.2cm][l]{$\bullet$ $\rnumDX{[2]}{}$:} $r$ transmits $\overline{W}_j\!\in\!\SRPQx{2}{1}$. The movement process is symmetric to $\rnumDX{[1]}{}$.

\noindent \makebox[1.2cm][l]{$\bullet$ $\rnumCX{}$:} $r$ transmits $[\overline{W}_i\!+\!\overline{W}_j]$ from $\overline{W}_i\!\in\!\SRPQx{1}{2}$ and $\overline{W}_j\!\in\!\SRPQx{2}{1}$. The movement process is as follows.
\tableSTARTreduce
\begin{table}[H]
    \centering
    \subfloat[]{
        \centering
        {\footnotesize
        \begin{tabular}{|c|c|c|}
        \hline
        $\SRPQx{2}{1}\!\!\xrightarrow{\overline{W}_j}$ & $\overline{d_1}d_2$ & {$\xrightarrow{Y_j(\equiv \overline{W}_j)}\SRPQd{2}$} \\[+3pt]
        \hline
        $\SRPQx{1}{2}\!\!\xrightarrow{\overline{W}_i}$ & $d_1\overline{d_2}$ & {$\xrightarrow{X_i(\equiv \overline{W}_i)}\SRPQd{1}$} \\[+3pt]
        \hline
        \multirow{4}{*}{$\begin{aligned} & \SRPQx{1}{2}\!\!\xrightarrow{\overline{W}_i}, \\ & \SRPQx{2}{1}\!\!\xrightarrow{\overline{W}_j} \end{aligned}$} & \multirow{4}{*}{$d_1 d_2$} & \multirow{2}{*}{$\xrightarrow{X_i(\equiv \overline{W}_i)}\SRPQd{1}$,}   \\[+3pt]
        & & \\
        & & \multirow{2}{*}{$\xrightarrow{Y_j(\equiv \overline{W}_j)}\SRPQd{2}$} \\
        & & \\
        \hline
        \end{tabular}
        }\label{tab:SRP:rCX**}
    }
\end{table}
\tableENDreduce
\begin{itemize}
\item[-] {\bf Departure}: From the property for $\overline{W}_i\!\in\!\SRPQx{1}{2}$, we know that $\overline{W}_i$ is known by $d_2$ but unknown to $d_1$. Symmetrically, $\overline{W}_j\!\in\!\SRPQx{2}{1}$ is known by $d_1$ but unknown to $d_2$. As result, whenever $d_1$ (resp. $d_2$) receives the mixture, $d_1$ (resp. $d_2$) can use the known $\overline{W}_j$ (resp. $\overline{W}_i$) and the received $[\overline{W}_i\!+\!\overline{W}_j]$ to extract $\overline{W}_i$ (resp. $\overline{W}_j$). Therefore, we must remove $\overline{W}_i$ from $\SRPQx{1}{2}$ whenever $d_1$ the mixture and remove $\overline{W}_j$ from $\SRPQx{2}{1}$ whenever $d_2$ receives.
\item[-] {\bf Insertion}: From the above discussions, we have observed that whenever $d_1$ (resp. $d_2$) receives the mixture, $d_1$ (resp. $d_2$) can extract $\overline{W}_i$ (resp. $\overline{W}_j$). We first focus on the case when $d_1$ receives the mixture. For those $d_1\overline{d_2}$ and $d_1 d_2$, we can use the same arguments for $\overline{W}_i$ as described in the Insertion process of $\rnumDX{[1]}{}$. Following these case studies, one can see that a session-$1$ packet $X_i$ that was unknown to $d_1$ can be newly decoded whenever $d_1$ receives $\overline{W}_i$. The reception status when $d_2$ receives the mixture can be followed symmetrically such that $d_2$ can always decode a new session-$2$ packet $Y_j$ that was unknown before.
\end{itemize}

\section{LNC encoding operations, Packet Movement Process, and Queue Invariance for newly added $s$-variables $\snumSX{k}{l}$ in  \PropRef{prop:SRP:sim-innerV2}}\label{app:SRP:queue_invariance:V2}

In the following, we will describe the newly added $6$ self-packets-XOR operations and the corresponding packet movement process of \PropRef{prop:SRP:sim-innerV2} one by one, and then prove that the Queue Invariance explained in \SecRef{sec:SRP:Queue-Property} always holds.

Again, to simplify the analysis, we will ignore the null reception and we will exploit the following symmetry: For those variables $\snumSX{k}{l}$ whose superscript indicates the session information $k\!\in\!\{1,2\}$ (either session-$1$ or session-$2$), here we describe session-$1$ ($k\!=\!1$) only. Those variables with $k\!=\!2$ in the superscript will be symmetrically explained by simultaneously swapping (a) session-$1$ and session-$2$ in the superscript; (b) $X$ and $Y$; (c) $i$ and $j$; and (d) $d_1$ and $d_2$, if applicable.

\noindent \makebox[1.2cm][l]{$\bullet$ $\snumSX{1}{1}$:} The source $s$ transmits $[X\!+\!X_i]$ from $X\in\SRPQr{1}$ and $X_i\in\SRPQsx{1}{2}{}$. The movement process is as follows.
\tableSTARTreduce
\begin{table}[H]
    \centering
    \subfloat[]{
        \centering
        {\footnotesize
        \begin{tabular}{|c|c|c|}
        \hline
        $\SRPQsx{1}{2}{}\!\!\!\xrightarrow{X_i}$ & $\overline{d_1d_2}r$ & $\xrightarrow[\text{Case~1}]{X_i}\SRPQx{1}{2}$ \\[+3pt]
        \hline
        $\SRPQr{1}\!\!\!\xrightarrow{X}$ & $\overline{d_1}d_2\overline{r}$ & $\xrightarrow[\text{Case~1}]{X}\SRPQx{1}{2}$\\[+3pt]
        \hline
        \multirow{6}{*}{$\begin{aligned}& \SRPQr{1}\!\!\!\xrightarrow{X}, \\ & \SRPQsx{1}{2}{}\!\!\!\xrightarrow{X_i}\end{aligned}$} & $d_1\overline{d_2r}$ & {$\xrightarrow{X_i}\SRPQd{1}$, $\xrightarrow{X_i}\SRPQsxSP{1}{2}$} \\[+3pt]
        \cline{2-3}
        & $\overline{d_1}d_2r$ & {$\xrightarrow[\text{Case~1}]{X}\SRPQx{1}{2}$, $\xrightarrow[\text{Case~1}]{X_i}\SRPQx{1}{2}$} \\[+3pt]
        \cline{2-3}
        & $d_1\overline{d_2}r$ & {$\xrightarrow{X_i}\SRPQd{1}$, $\xrightarrow[\text{Case~2}]{X_i}\SRPQx{1}{2}$} \\[+3pt]
        \cline{2-3}
        & $d_1d_2\overline{r}$& {$\xrightarrow{X}\SRPQd{1}$, $\xrightarrow[\text{Case~2}]{X}\SRPQx{1}{2}$} \\[+3pt]
        \cline{2-3}
        & \multirow{2}{*}{$d_1d_2r$} & {if $\xrightarrow{X}\SRPQd{1}$, then $\xrightarrow[\text{Case~2}]{X}\SRPQx{1}{2}$,} \\
        & & {if $\xrightarrow{X_i}\SRPQd{1}$, then $\xrightarrow[\text{Case~2}]{X_i}\SRPQx{1}{2}$} \\[+3pt]
        \hline
        \end{tabular}
        }\label{tab:SRP:sSX1-1}
    }
\end{table}
\tableENDreduce
\begin{itemize}
\item[-] {\bf Departure}: The property for $X\!\in\!\SRPQr{1}$ is that $X$ must be unknown to any of $\{d_1,d_2\}$, even not flagged in $\RL{d_1,d_2}$. As a result, whenever the mixture $[X+X_i]$ is received by any of $\{d_1,d_2\}$, $X$ must be removed from $\SRPQr{1}$. Similarly, the property for $X_i\!\in\!\SRPQsx{1}{2}{}$ is that $X_i$ must be unknown to any of $\{d_1,r\}$, even not flagged in $\RL{d_1,r}$. As a result, whenever the mixture is received by any of $\{d_1,r\}$, $X_i$ must be removed from $\SRPQsx{1}{2}{}$.
\item[-] {\bf Insertion}: Whenever $r$ receives the mixture, $r$ can use the known $X$ and the received $[X\!+\!X_i]$ to extract $X_i$. Moreover, whenever $d_2$ receives the mixture, $d_2$ can use the known $X_i$ and the received $[X\!+\!X_i]$ to extract $X$. From the above observations, we describe one by one for each reception status.
    When the reception status is $\overline{d_1d_2}r$, now $X_i$ is known by both $d_2$ and $r$ but still unknown to $d_1$ while $X$ is still at $r$. This $X_i$ falls exactly into the first-case scenario of $\SRPQx{1}{2}$ and thus we move $X_i$ into $\SRPQx{1}{2}$ as the Case~1 insertion.
    When the reception status is $\overline{d_1}d_2\overline{r}$, now $X$ is known by both $d_2$ and $r$ but still unknown to $d_1$ while $X_i$ is still at $d_2$. As a result, we can move $X$ into $\SRPQx{1}{2}$ as the Case~1 insertion.
    When the reception status is $d_1\overline{d_2 r}$, we now have $X$ at $r$; $[X\!+\!X_i]$ at $d_1$; and $X_i$ at $d_2$. In this case, whenever $X_i$ or $X$ is further delivered, $d_1$ can decode both $X$ and $X_i$ simultaneously. We can thus treat $X_i$ as information-equivalent to $X$ or vice versa. But since $X_i$ is overheard by $d_2$, we chose to treat $X_i$ as already decoded ``in advance"; insert $X_i$ into $\SRPQd{1}$; and treat $X$ as not-yet decoded by $d_1$. For such $X$, note that now $r$ can perform the naive delivery to $d_1$. This exactly falls into the scenario of $\SRPQsxSP{1}{2}$ when we substitute $Y_i$ by $X_i$. Originally, $\SRPQsxSP{1}{2}$ holds packets of pure session-$2$ where such a session-$2$ packet $Y_i\!\in\!\SRPQsxSP{1}{2}$ is information equivalent to a session-$1$ packet not yet delivered to $d_1$. $X_i$ here plays the same role as $Y_i$ as we treat $X\equiv X_i$ and that $X$ is not yet delivered to $d_1$. As a result, we move $X_i$ into $\SRPQsxSP{1}{2}$.
    When the reception status is $\overline{d_1}d_2 r$, we have both $X$ and $X_i$ known by both $d_2$ and $r$ but still unknown to $d_1$. As a result, we can move both $X$ and $X_i$ into $\SRPQx{1}{2}$ as the Case~1 insertion.
    When the reception status is $d_1\overline{d_2}r$, we now have $\{X,X_i\}$ at $r$; $[X\!+\!X_i]$ at $d_1$; and $X_i$ at $d_2$. Following the discussion when $d_1\overline{d_2 r}$, we can treat either $X$ or $X_i$ as already decoded. But here we choose to treat $X_i$ as already decoded since $X_i$ is known by both $d_2$ and $r$. Such $X_i$ falls exactly into the second-case scenario of $\SRPQx{1}{2}$ when we substitute $Y_i$ by $X_i$. As a result, we move $X_i$ into $\SRPQd{1}$, and also into $\SRPQsxSP{1}{2}$ as the Case~2 insertion.
    When the reception status is $d_1 d_2\overline{r}$, we now have $X$ at $r$; $[X\!+\!X_i]$ at $d_1$; and $\{X,X_i\}$ at $d_2$. Similarly following the discussion when $d_1\overline{d_2 r}$, here we choose to treat $X$ as already decoded since $X$ is known by both $d_2$ and $r$. Similarly following the above discussions, we move $X$ into $\SRPQd{1}$, and also into $\SRPQsxSP{1}{2}$ as the Case~2 insertion.
    Finally when the reception status is $d_1 d_2 r$, we now have $\{X,X_i\}$ at $r$; $[X\!+\!X_i]$ at $d_1$; and $\{X,X_i\}$ at $d_2$. Following the similar discussion of when $d_1 d_2\overline{r}$, we know that we can treat either $X$ or $X_i$ as already decoded because both $X$ and $X_i$ are known by $d_2$ and $r$. As a result, if we treat $X$ as already decoded by $d_1$, then we move $X$ into $\SRPQx{1}{2}$ as the Case~2 insertion. On the other hand, if we treat $X_i$ as already decoded, then we move $X_i$ into $\SRPQx{1}{2}$ as the Case~2 insertion.
\end{itemize}

\noindent \makebox[1.2cm][l]{$\bullet$ $\snumSX{2}{1}$:} $s$ transmits $[Y\!+\!Y_j]$ from $Y\!\in\!\SRPQr{2}$ and $Y_j\!\in\!\SRPQsx{2}{1}{}$. The movement process is symmetric to $\snumSX{1}{1}$.

\noindent \makebox[1.2cm][l]{$\bullet$ $\snumSX{1}{2}$:} $s$ transmits $[X\!+\!\overline{W}_i]$ from $X\in\SRPQr{1}$ and $\overline{W}_i\in\SRPQsxSP{1}{2}$. The movement process is as follows.
\tableSTARTreduce
\begin{table}[H]
    \centering
    \subfloat[]{
        \centering
        {\footnotesize
        \begin{tabular}{|c|c|c|}
        \hline
        $\SRPQsxSP{1}{2}\!\!\!\xrightarrow{\overline{W}_i}$ & $\overline{d_1d_2}r$ & $\xrightarrow[\text{Case~2}]{\overline{W}_i}\SRPQx{1}{2}$ \\[+3pt]
        \hline
        \multirow{2}{*}{$\SRPQr{1}\!\!\!\xrightarrow{X}$} & $\overline{d_1}d_2\overline{r}$ & $\xrightarrow[\text{Case~1}]{X}\SRPQx{1}{2}$\\[+3pt]
        \cline{2-3}
        & $d_1\overline{d_2r}$ & {$\xrightarrow{X}\SRPQd{1}$} \\
        \hline
        \multirow{5}{*}{$\begin{aligned}& \SRPQr{1}\!\!\!\xrightarrow{X}, \\ & \SRPQsxSP{1}{2}\!\!\!\xrightarrow{\overline{W}_i}\end{aligned}$} & $\overline{d_1}d_2r$ & {$\xrightarrow[\text{Case~1}]{X}\SRPQx{1}{2}$, $\xrightarrow[\text{Case~2}]{\overline{W}_i}\SRPQx{1}{2}$} \\[+3pt]
        \cline{2-3}
        & $d_1\overline{d_2}r$ & {$\xrightarrow{X}\SRPQd{1}$, $\xrightarrow[\text{Case~2}]{\overline{W}_i}\SRPQx{1}{2}$} \\[+3pt]
        \cline{2-3}
        & $d_1d_2\overline{r}$ & {$\xrightarrow{X}\SRPQd{1}$, $\xrightarrow[\text{Case~2}]{X}\SRPQx{1}{2}$} \\[+3pt]
        \cline{2-3}
        & \multirow{2}{*}{$d_1d_2r$} & {if $\xrightarrow{X}\SRPQd{1}$, then $\xrightarrow[\text{Case~2}]{\overline{W}_i}\SRPQx{1}{2}$} \\
        & & {if $\xrightarrow{X_i(\equiv \overline{W}_i)}\SRPQd{1}$, then $\xrightarrow[\text{Case~1}]{X}\SRPQx{1}{2}$} \\[+3pt]
        \hline
        \end{tabular}
        }\label{tab:SRP:sSX1-2}
    }
\end{table}
\tableENDreduce
\begin{itemize}
\item[-] {\bf Departure}: The property for $X\!\in\!\SRPQr{1}$ is that $X$ must be unknown to any of $\{d_1,d_2\}$, even not flagged in $\RL{d_1,d_2}$. As a result, whenever the mixture $[X+\overline{W}_i]$ is received by any of $\{d_1,d_2\}$, $X$ must be removed from $\SRPQr{1}$. Similarly, one property for $\overline{W}_i\!\in\!\SRPQsxSP{1}{2}$ is that $\overline{W}_i$ must be unknown to any of $\{d_1,r\}$, even not in $\RL{r}$. As a result, whenever $\overline{W}_i$ is received by any of $\{d_1,r\}$, $\overline{W}_i$ must be removed from $\SRPQsxSP{1}{2}$. %{\bf Modified until here}

    Similarly, one property for $\overline{W}_i\!\in\!\SRPQsxSP{1}{2}$ is that $\overline{W}_i$ must be unknown to any of $\{d_1,r\}$ and for $r$ not allowed to even have $\overline{W}_i$ in a mixed form with any other packet. As a result, whenever the mixture is received by $r$, it must be removed from $\SRPQsxSP{1}{2}$. We now need to consider the case when the mixture is received by both $\{d_1,d_2\}$ but not $r$. To that end, first note that since $X\!\in\!\SRPQr{1}$ and $\overline{W}_i\!\in\!\SRPQsxSP{1}{2}$, we already have $\{X,X_i\}$ at $r$; $\{[X_i\!+\!\overline{W}_i]\}$ at $d_1$; and $\overline{W}_i$ at $d_2$, where $X_i\!\not\in\!\SRPQd{1}$ is the information-equivalent pure session-$1$ packet corresponding to $\overline{W}_i$ from the property of $\overline{W}_i\!\in\!\SRPQsxSP{1}{2}$. Now assume that the mixture is received only by both $d_1$ and $d_2$. We then have $\{X,X_i\}$ at $r$; $\{[X_i\!+\!\overline{W}_i],[X\!+\!\overline{W}_i]\}$ at $d_1$; and $\{\overline{W}_i,[X\!+\!\overline{W}_i]\}$ at $d_2$. Then $d_2$ can now use the known $\overline{W}_i$ and the received $[X\!+\!\overline{W}_i]$ to further extract $X$. In this case, whenever $\overline{W}_i$ or $X$ is delivered to $d_1$, it can decode $X$ and $X_i$ simultaneously. But notice that $d_1$ also knows $[X_i\!+\!X]$ by manipulating its received mixtures $\{[X_i\!+\!\overline{W}_i],[X\!+\!\overline{W}_i]\}$. Moreover, $X$ is known by both $\{d_2,r\}$ while $X_i$ is known by $r$ only. As a result, we chose to use $X$ further and thus treat $X$ as already decoded. The reason is because, for such $X$, this exactly falls into Case~2 of $\SRPQx{1}{2}$ where $W_i\!=\!X\!\in\!\SRPQd{1\cup2}$ is known by both $\{d_2,r\}$ and $d_1$ has $[X_i\!+\!W_i]\!=\![X_i\!+\!X]$ where $X_i\!\not\in\!\SRPQd{1}$. In a nutshell, when the reception status is $d_1 d_2\overline{r}$, we can treat $X$ as if $X\!\in\!\SRPQd{1}$. Therefore, $X$ must be removed from $\SRPQr{1}$.
\item[-] {\bf Insertion}: Whenever $r$ receives the mixture, $r$ can use the known $X$ and the received $[X\!+\!\overline{W}_i]$ to extract $\overline{W}_i$. Also, whenever $d_2$ receives the mixture, $d_2$ can use the known $\overline{W}_i$ and the received $[X\!+\!\overline{W}_i]$ to extract $X$. From these observations, we describe one by one for each reception status.
    When the reception status is $\overline{d_1d_2}r$, now $\overline{W}_i$ is known by both $\{d_2,r\}$ where $X$ is still at $r$. Since $\overline{W}_i$ was coming from $\SRPQsxSP{1}{2}$, $d_1$ also knows $[X_i\!+\!\overline{W}_i]$ for some $X_i\!\not\in\!\SRPQd{1}$ where $\overline{W}_i\!\in\!\SRPQd{1\cup2}$. For such $\overline{W}_i$, this exactly falls into Case~2 of $\SRPQx{1}{2}$ and thus we move $\overline{W}_i$ into $\SRPQx{1}{2}$ as the Case~2 insertion.
    When the reception status is $\overline{d_1}d_2\overline{r}$, now $X$ is known by both $\{d_2,r\}$ where $\overline{W}_i$ is still at $d_2$. For such $X$, this exactly falls into Case~1 of $\SRPQx{1}{2}$ and thus we move $X$ into $\SRPQx{1}{2}$ as the Case~1 insertion.
    When the reception status is $d_1\overline{d_2 r}$, we now have $\{X,X_i\}$ at $r$; $\{[X_i\!+\!\overline{W}_i],[X\!+\!\overline{W}_i]\}$ at $d_1$; and $\overline{W}_i$ at $d_2$. In this case, whenever $\overline{W}_i$ or $X$ is further delivered to $d_1$, it can decode both $X$ and $X_i$ simultaneously. But since $\overline{W}_i$ is overheard by $d_2$, we chose to treat $X$ as already decoded and insert $X$ into $\SRPQd{1}$, while still keeping $X_i\!\not\in\!\SRPQd{1}$ as not-yet decoded. Since $X_i\!\not\in\!\SRPQd{1}$ is kept intact and the mixture is received by $d_1$ only, in order for $d_1$ to further decode $X_i$, $d_1$ needs to have either $X_i$ in $r$ or $\overline{W}_i$ in $d_2$. Namely, the original scenario of $\overline{W}_i\!\in\!\SRPQsxSP{1}{2}$ is still kept intact. As a result, we just insert $X$ into $\SRPQd{1}$.
    When the reception status is $\overline{d_1}d_2 r$, we now have that both $X$ and $\overline{W}_i$ are known by both $\{d_2,r\}$ and thus both $X$ and $\overline{W}_i$ falls into Case~1 and Case~2 of $\SRPQx{1}{2}$, respectively. We thus move both $X$ and $\overline{W}_i$ into $\SRPQx{1}{2}$ as the Case~1 and Case~2 insertion, respectively.
    When the reception status is $d_1\overline{d_2}r$, we now have $\{X,X_i,\overline{W}_i\}$ at $r$; $\{[X_i\!+\!\overline{W}_i],[X\!+\!\overline{W}_i]\}$ at $d_1$; and $\overline{W}_i$ at $d_2$. Following the discussion when $d_1\overline{d_2 r}$, we can treat either $X$ or $X_i$ as already decoded. But here we chose to treat $X_i$ as already decoded since $\overline{W}_i$ is overheard by both $\{d_2,r\}$ and $d_1$ contains $[X\!+\!\overline{W}_i]$. Namely, by treating $X_i\!\in\!\SRPQd{1}$, we can switch the $\overline{W}_i$-associated pure session-$1$ packet from $X_i$ to $X\!\not\in\!\SRPQd{1}$ since $d_1$ now knows $[X\!+\!\overline{W}_i]$. This is exactly the same to Case~2 of $\SRPQx{1}{2}$ where $W_i\!=\!\overline{W}_i\!\in\!\SRPQd{1\cup2}$ is known by both $\{d_2,r\}$ and $d_1$ has $[X\!+\!W_i]$ where $X\!\not\in\!\SRPQd{1}$. As a result, we can further move $\overline{W}_i$ into $\SRPQx{1}{2}$ as the Case~2 insertion.
    When the reception status is $d_1 d_2\overline{r}$, we now have $\{X,X_i\}$ at $r$; $\{[X_i\!+\!\overline{W}_i],[X\!+\!\overline{W}_i]\}$ at $d_1$; and $\{X,\overline{W}_i\}$ at $d_2$. Following the {\bf Departure} discussion when $d_1\overline{d_2 r}$, we can choose to treat $X$ as already decoded and use $X$ as for Case~2 of $\SRPQx{1}{2}$ where $W_i\!=\!X\!\in\!\SRPQd{1\cup2}$ is known by both $\{d_2,r\}$ and $d_1$ has $[X_i\!+\!W_i]\!=\![X_i\!+\!X]$ where $X_i\!\not\in\!\SRPQd{1}$. As a result, we can further move $X$ into $\SRPQx{1}{2}$ as the Case~2 insertion.
    Finally when the reception status is $d_1 d_2 r$, we now have $\{X,X_i,\overline{W}_i\}$ at $r$; $\{[X_i\!+\!\overline{W}_i],[X\!+\!\overline{W}_i]\}$ at $d_1$; and $\{X,\overline{W}_i\}$ at $d_2$. From the previous discussions, we know that we can treat either $X$ or $X_i$ as already decoded where both $X$ and $\overline{W}_i$ are known by both $\{d_2,r\}$. If we treat $X$ as already decoded, then since $\overline{W}_i\!\in\!\SRPQd{1\cup2}$ was from $\SRPQsxSP{1}{2}$ and is known by both $\{d_2,r\}$, we can thus move $\overline{W}_i$ into $\SRPQx{1}{2}$ as the Case~2 insertion. On the other hand, if we treat $X_i$ as already decoded, then since $X\!\not\in\!\SRPQd{1}$ is known by both $\{d_2,r\}$, we can thus move $X$ into $\SRPQx{1}{2}$ as the Case~1 insertion.
\end{itemize}

\noindent \makebox[1.2cm][l]{$\bullet$ $\snumSX{2}{2}$:} $s$ transmits $[Y\!+\!\overline{W}_j]$ from $Y\in\SRPQr{2}$ and $\overline{W}_j\in\SRPQsxSP{2}{1}$. The movement process is symmetric to $\snumSX{1}{2}$.

\noindent \makebox[1.2cm][l]{$\bullet$ $\snumSX{1}{3}$:} $s$ transmits $[X_i\!+\!X^\ast_i]$ from $X_i\in\SRPQsx{1}{2}{}$ and $X^\ast_i(\equiv \overline{W}_i\in\SRPQsxSP{1}{2})$. The movement process is as follows.
\tableSTARTreduce
\begin{table}[H]
    \centering
    \subfloat[]{
        \centering
        {\footnotesize
        \begin{tabular}{|c|c|c|}
        \hline
        $\overline{d_1d_2}r$ & $\SRPQsx{1}{2}{}\!\!\!\xrightarrow{X_i}$ & $\xrightarrow[\text{Case~1}]{X_i}\SRPQx{1}{2}$ \\[+3pt]
        \hline
        $\overline{d_1}d_2\overline{r}$ & $\SRPQsxSP{1}{2}\!\!\!\xrightarrow{\overline{W}_i}$ & $\xrightarrow[\text{Case~1}]{X^\ast_i(\equiv \overline{W}_i)}\SRPQx{1}{2}$ \\[+3pt]
        \hline
        $d_1\overline{d_2r}$ & $\SRPQsx{1}{2}{}\!\!\!\xrightarrow{X_i}$ & {$\xrightarrow{X_i}\SRPQd{1}$} \\
        \hline
        $\overline{d_1}d_2r$ & \multirow{5}{*}{$\begin{aligned}& \SRPQsx{1}{2}{}\!\!\!\xrightarrow{X_i}, \\ & \SRPQsxSP{1}{2}\!\!\!\xrightarrow{\overline{W}_i}\end{aligned}$} & {$\xrightarrow[\text{Case~1}]{X_i}\SRPQx{1}{2}$, $\xrightarrow[\text{Case~1}]{X^\ast_i(\equiv \overline{W}_i)}\SRPQx{1}{2}$} \\[+3pt]
        \cline{1-1}\cline{3-3}
        $d_1\overline{d_2}r$ & & {$\xrightarrow{X_i}\SRPQd{1}$, $\xrightarrow[\text{Case~2}]{X_i}\SRPQx{1}{2}$} \\[+3pt]
        \cline{1-1}\cline{3-3}
        $d_1d_2\overline{r}$ & & {$\xrightarrow{X_i}\SRPQd{1}$, $\xrightarrow[\text{Case~1}]{X^\ast_i(\equiv \overline{W}_i)}\SRPQx{1}{2}$} \\[+3pt]
        \cline{1-1}\cline{3-3}
        \multirow{2}{*}{$d_1d_2r$} & & {if $\xrightarrow{X_i}\SRPQd{1}$, then $\xrightarrow[\text{Case~1}]{X^\ast_i(\equiv \overline{W}_i)}\SRPQx{1}{2}$} \\
        & & {if $\xrightarrow{X^\ast_i(\equiv \overline{W}_i)}\SRPQd{1}$, then
        $\xrightarrow[\text{Case~1}]{X_i}\SRPQx{1}{2}$} \\[+3pt]
        \hline
        \end{tabular}
        }\label{tab:SRP:sSX1-3}
    }
\end{table}
\tableENDreduce
\begin{itemize}
\item[-] {\bf Departure}: One property for $X_i\!\in\!\SRPQsx{1}{2}{}$ is that $X_i$ must be unknown to any of $\{d_1,r\}$, not even in a mixed form with any other packet. As a result, whenever the mixture is received by any of $\{d_1,r\}$, it must be removed from $\SRPQsx{1}{2}{}$. Similarly, one property for $\overline{W}_i\!\in\!\SRPQsxSP{1}{2}$ is that there exists a pure session-$1$ packet $X^\ast_i\!\not\in\!\SRPQd{1}$ that is information-equivalent to $\overline{W}_i$ and is known by $r$ only. Note that whenever the mixture $[X_i\!+\!X^\ast_i]$ is received by $d_2$, it can use the known $X_i$ and the received $[X_i\!+\!X^\ast_i]$ to extract the pure $X^\ast_i$. As a result, $\overline{W}_i$ must be removed from $\SRPQsxSP{1}{2}$. We now need to consider the case when the mixture is received by both $\{d_1,r\}$ but not $d_2$. To that end, first note that since $X_i\!\in\!\SRPQsx{1}{2}{}$ and $\overline{W}_i\!\in\!\SRPQsxSP{1}{2}$, we have $X^\ast_i$ at $r$; $[X^\ast_i\!+\!\overline{W}_i]$ at $d_1$; and $\{X_i,\overline{W}_i\}$ at $d_2$, where $X^\ast_i\!\not\in\!\SRPQd{1}$ is the information-equivalent pure session-$1$ packet corresponding to $\overline{W}_i$ from the property of $\overline{W}_i\!\in\!\SRPQsxSP{1}{2}$. Now assume that the mixture is received only by both $d_1$ and $r$. We then have $\{X^\ast_i,[X_i\!+\!X^\ast_i]\}$ at $r$; $\{[X^\ast_i\!+\!\overline{W}_i],[X_i\!+\!X^\ast_i]\}$ at $d_1$; and $\{X_i,\overline{W}_i\}$ still at $d_2$. Then $r$ can now use the known $X^\ast_i$ and the received $[X_i\!+\!X^\ast_i]$ to further extract $X_i$. In this case, whenever $X_i$ or $X^\ast_i$ is delivered to $d_1$, it can decode both $X_i$ and $X^\ast_i$ simultaneously since $d_1$ has received $[X_i\!+\!X^\ast_i]$. Moreover, $X_i$ is known by both $\{d_2,r\}$ while $X^\ast_i$ is known by $r$ only. As a result, we chose to use $X_i$ further and thus treat $X_i$ as already decoded. The reason is because, for such $X_i$, this exactly falls into Case~2 of $\SRPQx{1}{2}$ where $W_i\!=\!X_i\!\in\!\SRPQd{1\cup2}$ is known by both $\{d_2,r\}$ and $d_1$ has $[X^\ast_i\!+\!W_i]\!=\![X^\ast_i\!+\!X_i]$ where $X^\ast_i\!\not\in\!\SRPQd{1}$. In a nutshell, when the reception status is $d_1\overline{d_2}r$, we can replace $\overline{W}_i$ by $X_i\!\in\!\SRPQd{1}$ for decoding $X^\ast_i$ later. Therefore, $\overline{W}_i$ must be removed from $\SRPQsxSP{1}{2}$.
\item[-] {\bf Insertion}: Whenever $r$ receives the mixture, $r$ can use the known $X^\ast_i$ and the received $[X_i\!+\!X^\ast_i]$ to extract $X_i$. Also, whenever $d_2$ receives the mixture, $d_2$ can use the known $X_i$ and the received $[X_i\!+\!X^\ast_i]$ to extract $X^\ast_i$. From these observations, we describe one by one for each reception status.
    When the reception status is $\overline{d_1d_2}r$, now $X_i$ is known by both $\{d_2,r\}$ where $X^\ast_i$ is still at $r$.
    %Since $\overline{W}_i$ was coming from $\SRPQsxSP{1}{2}$, $d_1$ also knows $[X_i\!+\!\overline{W}_i]$ for some $X_i\!\not\in\!\SRPQd{1}$ where $\overline{W}_i\!\in\!\SRPQd{1\cup2}$.
    For such $X_i$, this exactly falls into Case~1 of $\SRPQx{1}{2}$ and thus we move $X_i$ into $\SRPQx{1}{2}$ as the Case~1 insertion.
    When the reception status is $\overline{d_1}d_2\overline{r}$, now $X^\ast_i$ is known by both $\{d_2,r\}$ where $X_i$ is still at $d_2$. For such $X^\ast_i$, this exactly falls into Case~1 of $\SRPQx{1}{2}$ and thus we move $X^\ast_i$ into $\SRPQx{1}{2}$ as the Case~1 insertion.
    When the reception status is $d_1\overline{d_2 r}$, we now have $X^\ast_i$ at $r$; $\{[X^\ast_i\!+\!\overline{W}_i],[X_i\!+\!X^\ast_i]\}$ at $d_1$; and $\{X_i,\overline{W}_i\}$ at $d_2$. In this case, whenever $X_i$ or $X^\ast_i$ is further delivered to $d_1$, it can decode both $X_i$ and $X^\ast_i$ simultaneously. We then chose to treat $X_i$ as already decoded and insert $X_i$ into $\SRPQd{1}$, while still keeping $X^\ast_i\!\not\in\!\SRPQd{1}$ as not-yet decoded. Since $X^\ast_i\!\not\in\!\SRPQd{1}$ is kept intact and the mixture $[X^\ast_i\!+\!\overline{W}_i]$ was known by $d_1$ before, in order for $d_1$ to further decode $X^\ast_i$, $d_1$ needs to have either $X^\ast_i$ in $r$ or $\overline{W}_i$ in $d_2$. Namely, the original scenario of $\overline{W}_i\!\in\!\SRPQsxSP{1}{2}$ is still kept intact. As a result, we just insert $X_i$ into $\SRPQd{1}$.
    When the reception status is $\overline{d_1}d_2 r$, we now have that both $X_i$ and $X^\ast_i$ are known by both $\{d_2,r\}$ and thus both $X_i$ and $X^\ast_i$ falls into Case~1 of $\SRPQx{1}{2}$. We thus move both $X_i$ and $\overline{W}_i$ into $\SRPQx{1}{2}$ as the Case~1 insertions.
    When the reception status is $d_1\overline{d_2}r$, we now have $\{X_i,X^\ast_i\}$ at $r$; $\{[X^\ast_i\!+\!\overline{W}_i],[X_i\!+\!X^\ast_i]\}$ at $d_1$; and $\{X_i,\overline{W}_i\}$ at $d_2$. Following the {\bf Departure} discussion when $d_1\overline{d_2}r$, we can choose to treat $X_i$ as already decoded and use $X_i$ as for Case~2 of $\SRPQx{1}{2}$ where $W_i\!=\!_i\!\in\!\SRPQd{1\cup2}$ is known by both $\{d_2,r\}$ and $d_1$ has $[X^\ast_i\!+\!W_i]\!=\![X_i\!+\!X^\ast_i]$ where $X^\ast_i\!\not\in\!\SRPQd{1}$. As a result, we can further move $X_i$ into $\SRPQx{1}{2}$ as the Case~2 insertion.
    When the reception status is $d_1 d_2\overline{r}$, we now have $X^\ast_i$ at $r$; $\{[X^\ast_i\!+\!\overline{W}_i],[X_i\!+\!X^\ast_i]\}$ at $d_1$; and $\{X_i,\overline{W}_i,X^\ast_i\}$ at $d_2$. Following the discussion when $d_1\overline{d_2 r}$, we can treat either $X_i$ or $X^\ast_i$ as already decoded. But here we chose to treat $X_i$ as already decoded since $X^\ast_i$ is now overheard by both $\{d_2,r\}$ and $d_1$ contains $[X_i\!+\!X^\ast_i]$. Namely, by treating $X_i\!\in\!\SRPQd{1}$, we can simply focus on delivering $X^\ast_i\!\not\in\!\SRPQd{1}$ to $d_1$ that is known by both $\{d_2,r\}$. This is exactly the same to Case~1 of $\SRPQx{1}{2}$. As a result, we can further move $X^\ast_i$ into $\SRPQx{1}{2}$ as the Case~1 insertion.
    Finally when the reception status is $d_1 d_2 r$, we now have $\{X_i,X^\ast_i\}$ at $r$; $\{[X^\ast_i\!+\!\overline{W}_i],[X_i\!+\!X^\ast_i]\}$ at $d_1$; and $\{X_i,\overline{W}_i,X^\ast_i\}$ at $d_2$. From the previous discussions, we know that we can treat either $X_i$ or $X^\ast_i$ as already decoded where both $X_i$ and $X^\ast_i$ are known by both $\{d_2,r\}$. If we treat $X_i$ as already decoded, we can simply move $X^\ast_i\!\not\in\!\SRPQd{1}$ into $\SRPQx{1}{2}$ as the Case~1 insertion. Similarly, if we treat $X^\ast_i$ as already decoded, then since $X_i\!\not\in\!\SRPQd{1}$ is known by both $\{d_2,r\}$, we can thus move $X_i$ into $\SRPQx{1}{2}$ as the Case~1 insertion.
\end{itemize}

\noindent \makebox[1.2cm][l]{$\bullet$ $\snumSX{2}{3}$:} $s$ transmits $[Y_i\!+\!Y^\ast_j]$ from $Y_i\in\SRPQsx{2}{1}{}$ and $Y^\ast_j(\equiv W_j\in\SRPQsxSP{2}{1})$. The movement process is symmetric to $\snumSX{1}{3}$.

In the following Table~\ref{tab:SRP:queue_in-out_V2}, we also described for each queue, the associated LNC operations that moves packet into and takes packets out of in the general LNC inner bound of \PropRef{prop:SRP:sim-innerV2}. Note that $r$-variables are the same as $w$-variables where the superscript $(h), h\!\in\!\{s,r\}$ is by $(r)$, and thus they represent $w$-variables accordingly.

\setlength{\tabcolsep}{3pt}
\begin{table}[H]
    \caption{Summary of the associated LNC operations in \PropRef{prop:SRP:sim-innerV2} \\ including newly added $\snumSX{k}{l}$ operations}
    \centering
    {\footnotesize
    \renewcommand\arraystretch{1.2}
    \begin{tabular}{|c|c|c|}
        \hline
        \multicolumn{1}{|c|}{\bf{LNC operations $\mapsto$}} & \multicolumn{1}{c|}{\bf{Queue}} & \multicolumn{1}{c|}{\bf{$\mapsto$ LNC operations}} \\
        \hline
        \hline
        & $\SRPQe{1}$ & {$\snumUC{1}$, $\snumPM{1}$} \\[+3pt]
        \hline
        \multirow{2}{*}{$\snumUC{1}$, $\snumPM{1}$} & \multirow{2}{*}{$\SRPQr{1}$} & {$\snumPM{2}$, $\snumAM{1}$, $\rnumUC{1}$} \\
        & & {$\snumSX{1}{1}$, $\snumSX{1}{2}$} \\
        \hline
        {$\snumPM{1}$} & $\SRPQb{1}{2}$ & {$\snumRC{1}$}  \\[+3pt]
        \hline
        \multirow{3}{*}{$\snumUC{1}$, $\snumRC{1}$} & \multirow{3}{*}{$\SRPQsx{1}{2}{}$} & {$\snumAM{2}$, $\snumDX{1}{}$} \\
        & & {$\snumCX{1}$, $\snumCX{2}$, $\snumCX{5}$} \\
        & & {$\snumSX{1}{1}$, $\snumSX{1}{3}$} \\[+3pt]
        \hline
        \multirow{3}{*}{$\snumRC{2}$, $\snumSX{1}{1}$} & \multirow{3}{*}{$\SRPQsxSP{1}{2}$} & {$\snumDX{(1)}{}$, $\snumCX{3}$} \\
        & & {$\snumCX{4}$, $\snumCX{7}$, $\rnumDX{(1)}{}$} \\
        & & {$\snumSX{1}{2}$, $\snumSX{1}{3}$} \\[+3pt]
        \hline
        {$\snumUC{1}$, $\snumAM{2}$, $\snumRC{1}$, $\snumDX{1}{}$} & \multirow{3}{*}{$\SRPQx{1}{2}$ (Case~1)} & \multirow{8}{*}{$\begin{aligned} \snumCX{6}&, \snumCX{8} \\ \rnumDX{[1]}{}&, \rnumCX{} \end{aligned}$} \\
        {$\snumCX{5}$, $\rnumUC{1}$, $\rnumDX{(1)}{}$, $\rnumRC$} & & {} \\
        {$\snumSX{1}{1}$, $\snumSX{1}{3}$, $\snumSX{1}{2}$} & & {} \\[+3pt]
        \cline{1-2}
        {$\snumAM{2}$, $\snumRC{2}$, $\snumDX{(1)}{}$} & \multirow{3}{*}{$\SRPQx{1}{2}$ (Case~2)} & {} \\
        {$\snumCX{7}$, $\rnumRC$} & & {} \\
        {$\snumSX{1}{1}$, $\snumSX{1}{2}$, $\snumSX{1}{3}$} & & {} \\[+3pt]
        \cline{1-2}
        {$\snumCX{1}$, $\snumCX{2}$} & \multirow{2}{*}{$\SRPQx{1}{2}$ (Case~3)} & {} \\
        {$\snumCX{3}$, $\snumCX{4}$, $\rnumOX{}$} & & {} \\[+3pt]
        \hline
        {$\snumUC{1}$, $\snumAM{1}$, $\snumRC{1}$, $\snumRC{2}$} & \multirow{5}{*}{$\SRPQd{1}$} & \\
        {$\snumDX{1}{}$, $\snumDX{(1)}{}$, $\{\snumCX{1}\,\text{to}\,\snumCX{8}\}$} & & \\
        {$\rnumUC{1}$, $\rnumDX{(1)}{}$, $\rnumDX{[1]}{}$} & & \\
        {$\rnumRC$, $\rnumOX{}$, $\rnumCX{}$} & & \\
        {$\snumSX{1}{1}$, $\snumSX{1}{2}$, $\snumSX{1}{3}$} & & \\[+3pt]
        \hline
        \hline
        {$\snumPM{1}$, $\snumPM{2}$, $\snumAM{1}$, $\snumAM{2}$} & \multirow{2}{*}{$\SRPQm$} & \multirow{2}{*}{$\rnumRC$} \\
        {$\snumRC{1}$, $\snumRC{2}$} & & \\[+3pt]
        \hline
        {$\snumCX{1}$, $\snumCX{2}$, $\snumCX{3}$, $\snumCX{4}$} & $\SRPQstar$ & {$\rnumOX{}$} \\[+3pt]
        \hline
        \hline
        & $\SRPQe{2}$ & {$\snumUC{2}$, $\snumPM{2}$} \\[+3pt]
        \hline
        \multirow{2}{*}{$\snumUC{2}$, $\snumPM{2}$} & \multirow{2}{*}{$\SRPQr{2}$} & {$\snumPM{1}$, $\snumAM{2}$, $\rnumUC{2}$} \\
        & & {$\snumSX{2}{1}$, $\snumSX{2}{2}$} \\[+3pt]
        \hline
        {$\snumPM{2}$} & $\SRPQb{2}{1}$ & {$\snumRC{2}$} \\[+3pt]
        \hline
        \multirow{3}{*}{$\snumUC{2}$, $\snumRC{2}$} & \multirow{3}{*}{$\SRPQsx{2}{1}{}$} & {$\snumAM{1}$, $\snumDX{2}{}$} \\
        & & {$\snumCX{1}$, $\snumCX{3}$, $\snumCX{6}$} \\
        & & {$\snumSX{2}{1}$, $\snumSX{2}{3}$} \\[+3pt]
        \hline
        \multirow{3}{*}{$\snumRC{1}$, $\snumSX{2}{1}$} & \multirow{3}{*}{$\SRPQsxSP{2}{1}$} & {$\snumDX{(2)}{}$, $\snumCX{2}$} \\
        & & {$\snumCX{4}$, $\snumCX{8}$, $\rnumDX{(2)}{}$} \\
        & & {$\snumSX{2}{2}$, $\snumSX{2}{3}$} \\[+3pt]
        \hline
        {$\snumUC{2}$, $\snumAM{1}$, $\snumRC{2}$, $\snumDX{2}{}$} & \multirow{3}{*}{$\SRPQx{2}{1}$ (Case~1)} & \multirow{8}{*}{$\begin{aligned} \snumCX{5}&, \snumCX{7} \\ \rnumDX{[2]}{}&, \rnumCX{} \end{aligned}$} \\
        {$\snumCX{6}$, $\rnumUC{2}$, $\rnumDX{(2)}{}$, $\rnumRC$} & & {} \\
        {$\snumSX{2}{1}$, $\snumSX{2}{2}$, $\snumSX{2}{3}$} & & {} \\[+3pt]
        \cline{1-2}
        {$\snumAM{1}$, $\snumRC{1}$, $\snumDX{(2)}{}$} & \multirow{3}{*}{$\SRPQx{2}{1}$ (Case~2)} & {} \\
        {$\snumCX{8}$, $\rnumRC$, $\snumSX{2}{1}$} & & {} \\
        {$\snumSX{2}{2}$, $\snumSX{2}{3}$} & & {} \\[+3pt]
        \cline{1-2}
        {$\snumCX{1}$, $\snumCX{2}$} & \multirow{2}{*}{$\SRPQx{2}{1}$ (Case~3)} & {} \\
        {$\snumCX{3}$, $\snumCX{4}$, $\rnumOX{}$} & & {} \\[+3pt]
        \hline
        {$\snumUC{2}$, $\snumAM{2}$, $\snumRC{1}$, $\snumRC{2}$} & \multirow{5}{*}{$\SRPQd{2}$} & \\
        {$\snumDX{2}{}$, $\snumDX{(2)}{}$, $\{\snumCX{1}\,\text{to}\,\snumCX{8}\}$} & & \\
        {$\rnumUC{2}$, $\rnumDX{(2)}{}$, $\rnumDX{[2]}{}$} & & \\
        {$\rnumRC$, $\rnumOX{}$, $\rnumCX{}$} & & \\
        {$\snumSX{2}{1}$, $\snumSX{2}{2}$, $\snumSX{2}{3}$} & & \\[+3pt]
        \hline
    \end{tabular}
    \label{tab:SRP:queue_in-out_V2}
    }
\end{table}
\setlength{\tabcolsep}{6pt}

\section{Detailed Description of Achievability Schemes in Fig.~\ref{fig:SRP:comp}}\label{app:SRP:Schemes}

In the following, we describe $(\RB{1},\RB{2})$ rate regions of each suboptimal achievability scheme used for the numerical evaluation in \SecRef{sec:SRP:Eval}.

%%-- Back up
%\noindent$\bullet$ {\bf \SRPschemeSEVEN:} The rate regions can be described by \PropRef{prop:SRP:sim-innerV2} with the variables $\{\snumSX{k}{l}\,(l\!=\!1,2,3) : \text{for all } k\in\{1,2\}\}$ hardwired to $0$.
%
%\noindent$\bullet$ {\bf \SRPschemeSIX:} The rate regions can be described by \PropRef{prop:SRP:sim-innerV2} such that in addition to the variables hardwired to $0$ at {\bf \SRPschemeSEVEN}, if the variables $\{\snumPM{k},\snumAM{k},\snumRC{k} : \text{for all } k\in\{1,2\}\}$ and $\{\wnumRC{h}:h\in\{s,r\}\}$ are further hardwired to $0$ as well.

%%-- back up
%\noindent$\bullet$ {\bf \SRPschemeFIVE:} The rate regions can be described by \PropRef{prop:SRP:sim-innerV2} such that in addition to the variables hardwired to $0$ at {\bf \SRPschemeSIX}, if the variables $\{\snumCX{l}\,(l\!=\!1,\cdots,8)\}$ and $\{\wnumOX{h}{},\wnumCX{h}{} : h\in\{s,r\}\}$ are further hardwired to $0$ as well. Namely, we completely shut down all the variables dealing with packet mixtures. After such hardwirings, \PropRef{prop:SRP:sim-innerV2} is further reduced to the following form:

%%-- Instead
\noindent$\bullet$ {\bf \SRPschemeFIVE:} The rate regions can be described by \PropRef{prop:SRP:sim-inner}, if the variables $\{\snumPM{k},\snumAM{k},\snumRC{k} : \text{for all } k\in\{1,2\}\}$, $\{\snumCX{l}\,(l\!=\!1,\cdots,8)\}$, $\{\rnumRC, \rnumOX{},\rnumCX{}\}$ are all hardwired to $0$. Namely, we completely shut down all the variables dealing with cross-packet-mixtures. After such hardwirings, \PropRef{prop:SRP:sim-inner} is further reduced to the following form:
\begin{align*}
1 & \geq \sum_{k\in\{1,2\}}\left( \snumUC{k} + \snumDX{k}{} + \rnumUC{k} + \rnumDX{[k]}{}\right),
\end{align*}
and consider any $i,j\in(1,2)$ satisfying $i\neq j$. For each $(i,j)$ pair (out of the two choices $(1,2)$ and $(2,1)$),
\begin{align*}
\RB{i} & \geq \snumUC{i}\cdot\pSRPsimT{s}{d_i,d_j,r}, \\
\snumUC{i}\cdot\prOa{s}{d_i d_j}{r} & \geq \rnumUC{i}\cdot\pSRPsimT{r}{d_i,d_j}, \\
\snumUC{i}\cdot\prOd{s}{d_i}{d_j}{r} & \geq \snumDX{i}{}\cdot\pSRPsimT{s}{d_i,r}, \\
\snumUC{i}\cdot\prOa{s}{d_i}{d_j r} + \snumDX{i}{}\cdot\pSRPsimT{s}{\overline{d_i}r} & + \rnumUC{i}\cdot\prOa{r}{d_i}{d_j} \geq \rnumDX{[i]}{}\cdot\pSRPsimT{r}{d_i}, \\
\Big( \snumUC{i} + \snumDX{i}{} \Big)\cdot\pSRPsimT{s}{d_i} + \Big(\rnumUC{i} & + \rnumDX{[i]}{}\Big)\cdot\pSRPsimT{r}{d_i} \geq \RB{i}.
\end{align*}

%{\color{red} Note that the above rate region description only works for the strong-relaying scenario in \DefRef{def:SRP:strong-relaying}, as it is reduced from \PropRef{prop:SRP:sim-inner}. Therefore, as discussed in \SecRef{sec:SRP:disc}, the largest marginal rates are simply computed by $\left(\frac{\pSRPsimT{s}{d_1,r}\cdot\pSRPsimT{r}{d_1}}{\pSRPsimT{s}{\overline{d_1}r}+\pSRPsimT{r}{d_1}},0\right)$ and $\left(0,\frac{\pSRPsimT{s}{d_2,r}\cdot\pSRPsimT{r}{d_2}}{\pSRPsimT{s}{\overline{d_2}r}+\pSRPsimT{r}{d_2}}\right)$.}

\noindent$\bullet$ {\bf \SRPschemeFOUR:} This scheme requires that all the packets go through $r$, and then $r$ performs $2$-user broadcast channel NC. The corresponding rate regions can be described as follows:
\begin{align*}
\frac{\RB{1}}{\pSRPsimT{r}{d_1}} + \frac{\RB{2}}{\pSRPsimT{r}{d_1,d_2}} & \leq 1 - \frac{\RB{1}+\RB{2}}{\pSRPsimT{s}{r}}, \\
\frac{\RB{1}}{\pSRPsimT{r}{d_1,d_2}} + \frac{\RB{2}}{\pSRPsimT{r}{d_2}} & \leq 1 - \frac{\RB{1}+\RB{2}}{\pSRPsimT{s}{r}}.
\end{align*}

\noindent$\bullet$ {\bf \SRPschemeTHREE:} This scheme requires that all the packets go through $r$ as well, but $r$ performs uncoded routing for the final delivery. The corresponding rate regions can be described as follows:
\begin{align*}
\frac{\RB{1}}{\pSRPsimT{r}{d_1}} + \frac{\RB{2}}{\pSRPsimT{r}{d_2}} \leq 1 - \frac{\RB{1}+\RB{2}}{\pSRPsimT{s}{r}}.
\end{align*}

\noindent$\bullet$ {\bf \SRPschemeTWO:} This scheme completely ignores the relay $r$ in the middle, and $s$ just performs $2$-user broadcast channel LNC of \cite{GeorgiadisTassiulas:NetCod09}. The corresponding rate regions can be described as follows:
\begin{align*}
\frac{\RB{1}}{\pSRPsimT{s}{d_1}} + \frac{\RB{2}}{\pSRPsimT{s}{d_1,d_2}} \leq 1, \\
\frac{\RB{1}}{\pSRPsimT{s}{d_1,d_2}} + \frac{\RB{2}}{\pSRPsimT{s}{d_2}} \leq 1.
\end{align*}

\noindent$\bullet$ {\bf \SRPschemeONE:} This scheme completely ignores the relay $r$ in the middle, and $s$ just performs uncoded routing. The corresponding rate regions can be described as follows:
\begin{align*}
\frac{\RB{1}}{\pSRPsimT{s}{d_1}} + \frac{\RB{2}}{\pSRPsimT{s}{d_2}} \leq 1.
\end{align*}

\bibliographystyle{IEEEtran}

\end{document}